\documentclass[preprint]{aa}
\usepackage{txfonts}
\usepackage{graphicx,subfigure}
\usepackage{epsfig}
\usepackage{epsf}
\usepackage{lscape}
\usepackage{xcolor}
\usepackage{natbib}
\usepackage{rotating}
\usepackage{tabularx}
\usepackage{varwidth}
\newcommand{\mic}{\,{\rm \mu m} }

\begin{document}
\title{Toward a better understanding of the mid-infrared emission in
  the LMC}

\author{D. Paradis \inst{1,2} 
  \and
  C. Mény \inst{1,2}
  \and
  K. Demyk \inst{1,2}
  \and
  A. Noriega-Crespo \inst{3}
  \and
I. Ristorcelli \inst{1,2}}  
\institute{IRAP, Université de Toulouse, CNRS, UPS, 9 Av. du Colonel Roche, BP 44346, F-31028, Toulouse, cedex
4, France 
\and
Space Telescope Science Institute, 3700 San Martin Drive, Baltimore,
MD 21218, USA}

\authorrunning{Paradis et al.}
\titlerunning{}
\date{}
\abstract
{The scarcity of high signal-to-noise spectroscopic data of the
  in the interstellar medium
  between 20 to 100 $\mic$ has led to the development of several dust
  models with distinct dust properties that are poorly constrained in this broad
  wavelength range. Some of them require the presence of graphites
  whereas others consider small amorphous or small aromatic
  carbon grains, with various dust sizes. }
  {In this paper we  aim to constrain for the first time the dust emission in
    the mid-to-far infrared domain, in the Large Magellanic Cloud (LMC), with the use of the Spitzer IRS and
  MIPS SED data, combined with Herschel data. We also consider ultraviolet (UV) extinction
predictions derived from modeling.} 
{We selected 10 regions observed as part of the SAGE-Spec
  program (PI: F. Kemper), to probe dust properties in various environments
  (diffuse, molecular and ionized regions). All data were smoothed to
  the 40$^{\prime \prime}$ angular resolution before
  extracting the dust emission spectra and photometric data. The
  Spectral Energy Distributions (SEDs)  were
  modeled with dust models available in the DustEM package, using the
  standard Mathis radiation field, as well as three additional
  radiation fields, with stellar clusters ages ranging from 4 Myr to 600 Myr.}
{Previous analyses of molecular clouds in the LMC have 
reproduced reasonably well the SEDs of the different phases of the clouds 
constructed from near- to far-infrared photometric data, using the DustEM models. 
However, it is only by using spectroscopic data and
  by changing the dust abundances in comparison with our Galaxy, that
the present study brings new constraints on the small grain component. 
  Standard dust models used to reproduce the Galactic diffuse medium are
  clearly not able to reproduce the dust emission in the mid-infrared
  wavelength domain. This analysis evidences the need of adjusting
  parameters describing the dust size distribution and shows a
  clear distinct behavior according to the type of environments. In addition, whereas the
  small grain emission always seems to be negligible at long
  wavelengths in our Galaxy,
  the contribution of this small dust component could be more important
  than expected, in the submillimeter-millimeter range, in the LMC averaged SED.}  
{Properties of the small dust component of the LMC are clearly
  different from those of our Galaxy. Its abundance, significantly
  enhanced, could be the result of large grains shattering due to
  strong shocks or turbulence. In addition, this grain component in the LMC 
systematically shows smaller grain size in the ionized regions compared to the diffuse
medium. Predictions of extinction curves show significantly distinct
behaviors depending on the dust models but also from one region to
another. Comparison of model predictions with the LMC mean extinction curve shows 
that no models gives satisfactory agreement using the Mathis radiation field while 
using a harder radiation field tends to improve the agreement. }

\keywords{ISM:dust, extinction - Infrared: ISM - Submillimeter: ISM}

\maketitle
\section{Introduction}
Nowadays it is well accepted that dust emission in the 
interstellar medium (ISM) can be divided into
three domains: the near-infrared (NIR, from 0.7 to 5 $\mic$) to
mid-infrared (MIR, from 5 to 40 $\mic$) dominated by Polycyclic Aromatic
Hydrocarbons (PAH) emission in the 3-20 $\mu$m range, the MIR to far-infrared (FIR, from 40 to 350 $\mic$) dominated by emission from very small particles/grains (potentially carbon grains,
denoted as VSGs) in the $\sim$ 20 - 100 $\mu$m range and the FIR to
submillimeter/millimeter ( submm/mm) emission dominated by big grains (silicates
or a mixture of carbon/silicate grains, denoted as BGs) above $\sim$ 100-200 $\mu$m range. The understanding of the
NIR-to-MIR regime has experienced a significant progress over the past 25 years with
the ISO spectroscopy data \citep{Trewhella00, Boulanger00} and then the
Spitzer data \citep[][for instance]{Meixner06,  Bernard08, Paradis11a, Tibbs11}. Thanks to the
Planck-Herschel missions, the FIR/submm/mm
regime has been extensively studied in the past decade for
galactic studies \citep[see for instance][]{Juvela11,
  Paradis12, Paradis14, Planck14XIV, Planck14XVII, Planck14XI, Meisner15, Juvela18} and extragalactic studies \citep[see
for instance][]{Planck11XVII, Galliano11, Dale12, Galametz12, Chastenet17}.
Nevertheless the  origin of the emission arising in the FIR/submm is still
not well constrained in terms of grain composition/size/shape. However, the MIR-to-FIR domain has always
suffered from a lack of data. Most of the data in this
wavelength domain come from photometric data in a few bands. For
instance, the Spitzer telescope provided
data at 24 $\mic$ and 70 $\mic$, very
similar to photometric data observed with IRAS (25 and 60
$\mic$). PACS spectroscopic data do not provide data below 
55 $\mic$. They do not give any information on the emission of dust between 20 and 
55 $\mic$, spectral range however crucial to constrain the very small particles/grains. 
Spitzer spectroscopic data were mainly centered on the
NIR-to-MIR emission. Only a very few programs focused on the
MIR-to-FIR domain. This was the case of the SAGE-spec Spitzer Legacy Program
\citep{Kemper10}, a spectroscopic follow-up to the SAGE-LMC
photometric survey of the Large Magellanic Cloud \citep{Meixner06} carried out with
the Spitzer Space Telescope. Extended regions in the diffuse medium and
HII regions (HII) were observed with the use of the
Infrared Spectrometer (IRS) staring mode from $\sim$5 to 38 $\mic$, as
well as the MIPS-SED mode, from 52 to 97 $\mic$. These data are crucial to bring constraints
on the dust at the origin of the MIR-to-FIR emission.  
Some IRS observations were also available for the Dwarf Galaxy Survey (DGS)
sample. \citet{RemyRuyer15} analyzed a sample of 98 low-metallicity
galaxies (from the Dwarf Galaxy and KINGFISH surveys) and modeled
their global SEDs from the NIR to the submm using the dust model described in
\citet{Galliano11}. They merged different data sets including IRS
spectra when available. For 11 sources, they included an additional
modified blackbody component at MIR-to-FIR wavelengths with temperatures in
the range 80 K -300K, that significantly improve the modeling of the
entire galaxies. They
attributed this component to hot HII regions added to
the total emission of the galaxies, even though, in this kind of regions
one would expect to observe an increase of emission over the
entire SED and not specifically in the MIR-to-FIR range.

The advantage of studying the LMC is that the IRS observations
spatially resolve different environments of the Galaxy which was not
possible in the DGS survey.
The LMC is one of the closest galaxy, at a distance of $\sim$50 kpc,
with a complex structure, including HI filaments, arcs, holes and shells
\citep{Kim98}. The small scale of the neutral atomic ISM is dominated by a turbulent and fractal
structure due to the dynamical feedback of the star
formation processes. The large scale of the HI disk evidences a symmetric and
rotational field. Most of LMC SED studies were based on the results
derived using a single dust model \citep[see for instance][]{Bernard08, Paradis09,
  Paradis11a, Galliano11, Galametz13, Stephens14, Roman-Duval17, Chastenet19}. Only
\citet{Chastenet17} and \citet{Paradis19} fitted the SED of
the LMC using two or more dust models from the DustEM package \citep{Compiegne11}. More
recently, \citet{Chastenet21} performed a comparative study on M101, to
derive the dust mass and show the dependence of the results with dust
models. None of the previous LMC studies had constraints in the
MIR-to-FIR range, and more specifically between 25 and 70
$\mic$. In addition, except \citet{Paradis11a} who investigated the
impact of another radiation field (RF) template, all analysis considered
the standard Mathis RF, adapted to the Milky-Way (MW) interstellar
RF, and never made any attempts to
modify its shape. The LMC studies showed an abundance of dust half that of the MW, 
which is explained by the low metallicity of the LMC in comparison with the MW. Its dust emission spectrum is significantly flatter in
the submm than in the MW \citep{Planck11XVII}. PAHs seem to be enhanced in molecular clouds,
as well as in the old stellar bar, but are potentially destroyed in
regions with high RF \citep{Paradis11a, Paradis19}. The PAH production by fragmentation could also
have a link with the metallicity of the Galaxy. On the other hand,
very small grains could be formed in HII regions \citep{Paradis19}. Large VSG potentially produced by the erosion of large grains could be responsible for the 70 $\mic$ excess
evidenced in the Magellanic Clouds \citep{Paradis09}.%The presence of an additional gas component
%undetected with gas tracers , also called ``dark gas'' has been
%pointed out in the LMC \citep{Bernard08}.

In this study, using the combination of
photometric and spectroscopic data in the NIR to FIR domain, we fit the spectral shape of dust
emission in different environments of the LMC (2 diffuse, 2 molecular
and 6 HII regions) with four differents dust models \citep{Jones13,
  Compiegne11, Draine07, Desert90}. The modeling has been done
with four interstellar radiation fields templates, and
allowing different parameters, such as the dust abundances and the dust
size distribution, to vary. We also analyze the extinction
curves produced from dust models.

After the description of the data sets (Section \ref{sec_obs}), we present
the method to extract the SED in each region in Section
\ref{sec_sed_construction}. In Sections \ref{sec_regions} and
\ref{sec_models} we give a short
description of the studied targets and of the dust
models we used as part of the DustEM package, then we describe the fitting results in
Section \ref{sec_fitting}. After the discussion in Section
\ref{sec_discussion}, we provide a summary of the results in Section \ref{sec_conclusions}. 
%In the LMC, the 
%It is an alternative way to a change in the very small grain dust size
%distribution. Their results indicate that in most of the cases, the
%modelling of the low-metallicity galaxies is significantly different than what is observed in our
%Galaxy
\section{Observations}
\label{sec_obs}
\subsection{Spitzer data}
\subsubsection{IRS staring and MIPS SED mode}
Spectroscopic data were obtained as part of the SAGE-Spec Spitzer
Legacy program (PID: 40159), a spectroscopic follow-up to the SAGE-LMC photometric
survey of the Large Magellanic Cloud. Extended regions (atomic/molecular and
HII regions) were observed in the IRS staring (between 5 $\mic$ and 38 $\mic$) and MIPS SED modes
(between 52 $\mic$ and 97 $\mic$). We use the last
  data release available, produced by the SAGE-Spec team. The
reduction of the data has been done in the past by the team using the standard pipeline
data as produced by the Spitzer Science Center. The individual
observations have been combined
into a spectral cube using CUBISM \citep{Smith07}. The MIPS SED
extended source observations have been reduced using the MIPS DAT v3.10
\citep{Gordon05}, and calibrated according to the prescription of
\citet{Lu08}. 

\subsubsection{MIPS photometry}
The SAGE-LMC survey \citep{Meixner06} observed the entire LMC using
IRAC \citep{Fazio04} and MIPS \citep{Rieke04}
instruments. In this work we combine the
spectroscopic data with MIPS photometry at 70 and 160 $\mic$, at
18$^{\prime \prime}$
and 40$^{\prime \prime}$ angular resolution. 

\subsection{Herschel data}
To trace dust in the far-IR and the submm we use the Herschel PACS (160
$\mic$, at 13$^{\prime \prime}$ angular resolution) and the SPIRE (250,
350 and 500 $\mic$, at 18$^{\prime \prime}$, 25$^{\prime \prime}$ and
36$^{\prime \prime}$ angular resolution, respectively) data, as part of the
Heritage program \citep{Meixner10}. We use the last version of the
data available on the ESA Herschel Science Archive\footnote{archives.esac.esa.int/hsa/whsa}. 

\subsection{Gas tracers}
\subsubsection{Atomic Hydrogen}
We use the \citet{Kim03} 21-cm map (spatial
resolution of 1$^{\prime}$) to trace the atomic gas, integrated in the
velocity range 190 km s$^{-1}$$<\rm V_{LSR}<386$ km s$^{-1}$. This map has been
done by combining interferometric data from the Australia Telescope Compact Array
(ATCA; 1$^{\prime}$), and the Parkes antenna \citep[15.3$^{\prime}$;][]{Staveleysmith03}. 
To derive the HI column density we apply the standard conversion
factor $X\rm_{HI}$, equal to $1.82\times 10^{18}$
$\rm H/cm^2/(K\,km\,s^{-1}),$ \citep{Spitzer78, Lee15} such as:
\begin{equation}
  \label{eq_hi}
  N_{\rm H}^{\rm HI}=X_{\rm HI}W_{\rm HI}
  \end{equation}
  with $W_{\rm HI}$ the integrated intensity map.
  
\subsubsection{Carbon monoxide}
To trace the molecular gas, we use the 22-m Mopra telescope data of
the Australia Telescope National Facility, at an angular resolution of
$45^{\prime \prime}$. This survey of the LMC has been done as part of
the MAGMA project \citep{Wong11}.
The integrated intensity map ($W\rm_{CO}$) is converted to molecular
column densities using the relation:
\begin{equation}
  \label{eq_co}
N_{\rm H_2}^{\rm CO}=X_{\rm CO}W_{\rm CO}
,\end{equation}
with $X\rm_{CO}$ being the CO-to-H$_2$ conversion factor. We use an
$X\rm_{CO}$ value equals to 4 $\times 10^{20}$ $\rm
H/cm^2/(K\,km\,s^{-1})$, as in \citet{Paradis19}. This value is a good
compromise taking into account the large dispersion of the $X\rm_{CO}$ values derived by different authors  
\citep[see for instance][]{Hughes10, Leroy11, Roman-Duval14}.

\subsubsection{Ionized Hydrogen}
The ionized gas is usually traced by the H$\alpha$ recombination
line. We therefore use the Southern H-Alpha Sky Survey Atlas
\citep[SHASSA,][]{Gaustad01} at an angular resolution of 48$^{\prime
  \prime}$. The $\rm H^+$ column density can be derived, assuming a
constant electron density $n_e$ along each line of sight, by applying
the relation \citep{Lagache99}:
\begin{equation}
\label{eq_halpha}
\frac{N{\rm_H^{H^+}}}{{\rm H\,cm^{-2}}}=1.37 \times 10^{18}
\frac{I_{\rm {H
    \alpha}}}{\rm R} \frac{n\rm_e}{\rm cm^{-3}}
,\end{equation}
Following \citet{Dickinson03}, 1 Rayleigh=2.25 pc cm$^{-6}$ for T$\rm
_e$=8000 K.
The electron density has
been derived in \citet{Paradis11a} for different regimes over the
entire LMC, and more recently in \citet{Paradis19} for the molecular clouds of
the LMC. The electron density can be as high as 3.98 cm$^{-3}$ for bright HII
regions, as low as 0.055 cm$^{-3}$
for the diffuse ionized gas of the LMC, and close to 1 cm$^{-3}$ for
the molecular clouds. 
A value of 1.52 cm$^{-3}$ has been
determined for typical HII regions. We adopt the appropriate electron
density depending on the H$\alpha$ brigthness, following \citet{Paradis11a} (see Sect. \ref{sec_regions}).
%Values are reported in Table \ref{table_regions}. 
% SSDR7 NCO=0 et NH+=0
% DEML 3.98
%SSDR1 et 5 NH+ negligeable. $^{\star}$
\begin{table}
  \caption{Characteristics of the studied regions : type of environment
    (column 2), Hydrogen column density (H/cm$^2$) in each phase of the gas
    (columns 3 to 5), and electron density in cm$^{-3}$(column 6). \label{table_regions}}
  \begin{center}
    \begin{tabular}{lccccc}
\hline
      \hline
      Region & Type & $N_H^{HI}$ & $N_H^{CO}$ & $N_H^{H \alpha}$ &
                                                                   $n_e$\tablefootmark{$\star$} \\
      &&($10^{21}$)&($10^{21}$)&($10^{21}$)& \\
      SSDR1 & Molecular &4.8& 7.3& 0.50& 1.52 \\
      SSDR5 & Molecular & 1.6&3.0&0.23 & 3.98\\
      \hline
      SSDR7 & Diffuse & 3.9&-&0.035&1.52 \\
      SSDR9 & Diffuse & 4.7&-&- &-\\
      \hline
      SSDR8 & Ionized& 1.7&- &1.7 & 3.98\\
      SSDR10 & Ionized & 0.68 & - &1.9 & 3.98\\
      DEML10 & Ionized & 3.1& 0.36& 1.9& 3.98\\
      DEML34 & Ionized & 2.9& 6.9& 6.8 &3.98 \\
      DEML86 & Ionized & 2.2 & 3.3 & 4.0& 3.98\\
      DEML323 & Ionized& 3.5& 0.20& 2.4 & 3.98\\
      \hline
      \end{tabular}
%\tablefoot{\textbf\tablefoottext{$\star$}{following
 %   \citet{Paradis11a}.}}
 %   $^{\star}$ following \citet{Paradis11a}
 \tablefoot{{$\star$}{following \citet{Paradis11a}.}}

    \end{center}
 \end{table} 

\subsection{Convolution}
All Spitzer and Herschel data are smoothed to a common angular resolution (40$^{\prime
  \prime}$) and reprojected on the same grid. The smoothing is
performed using a Gaussian kernel with $\theta_{\rm FWHM}$ $= \sqrt{ \left(
    (40^{\prime \prime})^2-(\theta{_{\rm FWHM}^{\rm d}})^2 \right) }$ 
, with $\theta{\rm_{FWHM}^{\rm d}}$ the original
resolution of the maps. For the IRS and MIPS SED data, the smoothing
is done on each plane of the cubes.    
For the gas tracers, for which the angular resolution is
slightly larger than 40$^{\prime \prime}$ (from 45$^{\prime \prime}$
to 1$^{\prime}$), the integrated intensity maps are only reprojected on the
same grid as the infrared data. The pixelization is 
therefore slightly oversampled but the impact on the gas estimates is very limited. 
Since the gas tracers are only used to determine the indicative gas column 
densities, the shape of the SED is not affected at all.

\section{SED construction}
\label{sec_sed_construction}
For the 24 extended regions observed as part of the SAGE-Spec program
we extract the spectral energy distribution by computing the median
brightness at each wavelength in a circular region enlarged by 2 pixels around the
central position of the source. We select 10 regions (SSDR1, SSDR5,
SSDR7, SSDR8, SSDR9, SSDR10, DEML10, DEML34, DEML86 and DEML323) for which the
IRS SS and LL spectra overlaid well at the same wavelengths and with
reasonable dispersion in the data. A short description of the regions
is given in Section \ref{sec_regions}.

We then compute the total column density for each region using
eq. \ref{eq_hi}, \ref{eq_co} and \ref{eq_halpha}, and we
normalize each SED to an HI column density $N_H$  equal to 1$\times10^{20}$ $\rm
H/cm^2$.

We consider absolute calibration uncertainties of 20$\%$ for the IRS
spectra (Protostars and Planets v.5), 15$\%$ for the MIPS SED data (MIPS Instrument Handbook\footnote{https://irsa.ipac.caltech.edu/data/SPITZER/docs/mips/mipsinstrumenthandbook/44/}), 10$\%$ for the MIPS photometric
data (MIPS Instrument
Handbook\footnote{https://irsa.ipac.caltech.edu/data/SPITZER/docs/mips/mipsinstrumenthandbook/42/}),
and 7$\%$ for the Herschel data \citep[][for PACS 160 $\mic$, and observer
manual v2.4 for SPIRE]{Balog14}. However, for SED modeling (see
Section \ref{sec_fitting_standard}) it is crucial to increase the weight of the far-IR to
submm data to
ensure an equal balance between the large amount of spectroscopic data
and the low number of photometric data in each SED. Indeed, 
the SEDs account for $\sim$400 spectrocopic data and 5 photometric
data (MIPS 70 data are only used to rescale photometric data, see 
below). We first tested the fitting procedure without 
changing the weight of the data. 
The results were not acceptable since the ${\chi}^2$ of the fits
were very good  but the SPIRE data were not well-reproduced. We therefore applied different
 weights in the photometric data to check the reliability of the fitting
 results. We obtained satisfactory results by increasing the weight of the SPIRE data by a factor of 50 (ie. a factor
of 150 for the three SPIRE data, which corresponds to a similar weight
between the spectral data and the photometric fluxes). In this way we ensure to have a good representation of the SED
over the entire wavelength range.

The SEDs have been normalized by computing the ratio between the integrated flux in the
MIPS-SED band and the MIPS 70 $\mic$
photometric data. The photometric data (from 70 to 500 $\mic$) have been
rescaled by mutliplying them by this factor. We can see, in a
few cases (mainly DEML34 and DEML86), a significant difference
between the MIPS and PACS 160 $\mic$ data. This discrepancy is not the
  result of non-linearity effect between MIPS and PACS since this
  effect appears for brightness above 50 MJy/sr at 160 $\mic$, significantly
  higher than our values. Whereas recent analyses of the original
Heritage maps \citep{Clark21} evidence missing dust in the periphery of the Magellanic
Clouds (mainly at shorter wavelengths), Herschel PACS
maps seem to over-estimate the brightness of
large-scale emission by 20-30$\%$, compared to absolutely calibrated all-sky
survey data. However, when the disagreement between
MIPS and PACS 160 $\mic$ is visible, our fits tend to reproduce the
MIPS 160 $\mic$
data. The results are therefore not affected. 
%Either we add a distinct weight associated to mid-IR to far-IR data and to far-IR to submm
%data, either we increase or decrease the error bars. We decide to
%arbitrary decrease the SPIRE calibration uncertainty from 7$\%$ to
%1$\%$. 
\section{Target description}
\label{sec_regions}
The location of each region is
given in \citet{Kemper10}. In Table \ref{table_regions} we present
the type of environment, i.e.``diffuse'', ``molecular'' or ``ionized''; and the Hydrogen
column density in each phase of the gas. The adopted value for the
electron density is also given in the table, in agreement with considerations
presented in \citet{Paradis11a}. 
While the SSDR designation for ``SAGE-Spec Diffuse Regions'' should
 correspond to diffuse regions (atomic and molecular) it appears that
 some of them (SSDR8 and SSDR10) represent ionized regions. Our
 sample has only two diffuse regions (SSDR7 and SSDR9), two molecular
 regions (SSDR1 and SSDR5) and 6 HII regions (SSDR8, SSDR10 and the four DEML regions).
 
 \section{Dust models}
 \label{sec_models}
We use the DustEM Wrapper package to model the
SEDs of the different regions, using the following dust models :
THEMIS \citep[][hereafter AJ13]{Jones13}, \citet{Compiegne11}
(hereafter MC11), \citet{Draine07} (hereafter DL07) and an improved
version of the \citet{Desert90} model (hereafter DBP90). Below is
a very brief description of the models (we encourage the readers to look at the original
papers corresponding to each model): \\ 
- The THEMIS (AJ13) model considers two dust components covered by an aromatic
mantle: a population of
carbonaceous grains consisting of large grains ($\it a$$<$200 nm) of amorphous and aliphatic nature (a-C:H), 
and smaller grains ( $\it a$$<$20 nm) of more aromatic nature (a-C), and
a population of large amorphous silicate grains ($\it a$$<$200 nm) containing nanometer
scale inclusions of FeS (large a-Sil). \\
- MC11 includes  PAHs (neutral and ionized), small ($\it a$$<$10 nm)
and large ($\it a$$>$10 nm) amorphous carbon grains (SamC and LamC), and
large amorphous silicates (aSil). \\
- The silicate-graphite-PAH model (DL07) assumes a mixture of carbonaceous
and amorphous silicates grains, including different amounts of
PAH material (neutral and ionized). \\
- In the adapted version of the \citet{Desert90} model we use, we substitute the original PAH component which suffers from
an incomplete description of the PAHs bands, by the
neutral PAH component taken from MC11. The small (VSG) and big grain
(BG) components are made of carbon and silicates grains.

\section{Fitting results}
\label{sec_fitting}
We perform modeling of the SED of the different regions by first allowing the standard parameters to vary,
then changing the parameters of the carbon dust size distribution. In
a second step we inject different radiation fields in the
modeling. The dust over gas mass ratio is discussed according to 
the different models.

\subsection{Standard free parameters}
\label{sec_fitting_standard}
We first fit the observations with the different dust models, allowing standard
parameters to vary: the abundances of the different dust components
($Y_{component}$), the intensity of the Solar neigborhood radiation field
\citep[$X_{ISRF}$,][]{Mathis83} and the intensity of a NIR-continuum
($I_{NIR Cont}$). These standard parameters (except the NIR-continuum)
were constrained from Galactic observations at high latitudes. 
Figure
\ref{fig_all_noslope} presents the results obtained with the different
models for each SED and for the 10 selected regions. Table \ref{table_chi2_all} gives the values of the reduced
$\chi^2$ for each SED as well as the values of $\chi^2$ divided by the
number of data in three ranges of wavelengths ([5-20]; ]20-97] and
[160-500]).  We can first note that, for all dust models, using the standard
parameters does not give a satisfactory modeling. For instance AJ13
shows an important underestimate of the model in the MIR domain
(and more specifically between 15 and 30 $\mic$) and an overestimate in the MIPS SED range (especially
in HII regions). These discrepancies result in high values
of reduced $\chi^2$ with AJ13 compared to other models. Even if
the other models are significantly better than AJ13, they all show an
imperfect description of the data in the MIR domain. For instance MC11 evidences some underestimate of the model
in the 20-50 $\mic$ range of diffuse regions, even if less important than with
AJ13. For the HII regions the SEDs are reasonably well represented (as
opposed to AJ13). DL07 shows the same behavior as AJ13 for the HII
regions in particular, but significantly less pronounced than with
AJ13. Concerning DBP90, either the 20-50 $\mic$ range is underestimate, or
the 60-100 $\mic$ range is overestimate.

Whatever the model we consider, we clearly see that the MIR-to-FIR
domain is not well described, showing the low-quality of modeling
using the standard parameters. Since the standard parameters are not
suitable to reproduce LMC observations, in the following section we try to
improve the various modeling by allowing the parameters of the size
distribution of the dust (responsible for
the MIR-to-FIR emission) to vary.

\begin{figure*}
  \begin{center}
\subfigure{\includegraphics[width=18cm]{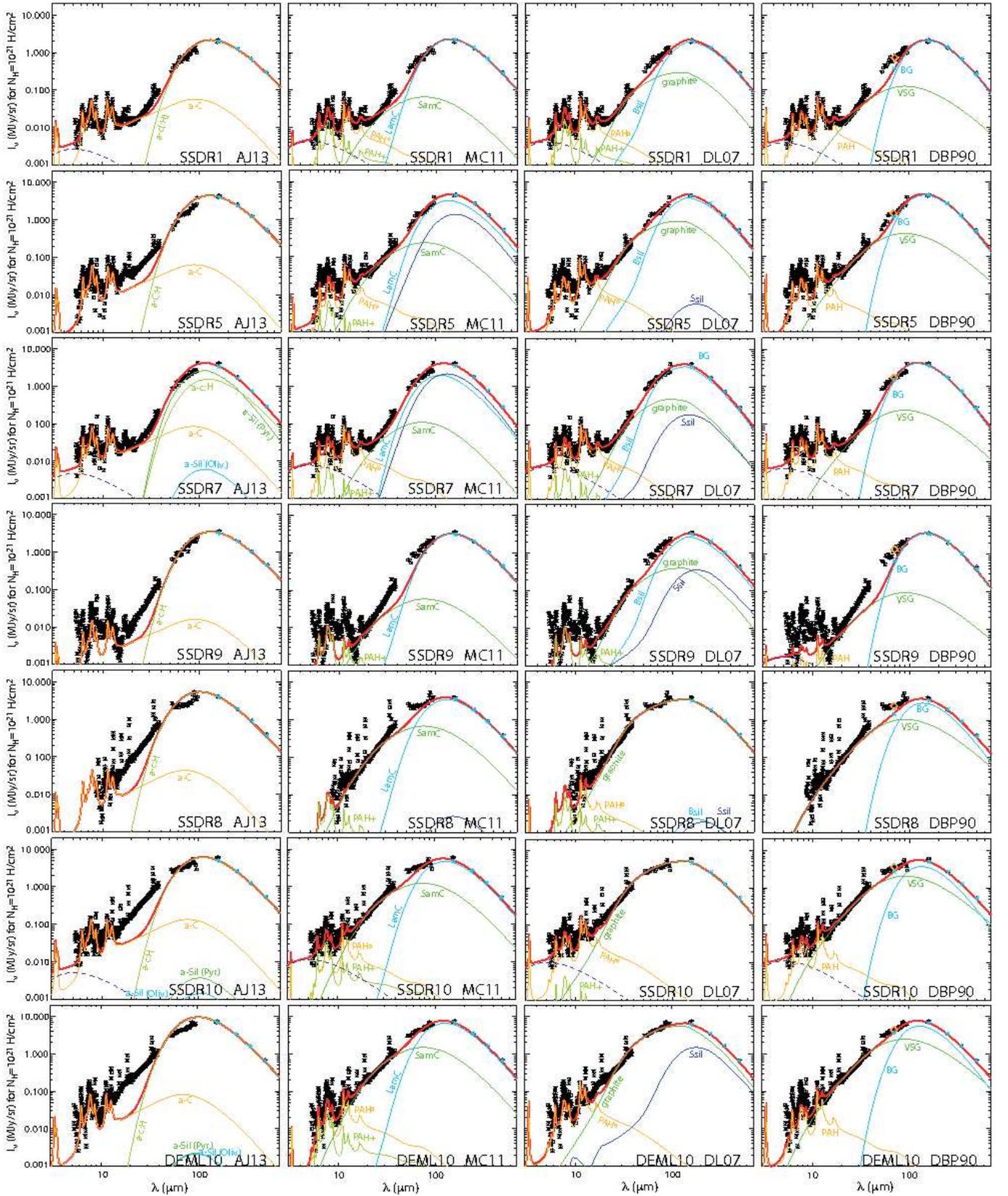}}
\caption{Modeling of the SEDs of the ten regions with different dust models and free
  standard parameters ($\rm X_{ISRF}$ and dust abundances), using the
  Mathis RF. The
  observations (Spitzer IRS SS and LL, MIPS SED, MIPS 160
  $\mic$, Herschel Photometric PACS 160 $\mic$ and SPIRE 250 $\mic$, 350 $\mic$
  and 500 $\mic$ data) are shown in black. The total modeled SED is shown
  as a red line. The other colored lines correspond to
  the different dust components of the models. The dashed line
  represents the additional NIR continuum. Blue asterisks show
  the color-corrected brightness derived from the models. The orange
  diamonds that are visible in the DBP90 panels show the MIPS 70 $\mic$
photometric data normalized to the integrated flux in the
MIPS-SED band. Each column shows the fit using different dust models
(from left to right: AJ13, MC11, DL07 and DBP90). Each row presents a
different region.  
The figure continues on the next page. \label{fig_all_noslope}}
\end{center}
\end{figure*}

\addtocounter{figure}{-1}
\begin{figure*}
  \addtocounter{subfigure}{1}
  \begin{center}
\subfigure{\includegraphics[width=18cm]{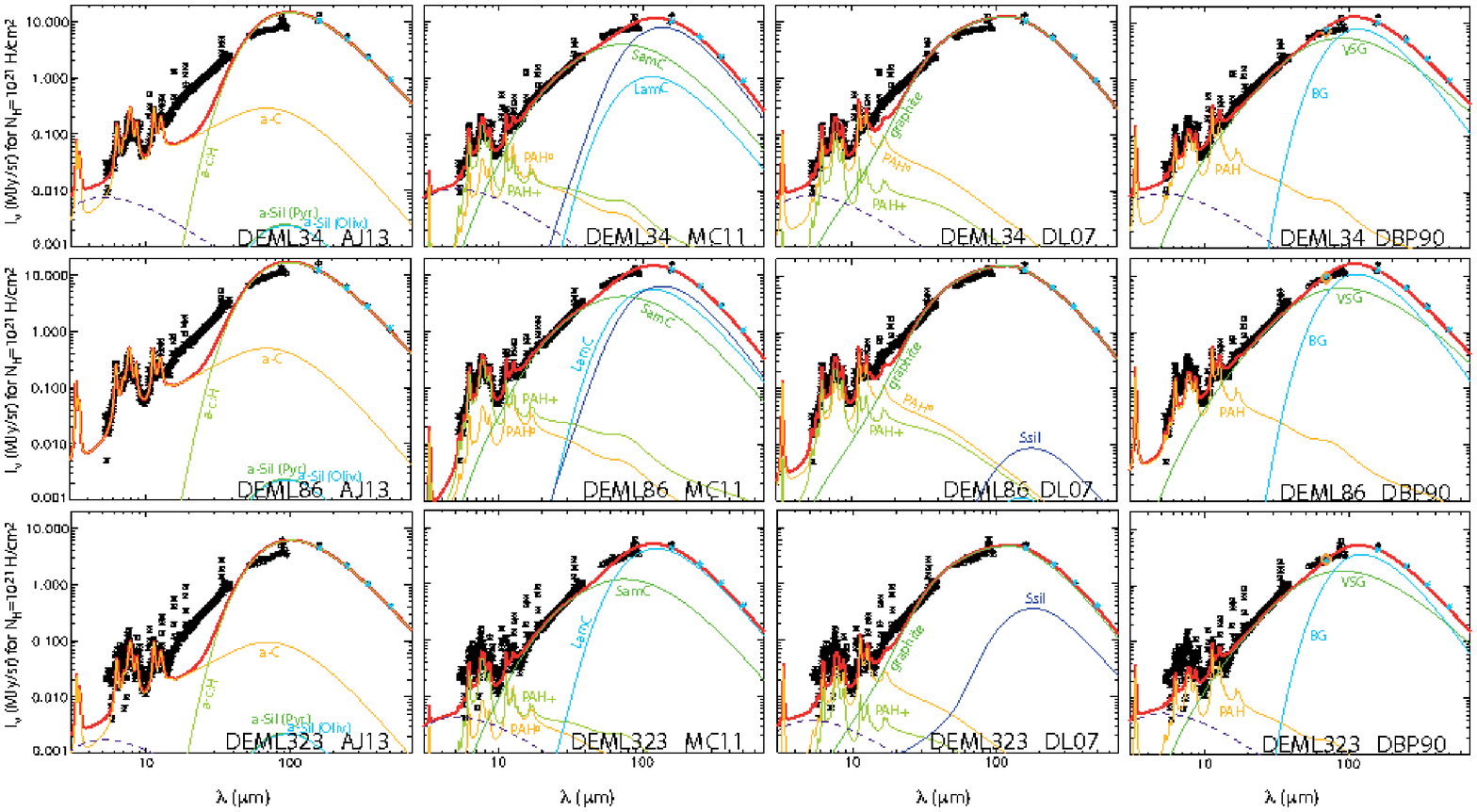}}
\end{center}
\caption{Continued}
\end{figure*}

%\begin{sidewaystable}%[!h]
  \begin{table*}
\caption{Best fit reduced $\chi^2$ obtained for the modeling of the
  full SED of the ten regions and for the four adopted dust
  models (between 5 $\mic$ and 500 $\mic$), using the Mathis
  RF. \label{table_chi2_all}}
\begin{center}
\resizebox{\textwidth}{!}{%
\begin{tabular}{lcccccccccccc}
\hline
  \hline
 Model  & \multicolumn{4}{c}{AJ13} &\multicolumn{4}{c }{AJ13 ($\alpha$) } &\multicolumn{4}{c}{AJ13 ($\alpha, a_{min}, a_{max}$)}\\
 \hline
  & $\chi^2/dof$ &\multicolumn{3}{c}{$\chi^2/N_{data}$}  &$\chi^2/dof$  & 
                  \multicolumn{3}{c}{$\chi^2/N_{data}$} & $\chi^2/dof$ &\multicolumn{3}{c}{$\chi^2/N_{data}$} \\
  \hline
  Region & $[5-500]$ & $[5-20] $& $]20-97]$&$[160-500]$ & $[5-500]$ &
                                                                $[5-20]$
                                                                  &$]20-97]$
                                                                    &$[160-500]$&$[5-500]$  & $[5-20]$&$]20-97]$&$[160-500]$ \\                                     
  \hline
SSDR1 & 2.92 & 2.80&3.05&2.51 & 2.45 & 2.46&2.27&3.51 & 2.46 & 2.47&2.25&4.05\\
 SSDR5 &10.92 &10.80&10.26&24.18 & 5.97 & 8.27&1.35&4.94 & 5.75 &
                                                                      8.24&0.84&0.81 \\
  \hline
  SSDR7 & 4.65&5.35&2.87&11.80 & 3.56 & 4.42&1.76&3.64 & - & -&-&-\\
   SSDR9 & 11.82&12.11&11.14&2.04 & 10.78 & 12.54&7.33&0.84 &10.70& 12.84&6.40&1.20 \\
\hline
  SSDR8 & 13.27 & 12.4&13.47&20.49 & 9.10&9.56&10.12&4.79 & 3.67 & 6.11&0.74&2.23 \\
  SSDR10 & 9.89 & 7.74&13.82&1.94 & 4.85 & 4.36&5.62&1.89 & 2.65 & 3.59&0.68&3.72  \\                                                 
DEML10 &13.73 & 13.88&13.22&3.58 & 7.77 & 8.65&5.90&2.33 & 4.28 & 6.15&0.52&1.48 \\
 DEML34 &9.23 & 6.03&14.14&35.36 & 4.98 & 3.18&7.38&29.53 &  1.59 & 2.08&0.47&3.45 \\
  DEML86 & 7.35& 4.75&11.22&31.41 & 3.75 &2.87&4.80&17.40 & 2.26& 3.02&0.55&6.02\\
   DEML323 & 10.97& 9.45&13.55&3.19 & 6.30 & 6.11&6.50&1.93 & 3.78 & 5.43&0.58&2.58 \\
  \hline
  \hline
  
Model & \multicolumn{4}{c}{MC11}  &\multicolumn{4}{c }{MC11 ($a_0$)}  &\multicolumn{4}{c}{MC11 ($a_0$, $a_{min}, a_{max}$)} \\
  \hline
  & $\chi^2/dof$ &\multicolumn{3}{c}{$\chi^2/N_{data}$}  &$\chi^2/dof$  & 
                  \multicolumn{3}{c}{$\chi^2/N_{data}$} & $\chi^2/dof$ &\multicolumn{3}{c}{$\chi^2/N_{data}$} \\
  \hline
  Region & $[5-500]$ & $[5-20]$&$]20-97]$&$[160-500]$ & $[5-500]$ &
                                                                $[5-20]$&$]20-97]$&$[160-500]$
                                                                &
                                                                $[5-500]$
  & $[5-20]$&$]20-97]$&$[160-500]$ \\                                     
  \hline
SSDR1 & 2.63& 2.44&2.65&8.42 & 1.63 & 2.22&0.46&1.01 & -&-&-&- \\
  SSDR5 & 5.20& 7.29&1.12&1.06 & 4.98 & 7.21&0.59&1.36 & -& -&-&- \\
  \hline
  SSDR7 & 3.19& 4.03&1.40&3.93  & 2.88 & 3.98&0.54&5.72 &-&-&-&-\\
    SSDR9 & 10.55& 12.50&6.61&4.62  & 7.38 & 10.89&0.65&2.04&-&-&-&-\\
\hline
  SSDR8 & 4.64& 5.99&2.36&20.22  & 3.40& 5.45&0.91&3.17&-&-&-&-\\
 SSDR10 & 2.98& 3.69&1.38&6.36 & 2.58 & 3.55&0.54&3.78 & -&-&-&-\\                                                 
DEML10 & 3.90 & 5.42&0.75&5.51  &3.87 & 5.42&0.70&4.05 & -&-&-&-\\
 DEML34 & 0.83 & 0.99&0.42&3.03  & 0.83& 0.99&0.41&3.13 & -&-&-&-\\
  DEML86 & 1.11& 1.38&0.43&3.97 &1.06 & 1.37&0.30&3.80 & - &-&-&-\\
  DEML323& 3.52 & 4.90&0.88&2.12 & 3.21 & 4.63&0.45&2.94 & -&-&-&-\\
  \hline
  \hline
  
  Model  & \multicolumn{4}{c }{DL07 } &\multicolumn{4}{c }{DL07 ($a_0$) } &
                                                  \multicolumn{4}{c}{DL07
                                                  ($a_0, a_{min}, a_{max}$)}\\
  \hline
  & $\chi^2/dof$ &\multicolumn{3}{c}{$\chi^2/N_{data}$}  &$\chi^2/dof$  & 
                  \multicolumn{3}{c}{$\chi^2/N_{data}$} & $\chi^2/dof$ &\multicolumn{3}{c}{$\chi^2/N_{data}$} \\
  \hline
  Region & $[5-500]$ & $[5-20]$&$]20-97]$&$[160-500]$ & $[5-500]$ &
                                                                $[5-20]$&$]20-97]$&$[160-500]$
                                                                &
                                                                $[5-500]$
  & $[5-20]$&$]20-97]$&$[160-500]$ \\                                     
  \hline
SSDR1 &2.13  & 2.27&1.30&15.73  & 1.65 & 2.22&0.45&2.06 &-&-&-&- \\
  SSDR5 & 5.29 & 7.46&0.98&2.64  & 5.21 &  7.39&0.80&4.76 &-&-&-&-\\
  \hline
  SSDR7 & 2.87 & 3.94&0.66&3.89 & 2.84 &3.96&0.50&4.09& -&-&-&-\\
 SSDR9 & 8.32& 11.52&1.40&25.26 & 7.35 &10.59&0.87&9.27& -&-&-&-\\
\hline
  SSDR8 & 4.27 & 6.78&1.23&4.66 & 4.17 &6.44&1.35&4.85& -&-&-&-\\
 SSDR10 & 3.16 & 4.06&1.23&4.78 & 2.78 & 3.58&0.90&9.05&-&- &-&-\\
DEML10 & 5.12 & 6.67&1.89&6.02  & 4.74 &6.06&1.71&12.64& -& -&-&- \\
DEML34 & 3.16 & 2.95&3.18&10.06 &1.37 &1.58&0.74&5.89& -& - &-&- \\
 DEML86 & 2.47 & 2.67&1.85&6.13  & 1.44 &1.85&0.51&3.37& -& -&-&- \\
 DEML323 &4.20 & 5.37&1.85&4.61 & 3.45 & 4.69&0.85&7.73& -&-&-&-\\
  \hline
  \hline
 Model  & \multicolumn{4}{c }{DBP90 } &\multicolumn{4}{c }{DBP90 ($\alpha$) } &
                                                  \multicolumn{4}{c}{DBP90 ($\alpha=-2.6$, $a_{min},
                                                                                  a_{max}$)} \\
  \hline
  & $\chi^2/dof$ &\multicolumn{3}{c}{$\chi^2/N_{data}$}  &$\chi^2/dof$  & 
               \multicolumn{3}{c}{$\chi^2/N_{data}$} & $\chi^2/dof$ &\multicolumn{3}{c}{$\chi^2/N_{data}$} \\
  \hline
 Region & $[5-500]$ & $[5-20]$&$]20-97]$&$[160-500]$ & $[5-500]$ &
                                                                $[5-20]$&$]20-97]$&$[160-500]$
                                                                &
                                                                $[5-500]$
  & $[5-20]$&$]20-97]$&$[160-500]$ \\                                     
  \hline
SSDR1 & 3.11 & 2.40&3.29&33.85 & 1.89 & 2.20&1.008.22& 1.67 & 2.21&0.59&0.84 \\
 SSDR5 &5.40 & 7.45&1.13&10.41 & 5.10 & 7.29&0.76&3.83 &4.92 &
                                                                   7.12&0.64&0.53\\
  \hline
 SSDR7 & 3.23& 3.89&1.81&4.79 & 2.88 & 3.89&0.82&3.53 &-&-&-&-\\
SSDR9 & 12.22& 13.52&7.76&59.52 & 7.99 & 11.39&1.06&16.73 &7.29&10.71&0.78&1.44\\
 \hline
  SSDR8 & 5.79& 7.54&2.81&23.62 & 3.49 & 5.86&0.73&2.10 &-&-&-&-\\
 SSDR10 & 2.94& 3.86&1.05&4.48 & 2.61  &3.58&0.62&4.14 & -& -&-&- \\
DEML10 & 4.10 & 5.81&0.65&5.28 & 4.07 & 5.68&0.82&4.32& 3.87 & 5.55&0.58&1.23 \\
DEML34  & 2.28 & 2.18&1.26&35.35 & 2.25 & 2.30&1.22&27.14&1.32 & 1.65&0.60&1.90 \\
 DEML86 & 2.76 & 3.35&0.90&20.96 & 2.76 & 3.29&0.89&24.04&2.08 & 2.84&0.42&4.77  \\
  DEML323 &4.11  & 5.74&1.11&1.21 & 3.74 & 5.16&0.83&8.09 & 3.36 &4.87&0.50&1.69 \\
\hline
\end{tabular}}
\end{center}
\tablefoot{The results using standard parameters are shown in column 2,
  $\alpha$ or $a_0$ in column 3, and using standard
  parameters with $\alpha$ or $a_0$, $a_{min}$ and $a_{max}$ in column
  4. The sub-columns indicate the values of $\chi^2/N_{data}$ computed
  in different wavelength ranges: 5-20 $\mic$; 20-97 $\mic$ and
  160-500 $\mic$. The '-' symbol indicates that the $\chi^2$ is not improved.}
\end{table*}

\subsection{Changing parameters of the dust size distribution of
  carbon grains}
\label{sec_vsgsize}
Depending on the model and the dust component, this size distribution
is governed by a power law or log-normal distribution. Below is the summary of
the original dust size distribution of interest, for each model:\\
- a-C (AJ13): a power-law distribution $dn/da\propto a^{\alpha}$ ($\alpha$=-5)
with an exponential tail, $D(a)=e^{- \left ( (a-a_t)/a_c \right
  )^\gamma}$ with $a_c$= 50 nm and for $a \geq a_t=10$ nm 
($D(a)=$1 otherwise), for grain sizes between 0.4 nm
($a_{min}$) and
4900 nm ($a_{max}$),\\
- SamC (MC11): a logarithmic normal distribution $dn/da \propto
e^{- log(a/a_0)^2/\sigma}$ (with $a_0$ the centre radius equal to 2 nm and $\sigma$ the
width of the distribution equal to 0.35), with a grain size between
0.6 nm and 20 nm \\
- Graphite (DL07): a logarithmic normal distribution with $a_0$=2 nm
and $\sigma$=0.55, with $a$ between 0.31 nm and 40 nm \\
- VSG (DBP90): a power-law distribution with $\alpha$=-2.6 for
grains in the range 1.2 nm and 15 nm. 

We first allow either $\alpha$ in the power-law, or $a_0$ in the
log-normal distribution to vary
in order to better fit the SED. In a second step, we also include $a_{min}$
and $a_{max}$ as free parameters for AJ13, MC11 and DL07. In the case of
DBP90, changing $a_{min}$ and $a_{max}$ is highly degenerate with
changing $\alpha$. Therefore, when allowing $a_{min}$ and $a_{max}$ to
vary in the range of the original value, the $\alpha$ parameter is fixed to the original value
(-2.6). For the other models, the dust size parameters are not
anti-correlated but are not completely independent either. The
degeneracy between parameters is something usual when fitting data
with models. When the degeneracy (or anti-correlation) is not very
pronounced, as it is the case with the dust size parameters for AJ13,
MC11 and DL07, the parameters can be left as free parameters. Indeed, changing
one parameter has a very limited impact on the others.

The values of the $\chi^ 2$ of the various fits are given in Table \ref{table_chi2_all}. All
$\chi^2$ are improved for all models when including
$\alpha$ or $a_0$ in the fits. The values of $\chi^2$ obtained when
allowing $\alpha$/$a_0$ and $a_{min}$ and $a_{max}$ to vary are only given when the fit is improved. The
inclusion of $a_{min}$ and $a_{max}$ in the fits does not improve the
modeling in the 160-500 $\mu$m range for MC11 or DL07 models. We
therefore only adopt free $a_0$ for the VSG size parameters for MC11
and DL07. For DBP90, the fits are almost all better with free sizes ($a_{min}$ and
$a_{max}$), compared to the fits with free $\alpha$. We therefore 
consider $\alpha=-2.6$ and free $a_{min}$ and $a_{max}$ in the following. For AJ13, the
fits are significantly improved in the 20-97 $\mic$ domain, when using
free $\alpha$, $a_{min}$ and $a_{max}$ (except for SSDR7). We
therefore adopt these three free parameters for AJ13.

Figure \ref{fig_all_best} presents the best fits for each model
(free $\alpha$,  $a_{min}$ and $a_{max}$  for AJ13, free
$a_0$ for MC11 and DL07, free $a_{min}$ and $a_{max}$ for DBP90). The values of the best fit parameters are given in Tables \ref{table_chi2_AJ13}, \ref{table_chi2_MC11},
\ref{table_chi2_DL07} and \ref{table_chi2_DBP90}. A null dust
abundance is not possible for computational reasons. In that case, if
a null abundance is required in the fit, its value is set to \textit{1.00$\times$10$^{-6}$}. 

%Since the standard parameters are not
%suitable to reproduce LMC observations, we will not discuss
%this case, but either focus on the different models in the framework of their best
%fit (see Figure \ref{fig_all_best} and Table \ref{table_chi2_all})
%with the use of the Mathis RF.
By allowing the VSG size parameters to vary we can see that the results
of the fits, in terms of $\chi^2$, are good for the diffuse and
molecular regions whatever the model (MC11, DL07 or DBP90). However, the HII regions are not well described with DL07 model when
compared to MC11 and DBP90 ones. Indeed, the model shows a lack of emission
in the MIPS SED range. Modeling with AJ13 gives lower quality
fits, especially for diffuse and molecular environments. For instance, AJ13 shows a lack of
emission in the IRS spectra between 20 and 40 $\mic$ approximately for
SSDR1, SSDR5 and SSDR9. For the HII regions the data are still not well
reproduced with AJ13 in comparison with the other models, but the fits are not unreasonable.
For MC11 the $\rm a_0$ parameter always increases (from 2.03 to 4.60) for
all regions when compared to the standard value of 2.0, whereas for DL07 this parameters always decreases for
the HII regions and increases for the two molecular regions. For the two atomic regions,
the trend is not clear. For DBP90, $\rm a_{min}$ is always larger than
the standard parameters for all regions, and $\rm a_{max}$  is also
larger for 6 of the 10 regions.

In some regions and for some models, we find that the best models do
not contain silicates (mainly with MC10 and occasionally with AJ13, see
Tables \ref{table_chi2_AJ13}, \ref{table_chi2_MC11}, \ref{table_60Myr_AJ13} and \ref{table_60Myr_MC11}). This is possible because we did not impose
a lower limit value for the dust abundances (except for reasons of
computational limitations), but from a physical point of view this
result is surprising. This absence of silicates is not induced by 
the choice of the free parameters we adopted and in particular by the 
VSG size distribution parameters. Indeed, the lack of silicate component 
in some models was already visible with the use of standard parameters only. 
Moreover, we performed additional tests with the MC11 model by leaving 
free the power law parameters in the silicate grain size distribution. The absence 
of silicates in some cases is unchanged illustrating that
the choice of the free parameters does not seem to have any impact on
the fitting results. We note that the silicates are absent from the best fits only for
models including two populations of large grains, composed of carbon
and silicate. The potential absence of silicates in some models could
therefore reflect the predominance of carbon grain emission over
silicate component emission. However in the modeling, the
presence/absence of the carbon and/or silicate coarse-grained
components is constrained only by the slope of the FIR/submm emission
and this absence could also result from the lack of observational
constraints at longer wavelengths, i.e. from 500 $\mic$ to 1 mm for instance
combined with the fact that the emission of large grains in the FIR/submm is not 
very sensitive to the grain composition, the two emissions being somewhat 
degenerated. For that reason (sub)millimeter data with small uncertainties, 
which are not easily obtained with ground based telescopes 
are crucial to constrain the BG component. 

\begin{figure*}
  \begin{center}
\includegraphics[width=18cm]{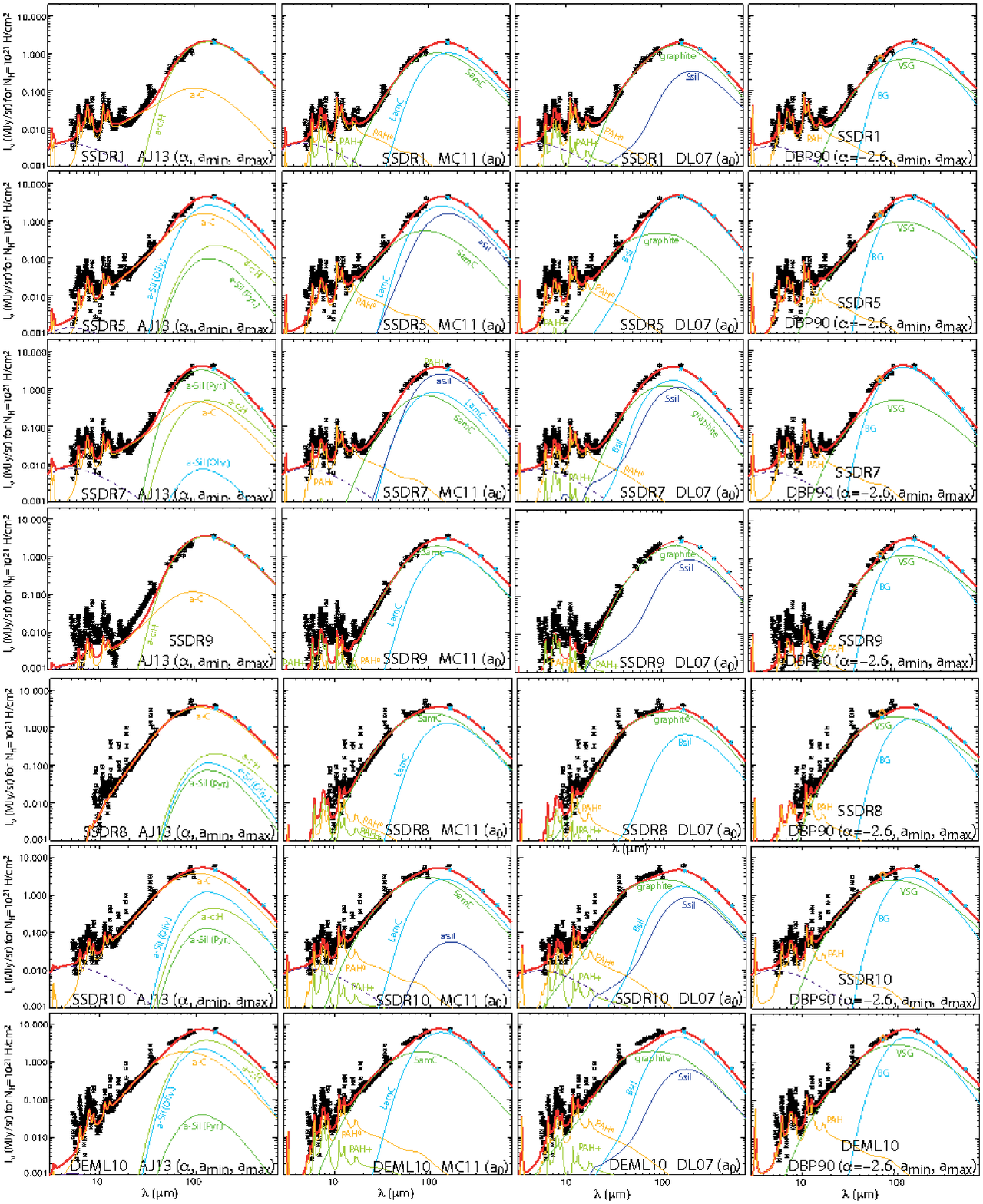}
\caption{Modeling of the SEDs of the ten regions with different dust models and free
  parameters ($\rm X_{ISRF}$, dust abundances and
  small grains dust size distribution), using the
  Mathis RF. The
  observations (Spitzer IRS SS and LL, MIPS SED, MIPS 160
  $\mic$, Herschel Photometric PACS 160 $\mic$ and SPIRE 250 $\mic$, 350 $\mic$
  and 500 $\mic$ data) are shown in black. The total modeled SED is shown
  as a red line. The other colored lines correspond to
  the different dust components of the models. The dashed line
  represents the additional NIR continuum. Blue asterisks show
  the color-corrected brightness derived from the models. The orange
  diamonds that are visible in the DBP90 panels show the MIPS 70 $\mic$
photometric data normalized to the integrated flux in the
MIPS-SED band. Each column shows the fit using different dust models
(from left to right: AJ13, MC11, DL07 and DBP90). Each row presents a
different region. The figure continues on the next page. 
\label{fig_all_best}}

\end{center}
\end{figure*}

\addtocounter{figure}{-1}
\begin{figure*}
  \addtocounter{subfigure}{1}
  \begin{center}
\subfigure{\includegraphics[width=18cm]{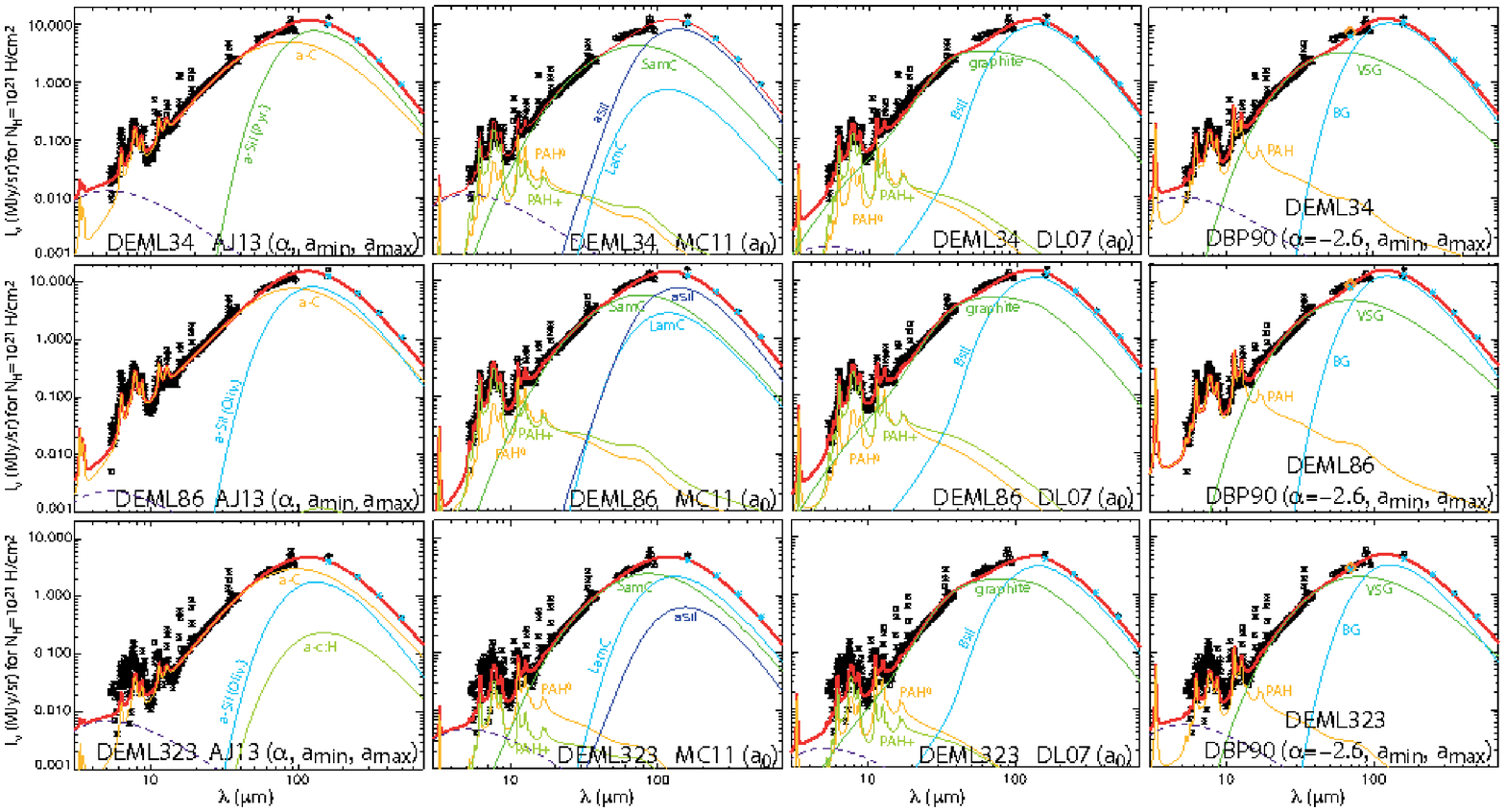}}
\end{center}
\caption{Continued}
\end{figure*}

\subsection{Changing the Radiation Field}
\begin{figure}
  \begin{center}
\includegraphics[width=8cm]{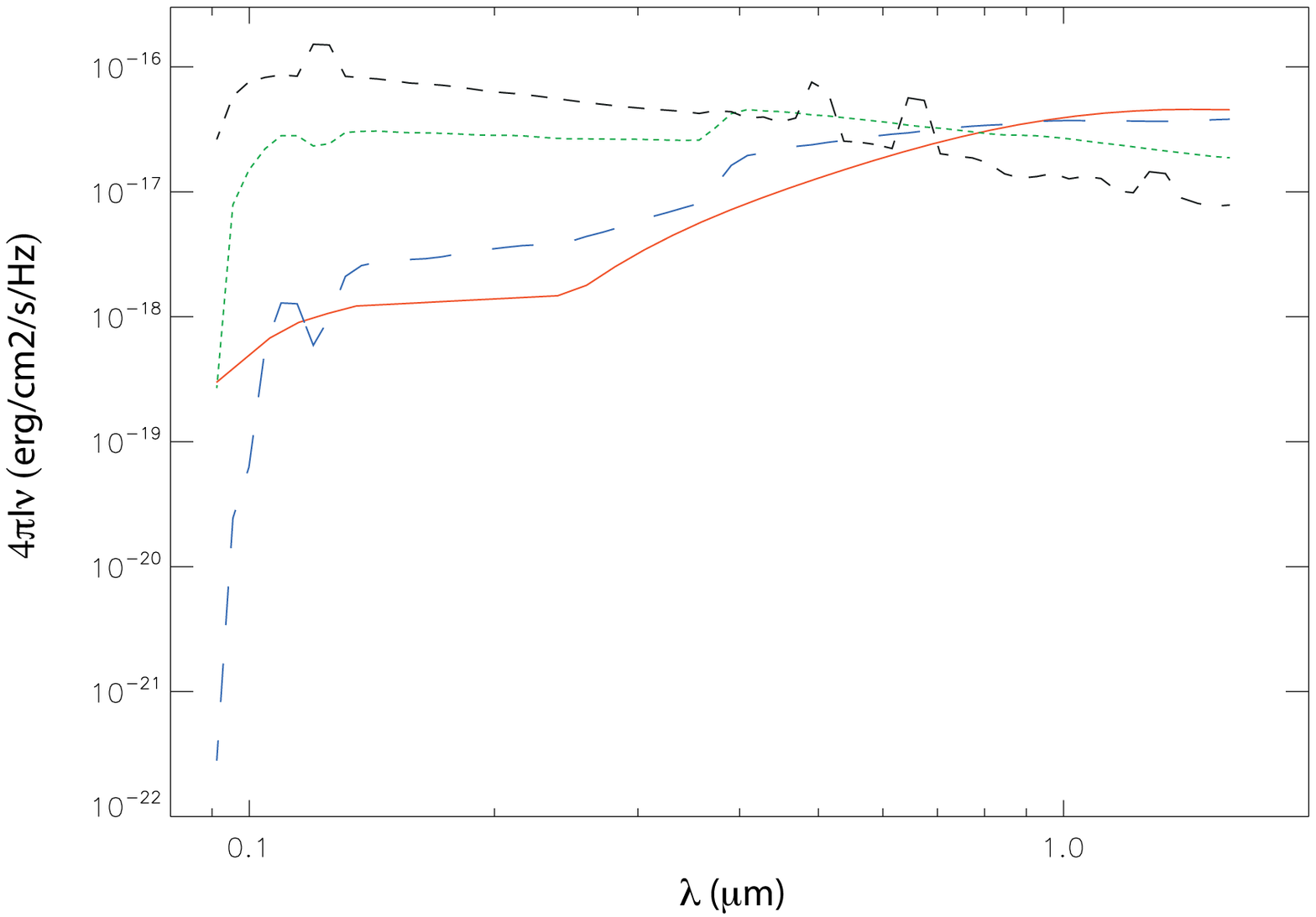}
\caption{Radiation field templates used for the SED modeling: Mathis (solid red line), 4-Myr
  (short dashed black line),
  60-Myr (dotted green line) and 600-Myr (long dashed blue line) stellar clusters.\label{fig_isrf}}
\end{center}
\end{figure}

Dust in galaxies can be illuminated by radiation field coming
from different stellar populations. The Solar neighborhood RF is the standard RF
used in most of modeling of dust emission. In order to see a possible
improvement of our best modeling, as well as to test the robustness of our
results, we also performed the modeling using other radiation
fields. \citet{Bica96} made a catalog of
504 star clusters and 120 stellar associations in the LMC using UBV
photometry. They studied groups of stellar clusters with ages from 10
Myr to 16 Gyr. \citet{Kawamura09} estimated that the youngest stellar
objects are less than 10 Myr. Using the GALEV code\footnote{see http://www.galev.org} we generated
UV/visible spectrum of stellar clusters with various ages: 4-Myr, 60-Myr and 600-Myr. We then make the format of the output files compatible
with the DustEM package (see Fig. \ref{fig_isrf}). Results of the modeling using the 4-Myr RF
are given in Tables \ref{table_chi2_AJ13}, \ref{table_chi2_MC11},
\ref{table_chi2_DL07} and \ref{table_chi2_DBP90}. Results of the fits 
using the 60-Myr and 600-Myr RFs and the corresponding
  figures are given in the appendix.
We can see that changing the RF in the modeling has a direct impact
on the values of dust abundances, and on the intensity of the RF in
particular (see the following sections), but has a very limited impact on the reduced
$\chi^2$. Therefore changing the RF does not improve the modeling,
nor make it worse.

\begin{figure*}
  \begin{center}
\includegraphics[width=16cm]{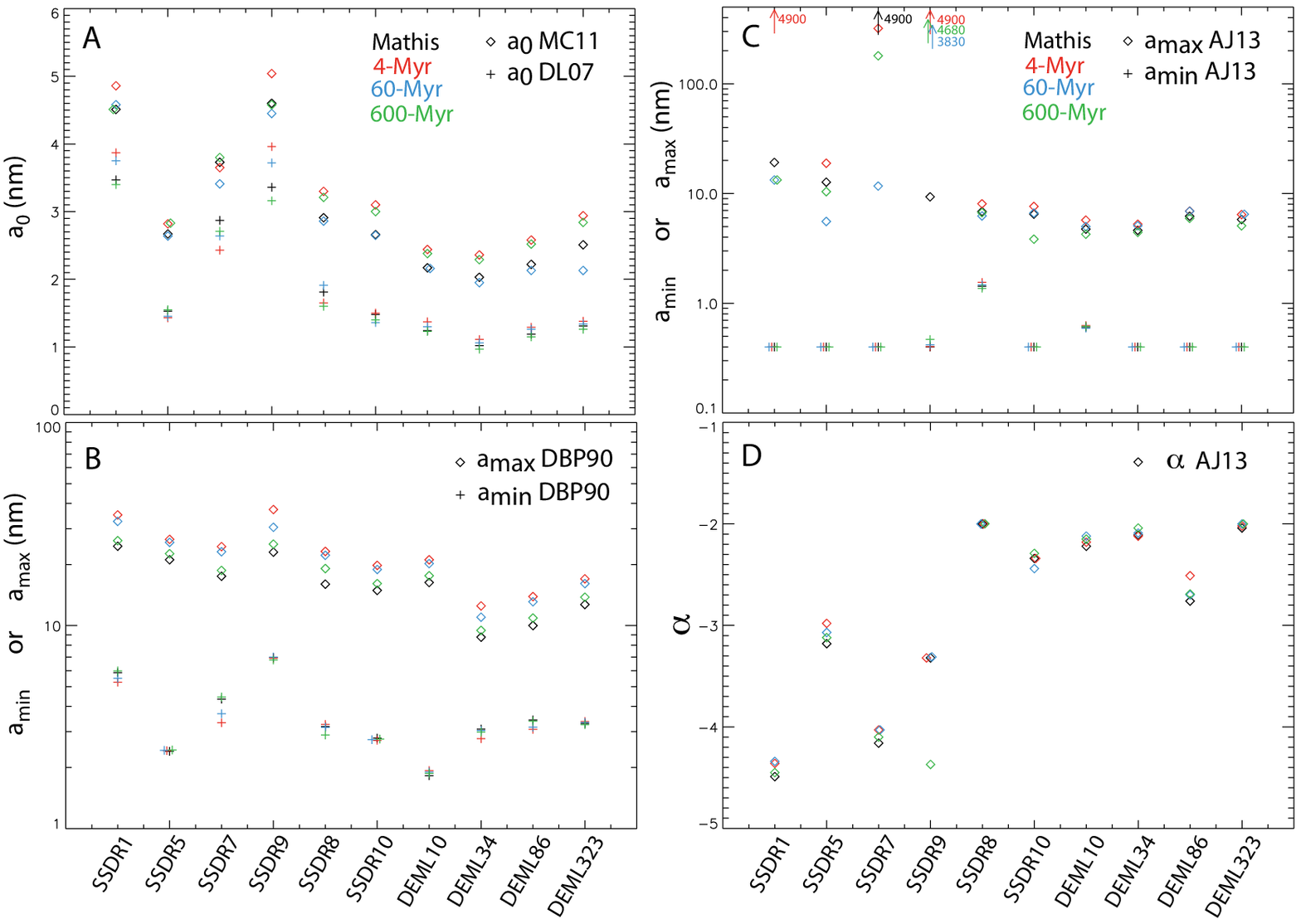}
\caption{Parameters of the small grain size distribution for each
  model, each
  region, and for the different RFs: Mathis (black), 4-Myr (red) , 60-Myr
  (blue) and 600-Myr(green). Panel A:
  $a_0$ (in nm) for
  MC11 and DL07; Panel B: $a_{min}$ and $a_{max}$ (in nm) for
  DBP90; Panel C: $a_{min}$ and $a_{max}$ (in nm) for AJ13 and Panel
  D: $\alpha$ for AJ13. Arrows indicate high values of the $a_{max}$
  parameter. \label{fig_alpha}}
\end{center}
\end{figure*}

In addition, the impact of the RFs on the VSG dust size parameters is presented in
Fig. \ref{fig_alpha}. The values of $a_0$ or $a_{min}$, $a_{max}$ can have
some little variations but stay globally stable for MC11, DL07 and
DBP90. For AJ13, the impact on $a_{min}$ is negligible, whereas
  $a_{max}$ is significantly affected for SSDR1, SSDR7 and
  SSDR9. This means that the $a_{max}$ values are not well constrained. The value of $\alpha$ do not vary much with the RF, except for
  SSDR9. However, the modeling of SSDR9 is of bad
  quality, whetever the RF. Therefore the analyis of this region using
  AJ13 should be considered very cautiously. 
% As visible, the general behavior is unchanged whatever
%the RF.

%The impact of
%the different RF is discussed in the following sections.

 \begin{table*}%[!h]}
\caption{Best fit parameters for the ten regions obtained with AJ13, using the
  Mathis RF (top table) and 4-Myr RF (bottom table).\label{table_chi2_AJ13}}
%\begin{center}
%%\resizebox{\textwidth}{!}{%
% \begin{tabular}{m{1.3cm}m{0.8cm}m{0.6cm}m{1.4cm}m{1.4cm}m{1.4cm}m{1.4cm}m{0.8cm}m{0.5cm}m{0.8cm}m{0.4cm}%m{0.4cm}c}
\begin{center}
\resizebox{\textwidth}{!}{
%\begin{tabular}{m{1.3cm}m{0.8cm}m{0.6cm}m{1.4cm}m{1.4cm}m{1.4cm}m{1.4cm}m{1.4cm}m{1.cm}m{0.7cm}m{0.4cm}m{0.8cm}c}
 \begin{tabular}{*{13}{V{12cm}}}

  \hline
  \hline
\multicolumn{13}{c}{AJ13 ($\alpha, a_{min}, a_{max}$) - Mathis RF}\\
\hline
  Region              & $\chi^2/dof$ &$X_{ISRF}$  &$Y_{a-C}$  & $Y_{a-C:H}$
                 &$Y_{Pyr.}$ &$Y_{Oliv.}$ & $Y_{dust,tot}$ & $I_{NIR\,cont.}$ &
                                                          $\alpha$ &
                                                                       $a_{min}$&$a_{max}$
         & $\frac{Y_{a-C}}{Y_{Sil.}+Y_{a-C:H}}$\\
         &&&&&&&($10^{-2}$)&($10^{-3}$)&&($10^{-1}$)& &($10^{-1}$)\\
  \hline
  MW & - & 1.00 &1.70$\times 10^{-3}$&6.30$\times
                                         10^{-4}$&2.55$\times
                                                   10^{-3}$&2.55$\times
                                                             10^{-3}$&0.74&-&-5.00&4.00&4900&0.34\\
  \hline
   SSDR1 & 2.46&4.77 &2.25$\times 10^{-4}$&
                                1.57$\times 10^{-3}$&1.36$\times 10^{-6}$&1.39$\times 10^{-6}$&0.18&3.99&-4.49&4.00&19.2
         & 1.43\\
  SSDR5 & 5.76& 1.83&2.42$\times 10^{-3}$ &3.79$\times 10^{-4}$
                  &3.65$\times 10^{-4}$&1.00$\times
                                         10^{-2}$&1.31&1.43&-3.18&4.00&12.7
                & 2.25\\
  \hline
  SSDR7 & 3.56& 5.47&
                      4.14$\times 10^{-4}$&4.30$\times 10^{-4}$&4.50$\times 10^{-3}$&1.48$\times 10^{-6}$&0.53&7.57&-4.16&4.00&4900&
  0.840 \\
    SSDR9 & 10.7 &6.11&8.11$\times 10^{-5}$ &2.21$\times 10^{-3}$ &\textit{1.00$\times 10^{-6}$}
                               &\textit{1.00$\times 10^{-6}$}&0.23&1.25&-3.32&4.00&9.33
         & 0.368\\
\hline
  SSDR8 & 3.67 &2.46 & 4.43$\times 10^{-3}$ &2.78$\times 10^{-4}$ &2.12$\times 10^{-4}$
                               &3.46$\times 10^{-4}$&0.53&0.00&-2.00&14.3&6.86 & 5.30\\
  SSDR10 & 2.65 &2.81 &5.10$\times 10^{-3}$ &5.48$\times 10^{-4}$ &
                                          3.37$\times 10^{-4}$&3.26$\times 10^{-3}$&0.92&12.6&-2.34&4.00&6.49
         & 12.3\\
  DEML10 & 4.28 & 6.72& 1.52$\times 10^{-3}$&
                                 2.24$\times 10^{-3}$&4.86$\times 10^{-5}$&2.88$\times 10^{-3}$&0.67&1.08&-2.22&6.06&4.74& 2.94\\
  DEML34 & 1.59&3.11 &8.80$\times 10^{-3}$ &
                                \textit{1.00$\times 10^{-6}$}&1.82$\times 10^{-2}$&\textit{1.00$\times 10^{-6}$}&2.70&13.3&-2.11&4.00&4.62
         & 4.83\\
 DEML86 & 2.26& 4.43&8.11$\times 10^{-3}$ &\textit{1.00$\times 10^{-6}$} &\textit{1.00$\times 10^{-6}$}&1.52$\times 10^{-2}$&2.33&2.34&-2.76&4.00&6.21&5.33\\
   DEML323 & 3.78& 3.35& 3.63$\times 10^{-3}$& 2.50$\times 10^{-4}$&1.23$\times 10^{-6}$&4.20$\times 10^{-3}$&0.81&7.02&-2.04&4.00&5.80&8.16\\
  \hline
  \hline
  \multicolumn{13}{c}{AJ13 ($\alpha, a_{min}, a_{max}$) - 4-Myr RF}\\

\hline
  Region              & $\chi^2/dof$ &$X_{ISRF}$  &$Y_{a-C}$  & $Y_{a-C:H}$
                 &$Y_{Pyr.}$ &$Y_{Oliv.}$ & $Y_{dust,tot}$ & $I_{NIR\,cont.}$ &
                                                          $\alpha$ &
                                                                       $a_{min}$&$a_{max}$&$\frac{Y_{a-C}}{Y_{Sil.}+Y_{a-C:H}}$
  \\
      &&&&&&&($10^{-2}$)&($10^{-1}$)&&($10^{-1}$)&&($10^{-1}$)\\
  \hline
   SSDR1 & 2.42&0.22 &1.03$\times 10^{-4}$ & 1.74$\times 10^{-3}$&\textit{1.00$\times 10^{-6}$}&\textit{1.00$\times 10^{-6}$}&0.18&3.35&-4.36&4.00&4900&0.591\\
  SSDR5 & 6.23& 0.037&3.77$\times 10^{-3}$& \textit{1.00$\times
                                            10^{-6}$} &5.54$\times
                                                        10^{-3}$&5.76$\times
                                                                  10^{-3}$&1.51&
                                                                                 1.20&-2.98&4.00&18.9&3.34\\
  \hline
  SSDR7 & 3.52& 0.16&
                      2.97$\times 10^{-4}$&3.32$\times 10^{-4}$&6.33$\times 10^{-3}$&2.71$\times 10^{-4}$&0.72&6.78&-4.03&4.00&320&0.428\\
    SSDR9 & 10.5 &0.30&9.39$\times 10^{-5}$ &2.15$\times 10^{-3}$ &4.12$\times 10^{-4}$ &2.58$\times 10^{-6}$&0.27&1.35&-3.32&4.03&4900&0.366\\
\hline
  SSDR8 & 3.69 &0.10 & 2.19$\times 10^{-3}$ &9.64$\times 10^{-4 }$&5.81$\times 10^{-5}$ &1.89$\times 10^{-3}$&0.51&0.00&-2.00&15.5&8.06&7.52\\
  SSDR10 & 2.67&0.098&3.07$\times 10^{-3}$ &1.06$\times 10^{-3}$ &5.60$\times 10^{-3}$&2.27$\times 10^{-4}$&1.00&11.3&-2.34&4.00&7.63&4.46\\
  DEML10 & 4.37 & 0.16& 1.48$\times 10^{-3}$& 2.07$\times 10^{-3}$&2.90$\times 10^{-3}$&5.31$\times 10^{-3}$&1.27&0.232&-2.18&6.26&5.73&1.44\\
  DEML34 & 1.57&0.11 &5.09$\times 10^{-3}$ & \textit{1.00$\times 10^{-6}$}&2.48$\times 10^{-2}$&1.17$\times 10^{-6}$&2.99&10.6&-2.12&4.00&5.26&2.05\\
 DEML86 & 2.42& 0.14&6.27$\times 10^{-3}$ &1.78$\times 10^{-6}$ &\textit{1.00$\times 10^{-6}$}&1.86$\times 10^{-2}$&2.49&0.00&-2.51&4.00&6.93&3.37\\
   DEML323 & 3.83& 0.13& 1.80$\times 10^{-3}$&
                                  7.25$\times 10^{-4}$&1.88$\times 10^{-3}$&3.98$\times 10^{-3}$&0.84&6.33&-2.02&4.00&6.43& 2.73\\
  \hline
\hline

 \end{tabular}
 }
\end{center}
\tablefoot{ Value of the
  reduced $\chi^2$ is given in column 2, intensity
  of the RF in column 3,
  abundances of the different dust components in columns 4 to 7, total dust
  abundance in column 8, intensity
  of the NIR continuum in column 9, small grain size
  parameters in columns 10 to 12 (with $a_{min}$ and $a_{max}$ in nm),
  and ratio of the a-C component abundance over the total Big grains
  abundance in column 13. Bottom table: same as the top table
  but using a 4-Myr RF. A null abundance is not possible for
  computational reasons, and is set to \textit{1.00$\times$10$^6$}
  instead. Standard parameters for our Galaxy are also given for comparison.}
\end{table*}%[!h]}

\begin{table*}%[!h]}
\caption{Best fit parameters for the ten regions obtained with MC11, using the
  Mathis RF (top table) and 4-Myr RF (bottom table). \label{table_chi2_MC11}}
\begin{center}
\resizebox{\textwidth}{!}{
%\begin{tabular}{m{1.3cm}m{0.8cm}m{0.6cm}m{1.4cm}m{1.4cm}m{1.4cm}m{1.4cm}m{1.4cm}m{1.cm}m{0.7cm}m{0.4cm}m{0.8cm}c}
 \begin{tabular}{*{13}{V{12cm}}}
%  \hline
%\multicolumn{10}{c}{MC11} \\ 
%  \hline  
%  Region  & $\chi^2/dof$ &$X_{ISRF}$  &$Y_{PAH^0}$  & $Y_{PAH^+}$
%                 &$Y_{SamC}$ &$Y_{LamC}$& $Y_{aSil}$  &
%                                                        $I_{NIR\,cont.}$ 
%  & $\frac{Y_{SamC}}{(Y_{LamC}+Y_{aSil}}$\\
%    \hline
%  SSDR1 & 2.63& 1.32& 3.10E-4& 2.11E-4&2.28E-4&4.26E-3&1.00E-6&4.28E-3
%  & 5.35E-2\\
%  SSDR5 & 5.20&1.83 & 3.23E-4& 6.33E-5&5.83E-3&4.52E-3&1.10E-2&1.10E-3
%  & 3.75E-1 \\
%  SSDR7 & 3.19& 5.55& 1.48E-4& 2.61E-5&8.92E-5&1.13E-3&6.84E-3&7.84E-3
%  &1.11E-2\\
  %SSDR8 & 3.21E-1\\
 %SSDR9 &3.44E-2\\ 
%  SSDR10 & 2.98& 2.26& 3.02E-4& 2.60E-4&2.50E-3&5.79E-3&1.00E-6&1.15E-2&4.32E-1\\
%  DEML10 & 3.90& 2.61& 3.30E-4& 1.84E-4&2.59E-3&7.12E-3&1.00E-6&0.00&3.64E-1\\
%  DEML34 & 0.83& 4.43& 2.31E-4& 5.91E-4&4.11E-3&7.48E-4&3.12E-2&1.09E-2&1.29E-1\\
%  DEM86 & 1.11& 4.58&2.92E-4 &
%                               1.24E-3&4.18E-3&3.86E-3&2.43E-2&0.00 &1.48E-1\\
%  DEML323 & 3.52& 2.92&8.10E-5 &2.85E-4 &1.86E-3&4.17E-3&1.15E-6&4.34E-3&4.46E-1\\
  \hline
  \hline
\multicolumn{13}{c }{MC11 ($a_0$) - Mathis RF } \\
 \hline
  Region  & $\chi^2/dof$ &$X_{ISRF}$  &$Y_{PAH^0}$  & $Y_{PAH^+}$
                 &$Y_{SamC}$ &$Y_{LamC}$& $Y_{aSil}$  & 
                                                          $Y_{dust,tot}$ &
                                                 $I_{NIR\,cont.}$ &
                                                                    $a_0$&
                                                                           $\frac{Y_{SamC}}{(Y_{LamC}+Y_{aSil})}$&$\frac{(Y_{PAH^0}+Y_{PAH^+})}{(Y_{LamC}+Y_{aSil})}$\\
           &&&&&&&&($10^{-2}$)&($10^{-3}$)&& ($10^{-1}$) &($10^{-2}$) \\
  \hline
  MW & - & 1.00&7.80$\times 10^{-4}$&7.80$\times 10^{-4}$&1.65$\times 10^{-4}$&1.45$\times 10^{-3}$&6.70$\times 10^{-3}$&0.99&-&2.0&0.20&19.14\\
  \hline
  SSDR1 & 1.64 & 0.65& 8.74$\times 10^{-4}$&
                                2.83$\times 10^{-4}$&3.33$\times 10^{-3}$&3.50$\times 10^{-3}$&1.92$\times 10^{-5}$&0.80&4.06&4.51 &9.46&32.9\\
  SSDR5 & 4.98& 1.64& 4.71$\times 10^{-4}$&
                               5.95$\times 10^{-6}$&1.09$\times
                                                     10^{-3}$&3.93$\times
                                                               10^{-3}$&1.38$\times
                                                                         10^{-2}$&1.90&0.963&
                                                                                              2.67&0.615&2.69\\
  \hline
  SSDR7 & 2.88&4.71 &2.18$\times 10^{-4 }$&
                               4.59$\times 10^{-6}$&3.77$\times 10^{-4}$&5.36$\times 10^{-4}$&9.09$\times 10^{-3}$&1.02&7.62&3.73&0.392&2.31\\
  SSDR9  &7.38 &0.65 &
                               1.09$\times 10^{-4}$&2.38$\times 10^{-4}$&6.09$\times 10^{-3}$&4.60$\times 10^{-3}$&1.01$\times 10^{-6}$
   &1.10& 0.402&4.60&13.2&7.54\\
 \hline
  SSDR8 & 3.40&1.01 &1.85$\times 10^{-4}$ &
                               2.48$\times 10^{-4}$&7.59$\times 10^{-3}$&3.16$\times 10^{-3}$&\textit{1.00$\times 10^{-6}$}&1.11&0.00&2.91&24.0&13.7\\
 SSDR10 & 2.57& 1.41&8.93$\times 10^{-4}$ &1.78$\times 10^{-4}$ &7.15$\times 10^{-3}$&4.71$\times 10^{-3}$&5.83$\times 10^{-4}$&1.35&11.4&2.66&13.5&20.2\\
  DEML10 & 3.87& 2.38& 4.52$\times 10^{-4}$& 1.48$\times 10^{-4}$&3.25$\times 10^{-3}$&7.07$\times 10^{-3}$&\textit{1.00$\times 10^{-6}$}&1.09&0.00&2.17&4.60&8.49\\
  DEML34 & 0.83&4.38 &2.59$\times 10^{-4}$ &5.86$\times 10^{-4}$
                  &4.28$\times 10^{-3}$&5.12$\times 10^{-4}$&3.20$\times 10^{-2}$&3.76&0.109&2.03& 1.32&2.60\\
  DEML86 & 1.06 &4.22 &4.93$\times 10^{-4}$ &1.25$\times 10^{-3}$ &5.45$\times 10^{-3}$&2.01$\times 10^{-3}$&3.05$\times 10^{-2}$&3.97&0.00&2.22&1.68&5.36\\
  DEML323 & 3.21& 2.19&2.93$\times 10^{-4}$ &2.68$\times 10^{-4}$ &3.92$\times 10^{-3}$&2.65$\times 10^{-3}$&4.36$\times 10^{-3}$&1.15&4.76&2.51&5.59&8.00\\
  \hline
  $<$All$>$ & & & &&&&&&&3.00 &&\\
  $<$Diff.$>$ & & & &&&&&&&4.17 &&\\
  $<$Mol.$>$ & & & &&&&&&&3.59&&\\
  $<$HII$>$ & & & &&&&&&&2.42&& \\
  \hline
  \hline
\multicolumn{13}{c }{MC11 ($a_0$) - 4-Myr RF } \\
 \hline
  Region  & $\chi^2/dof$ &$X_{ISRF}$  &$Y_{PAH^0}$  & $Y_{PAH^+}$
                 &$Y_{SamC}$ &$Y_{LamC}$& $Y_{aSil}$  & $Y_{dust,tot}$ &
                                                        $I_{NIR\,cont.}$
   &  $a_0$& $\frac{Y_{SamC}}{(Y_{LamC}+Y_{aSil})}$&$\frac{(Y_{PAH^0}+Y_{PAH^+})}{(Y_{LamC}+Y_{aSil})}$\\
  & &&&&&&&($10^{-2}$)&($10^{-3}$)& &($10^{-1}$)&($10^{-2}$)\\
  \hline
  SSDR1 & 1.59 & 0.042& 3.08$\times 10^{-4}$&
                                2.57$\times 10^{-5}$&1.08$\times 10^{-3}$&4.32$\times 10^{-3}$&4.03$\times 10^{-3}$&0.98&3.3&4.86 &1.29&3.99\\
  SSDR5 & 4.95& 0.054& 2.62$\times 10^{-4}$&
                               \textit{1.00$\times
                                             10^{-6}$}&7.70$\times
                                                        10^{-4}$&4.62$\times
                                                                  10^{-3}$&1.63$\times
                                                                            10^{-2}$&2.20&0.309&
                                                                                                 2.82&0.368&1.26\\
  \hline
  SSDR7 & 2.88&0.099 &2.00$\times 10^{-4}$ &
                                  \textit{1.00$\times 10^{-6}$}&3.55$\times 10^{-4}$&4.37$\times 10^{-4}$&1.11$\times 10^{-2}$&1.21&6.25&3.65&0.308&1.74\\
  SSDR9  &7.35 &0.064&
                               3.30$\times 10^{-5}$&3.04$\times 10^{-5}$&1.21$\times 10^{-3}$&7.64$\times 10^{-3}$&1.77$\times 10^{-6}$
   & 0.89&0.208&5.04&1.58&0.829\\
\hline
  SSDR8 & 3.50&0.10 &3.80$\times 10^{-5}$ &
                               3.98$\times 10^{-5}$&1.82$\times 10^{-3}$&5.16$\times 10^{-3}$&2.27$\times 10^{-6}$&0.71&0.00&3.30&3.53&1.51\\
   SSDR10 & 2.58& 0.12&2.20$\times 10^{-4}$ &\textit{1.00$\times 10^{-6}$} &2.21$\times 10^{-3}$&6.80$\times 10^{-3}$&\textit{1.00$\times 10^{-6}$}&0.92&10.0&3.10&3.25&3.25\\
  DEML10 & 3.90& 0.21& 9.90$\times 10^{-5}$& 1.31$\times 10^{-5}$&9.71$\times 10^{-4}$&8.11$\times 10^{-3}$&\textit{1.00$\times 10^{-6}$}&0.92&0.00&2.44&1.20&1.47\\
  DEML34 & 0.89&0.15&1.62$\times 10^{-4}$ &2.83$\times 10^{-4}$
                  &3.56$\times 10^{-3}$&6.73$\times 10^{-3}$&1.45$\times 10^{-2}$&2.52&8.76&2.36&1.68&2.10 \\
  DEML86 & 1.10 &0.10 &4.24$\times 10^{-4}$ &7.48$\times 10^{-4 }$&6.28$\times 10^{-3}$&2.01$\times 10^{-3}$&2.57$\times 10^{-3}$&1.20&0.00&2.58&13.7&25.59\\
  DEML323 & 3.24& 0.13&1.15$\times 10^{-4}$ &5.46$\times 10^{-5}$ &1.85$\times 10^{-3}$&4.17$\times 10^{-3}$&2.98$\times 10^{-3}$&0.92&4.03&2.94&2.59&2.37\\
  \hline
  $<$All$>$ & & & &&&&&&&3.31&& \\
  $<$Diff.$>$ & & & &&&&&&&4.35&&\\
  $<$Mol.$>$ & & & &&&&&&&3.84&&\\
  $<$HII$>$ & & & &&&&&&&2.79&& \\
  \hline
  \hline
 \end{tabular}
 }
\end{center}
\tablefoot{Value of the
  reduced $\chi^2$ is given in column 2, intensity
  of the RF in column 3,
  abundances of the different dust components in columns 4 to 8, total dust
  abundance in column 9, intensity
  of the NIR continuum in column 10, small grain size
  parameter $a_0 $ in nm in column 11, ratio of the small amorphous carbon
  abundance over the total Big grain abundance in column 12,
  and ratio of the total PAH abundance over the total Big
    grain abundance in column 13. Bottom table: same as the top table
  but using a 4-Myr RF. Average values of $a_0$ in nm, deduced from all
  regions, from diffuse (SSDR7 and SSDR9), molecular (SSDR1 and SSDR5) and
  ionized region (SSDR8, SSDR10, DEML10, DEML34, DEML86 and DEML323)
  are also given.  A null abundance is not possible for
  computational reason, and is set to \textit{1.00$\times$10$^6$} instead. Standard parameters for our Galaxy are also given for comparison. }
\end{table*}%[!h]}

\begin{table*}%[!h]}
\caption{Best fit parameters for the ten regions obtained with DL07, using the
  Mathis RF (top table) and 4 My RF (bottom table). \label{table_chi2_DL07}}
\begin{center}
%\resizebox{\textwidth}{!}{%
   %    \begin{tabular}{m{1.3cm}m{0.8cm}m{0.6cm}m{1.4cm}m{1.4cm}m{1.4cm}m{1.4cm}m{1.4cm}m{0.7cm}m{1.4cm}m{0.4cm}m{0.8cm}c}
  \resizebox{\textwidth}{!}{
%\begin{tabular}{m{1.3cm}m{0.8cm}m{0.6cm}m{1.4cm}m{1.4cm}m{1.4cm}m{1.4cm}m{1.4cm}m{1.cm}m{0.7cm}m{0.4cm}m{0.8cm}c}
 \begin{tabular}{*{13}{V{12cm}}}

\hline
%  \hline
%\multicolumn{12}{c}{DL07} \\ 
% \hline  
%Region  & $\chi^2/dof$ &$X_{ISRF}$  &$Y_{PAH^0}$  & $Y_{PAH^+}$
%                 &$Y_{graph.}$ &$Y_{Bsil}$ & $Y_{Ssil}$&  $I_{NIR\,cont.}$ &  - &-&-\\
%    \hline
%  SSDR1 & 2.13& 0.77& 6.53$\times 10^{-4& 3.56$\times 10^{-4&1.96$\times 10^{-3&8.72$\times 10^{-3&1.00$\times 10^{-6&4.17$\times 10^{-3&-&-&-\\
%  SSDR5 & 5.29& 0.97& 9.54$\times 10^{-4& 1.00$\times 10^{-6&4.80$\times 10^{-3&1.41$\times 10^{-2&8.93$\times 10^{-5&5.27$\times 10^{-4&-&-&-\\
%  SSDR7 & 2.87& 2.98& 3.32$\times 10^{-4&4.52$\times 10^{-5 &8.46$\times 10^{-4&5.20$\times 10^{-3&1.15$\times 10^{-3&7.54$\times 10^{-3&-&-&-\\
%  SSDR10 & 3.16& 0.47&3.75$\times 10^{-3 &1.81$\times 10^{-4 &5.71$\times 10^{-2&1.00$\times 10^{-6&1.00$\times 10^{-6&1.05$\times 10^{-2&-&-&-\\
%  DEML10 & 5.12& 0.99& 1.86$\times 10^{-3&1.01$\times 10^{-6 &2.84$\times 10^{-2&1.65$\times 10^{-6&2.36$\times 10^{-2&0.00&-&-&-\\
%  DEML34 &3.16 & 0.99& 3.66$\times 10^{-3&1.55$\times 10^{-3&6.48$\times 10^{-2&1.00$\times 10^{-6&1.00$\times 10^{-6&8.26$\times 10^{-3&-&-&-\\
%  DEM86 & 2.47& 1.04& 4.11$\times 10^{-3&4.15$\times 10^{-3 &7.45$\times 10^{-2&4.09$\times 10^{-6&1.32$\times 10^{-4&0.00 &-&-&-\\
%  DEML323 & 4.20& 0.98&1.14$\times 10^{-3 &4.26$\times 10^{-4 &2.45$\times 10^{-2&1.01$\times 10^{-6&5.83$\times 10^{-3&3.74$\times 10^{-3&-&-&-\\
  \hline
\multicolumn{13}{c }{DL07 ($a_0$ - Mathis RF)} \\
 \hline
 Region  & $\chi^2/dof$ &$X_{ISRF}$  &$Y_{PAH^0}$  & $Y_{PAH^+}$
                 &$Y_{graph.}$ &$Y_{Bsil}$ & $Y_{Ssil}$& $Y_{dust,tot}$ &
                                                         $I_{NIR\,cont.}$
                                                                           &
                                                                             $a_0$&
                                                                                    $\frac{Y_{graph.}}{(Y_{Bsil}+Y_{Ssil})}$&$\frac{(Y_{PAH^0}+Y_{PAH^+})}{(Y_{Bsil}+Y_{Ssil})}$\\
         &&&&&&&&($10^{-2}$)&&&($10^{-1}$)&\textbf{($10^{-2}$)}\\
  \hline
  MW & -&1.00&4.97$\times 10^{-4}$&4.97$\times 10^{-4}$&1.66$\times 10^{-4}$&7.64$\times 10^{-3}$&2.21$\times 10^{-3}$&1.10&-&2.00&0.17&10.09\\
  \hline
  SSDR1 & 1.65& 0.38& 1.48$\times 10^{-3}$&6.55$\times 10^{-4}$ &1.34$\times 10^{-2}$&\textit{1.00$\times 10^{-6}$}&1.18$\times 10^{-2}$&2.70&3.99$\times 10^{-3}$&3.47&11.4&18.1\\
  SSDR5 & 5.21&1.16 &7.29$\times 10^{-4}$ & 2.47$\times
                                            10^{-5}$&2.64$\times
                                                      10^{-3}$&1.42$\times
                                                                10^{-2}$&\textit{1.00$\times
                                                                          10^{-6}$}&1.76&5.56$\times
                                                                                          10^{-4}$&1.53&1.86&5.31\\
  \hline
  SSDR7 & 2.84& 3.07& 3.29$\times 10^{-4}$&
                               4.18$\times 10^{-5}$&1.59$\times 10^{-3}$&2.45$\times 10^{-3}$&6.77$\times 10^{-3}$&1.11&7.23$\times 10^{-3}$&2.87&1.72&4.02\\
   SSDR9 & 7.35& 0.60& 4.66$\times 10^{-5}$& 3.31$\times 10^{-4}$&1.24$\times 10^{-2}$&1.27$\times 10^{-6}$&2.35$\times 10^{-2}$&3.63&3.56$\times 10^{-4}$&3.36&5.28&1.61\\
\hline
  SSDR8 & 4.17& 0.43& 4.74$\times 10^{-4}$& 3.18$\times 10^{-4}$&3.59$\times 10^{-2}$&5.07$\times 10^{-3}$&3.01$\times 10^{-5}$&4.18&0.00&1.81&70.4&15.5\\
  SSDR10 & 2.78& 1.03& 1.22$\times 10^{-3}$& 3.04$\times 10^{-4}$&1.73$\times 10^{-2}$&6.21$\times 10^{-3}$&1.35$\times 10^{-2}$&3.85&9.97$\times 10^{-3}$&1.48&8.78&7.73\\
  DEML10 & 4.74&1.45 &7.61$\times 10^{-4}$ &1.51$\times 10^{-4}$ &1.06$\times 10^{-2}$&1.26$\times 10^{-2}$&7.32$\times 10^{-3}$&3.14&0.00&1.24&5.32&4.58\\
  DEML34 & 1.37 &2.21 &4.59$\times 10^{-4}$ &1.07$\times 10^{-3}$ &1.40$\times 10^{-2}$&1.83$\times 10^{-2}$&\textit{1.00$\times 10^{-6}$}&3.47&1.37$\times 10^{-3}$&1.02&7.65&8.35\\
  DEML86 & 1.44& 2.39&7.84$\times 10^{-4}$ &2.07$\times 10^{-3}$ &1.75$\times 10^{-2}$&2.00$\times 10^{-2}$&\textit{1.00$\times 10^{-6}$}&4.04&0.00&1.19&8.75&7.61\\
  DEML323 & 3.45& 1.48&3.69$\times 10^{-4}$ & 4.71$\times 10^{-4}$&9.74$\times 10^{-3}$&8.34$\times 10^{-3}$&\textit{1.00$\times 10^{-6}$}&1.89&2.14$\times 10^{-3}$&1.31&11.7&10.1\\
  \hline
  $<$All$>$ &&&&& &&&&&1.93&&\\
   $<$Diff.$>$ &&&&& &&&&&3.11&&\\
  $<$Mol.$>$ &&&&& &&&&&2.50&&\\
    $<$HII$>$ &&&&& &&&&&1.34&&\\
  \hline
  \hline
\multicolumn{13}{c }{DL07 ($a_0$ - 4-Myr RF} \\
 \hline
 Region  & $\chi^2/dof$ &$X_{ISRF}$  &$Y_{PAH^0}$  & $Y_{PAH^+}$
                 &$Y_{graph.}$ &$Y_{Bsil}$ & $Y_{Ssil}$&
                                                         $Y_{dust,tot}$ & 
                                                         $I_{NIR\,cont.}$
                                                                           &
                                                                             $a_0$&
                                                                                    $\frac{Y_{graph.}}{(Y_{Bsil}+Y_{Ssil})}$&$\frac{(Y_{PAH^0}+Y_{PAH^+})}{(Y_{Bsil}+Y_{Ssil})}$\\
  &&&&&&&&($10^{-2}$)&&&($10^{-1}$)&($10^{-2}$)\\
  \hline
  SSDR1 & 1.59& 0.017&7.16$\times 10^{-4}$&1.81$\times 10^{-4}$&5.35$\times 10^{-3}$&6.61$\times 10^{-3}$&7.40$\times 10^{-3}$&2.02&3.33$\times 10^{-3}$&3.68&3.82&6.40\\
  SSDR5 & 5.17&0.063&2.41$\times 10^{-4}$ & \textit{1.00$\times
                                            10^{-6}$}&7.81$\times
                                                       10^{-4}$&1.71$\times
                                                                 10^{-2}$&5.72$\times
                                                                           10^{-5}$&1.82&7.39$\times
                                                                                          10^{-8}$&1.43&0.455&1.41\\
  \hline
  SSDR7 & 2.84& 0.14& 1.50$\times 10^{-4}$&
                               2.81$\times 10^{-6}$&3.50$\times 10^{-4}$&6.16$\times 10^{-3}$&1.61$\times 10^{-3}$&0.83&6.32$\times 10^{-3}$&2.43&0.450&1.97\\
   SSDR9 & 7.13& 0.014& 2.49$\times 10^{-5}$& 2.50$\times 10^{-4}$&1.16$\times 10^{-2}$&2.57$\times 10^{-4}$&2.34$\times 10^{-2}$&3.55&1.37$\times 10^{-4}$&3.96&4.90&1.16\\
\hline
  SSDR8 & 4.76& 0.088& 2.24$\times 10^{-5}$& 2.63$\times 10^{-5}$&2.51$\times 10^{-3}$&9.38$\times 10^{-3}$&\textit{1.00$\times 10^{-6}$}&1.19&0.00&1.65&2.68&0.519\\
  SSDR10 & 2.73& 0.058& 3.93$\times 10^{-4}$&
                                   7.07$\times 10^{-5}$&5.97$\times 10^{-3}$&1.38$\times 10^{-2}$&3.07$\times 10^{-3}$&2.33&8.55$\times 10^{-3}$& 1.50&3.54&2.75\\
  DEML10 & 4.72&0.084&2.54$\times 10^{-4}$ &\textit{1.00$\times 10^{-6}$}&3.94$\times 10^{-3}$&1.71$\times 10^{-2}$&1.95$\times 10^{-3}$&2.32&0.00&1.37&2.07&1.34\\
  DEML34 & 1.23 &0.14 &1.47$\times 10^{-4}$ &3.01$\times 10^{-4}$ &5.24$\times 10^{-3}$&2.18$\times 10^{-2}$&\textit{1.00$\times 10^{-6}$}&2.75&0.00&1.11&2.40&2.05\\
  DEML86 & 1.56& 0.13&2.79$\times 10^{-4}$ &5.64$\times 10^{-4}$ &6.80$\times 10^{-3}$&2.51$\times 10^{-2}$&2.81$\times 10^{-6}$&3.27&0.00&1.29&2.71&3.36\\
  DEML323 & 3.41& 0.092&1.14$\times 10^{-4}$ & 1.23$\times 10^{-4}$&3.25$\times 10^{-3}$&1.07$\times 10^{-2}$&\textit{1.00$\times 10^{-6}$}&1.42&1.04$\times 10^{-3}$&1.38&3.04&2.21\\
  \hline
  $<$All$>$ &&&& &&&&&&1.98&&\\
   $<$Diff.$>$ &&&&& &&&&&3.20&&\\
  $<$Mol.$>$ &&&&& &&&&&2.56&&\\
    $<$HII$>$ &&&&& &&&&&1.38&&\\
  \hline
  \hline

 \end{tabular}
 }
\end{center}
\tablefoot{Value of the
  reduced $\chi^2$ is given in column 2, intensity
  of the RF in column 3,
  abundances of the different dust components in columns 4 to 8, total dust
  abundance in column 9, intensity
  of the NIR continuum in column 10, small grain size
  parameter $a_0$ in nm in column 11, ratio of the graphite grain
  abundance over the total silicate grains in column 12, and
    ratio of the total PAH abundance over the total Big
    grain abundance in column 13. Average values
  of $a_0$ in nm, deduced from all
  regions, from diffuse (SSDR7 and SSDR9), molecular (SSDR1 and SSDR5) and
  ionized region (SSDR8, SSDR10, DEML10, DEML34, DEML86 and DEML323)
  are also given.  A null abundance is not possible for
  computational reason, and is set to \textit{1.00$\times$10$^6$ }instead. Standard parameters for our Galaxy are also given for comparison. }
\end{table*}%[!h]}

\begin{table*}%[!h]}
\caption{Best fit parameters for the ten regions obtained with DBP90, using the
  Mathis RF (top and middle tables) and 4 My RF (bottom table).\label{table_chi2_DBP90}}
\begin{center}
%\resizebox{\textwidth}{!}{%
%\begin{tabular}{m{1.4cm}m{0.8cm}m{0.7cm}m{1.4cm}m{1.4cm}m{1.4cm}m{1.cm}m{1.2cm}m{0.8cm}m{0.7cm}m{0.7cm}cc}
\resizebox{\textwidth}{!}{
%\begin{tabular}{m{1.3cm}m{0.8cm}m{0.6cm}m{1.4cm}m{1.4cm}m{1.4cm}m{1.4cm}m{1.4cm}m{1.cm}m{0.7cm}m{0.4cm}m{0.8cm}c}
 \begin{tabular}{*{13}{V{12cm}}}
  \hline
%  \hline
%\multicolumn{10}{c}{DBP90} \\ 
%  \hline  
%Region  & $\chi^2/dof$ &$X_{ISRF}$  &$Y_{PAH^0}$  &$Y_{VSG}$ &$Y_{BG}$ &  $I_{NIR\,cont.}$ &  - &-&-\\
%    \hline
%  SSDR1 & 3.11&1.22 &3.42$\times 10^{-4 &4.18$\times 10^{-4 &6.99$\times 10^{-3&3.95$\times 10^{-3&-&-&-\\
%  SSDR5 & 5.40&1.78 &2.56$\times 10^{-4 & 9.43$\times 10^{-4&1.08$\times 10^{-2&1.96$\times 10^{-4&-&-&-\\
%  SSDR7 & 3.23& 4.94& 1.31$\times 10^{-4 & 1.92$\times 10^{-4&4.21$\times 10^{-3&6.60$\times 10^{-3&-&-&-\\
%  SSDR10 & 2.94& 3.02& 2.19$\times 10^{-4&2.75$\times 10^{-3 &5.76$\times 10^{-3&1.13$\times 10^{-2&-&-&-\\
%  DEML10 & 4.10& 3.64&1.80$\times 10^{-4 & 2.71$\times 10^{-3&7.19$\times 10^{-3&0.00&-&-&-\\
%  DEML34 & 2.28& 7.04&2.75$\times 10^{-4 &3.13$\times 10^{-3 &5.92$\times 10^{-3&9.50$\times 10^{-3&-&-&-\\
%  DEM86 & 2.76 &7.92 & 4.03$\times 10^{-4& 3.07$\times 10^{-3&7.28$\times 10^{-3&0.00&-&-&-\\
%  DEML323 & 4.11& 4.80&9.84$\times 10^{-5 &1.52$\times 10^{-3 &3.70$\times 10^{-3&5.35$\times 10^{-3&-&-&-\\
  \hline
\multicolumn{13}{c }{DBP90 ($\alpha$, $a_{min}=1.2\,nm$,
  $a_{max}=15\,nm$) - Mathis RF} \\
  \hline
 Region  & $\chi^2/dof$ &$X_{ISRF}$  &$Y_{PAH^0}$  &$Y_{VSG}$
                                                              &$Y_{BG}$
                                                                        & $Y_{dust,tot}$
                                                                        &
                                                                          $I_{NIR\,cont.}$
                                                                                            &
                                                                                              $\alpha$
                                                                                                 &$a_{min}$
                                                                                                   &
                                                                                                     $a_{max}$&
                                                                                                                $\frac{Y_{VSG}}{Y_{BG}}$&$\frac{Y_{PAH}}{Y_{BG}}$ \\
         &&&&&&($10^{-2}$)&&&&($\times 10$)&($10^{-1}$)&($10^{-2}$)\\
  \hline
  MW & - & 1.00&4.30$\times 10^{-4}$&4.70$\times 10^{-4}$&6.40$\times
                                                           10^{-3}$&0.73&-&-2.6&1.2&1.5&0.73&6.72\\
  \hline
  SSDR1 & 1.89&1.02 &4.94$\times 10^{-4}$ &1.23$\times 10^{-3}$ &6.53$\times 10^{-3}$&0.83&3.43$\times 10^{-3}$&-0.1&- &-&1.88&7.57\\
  SSDR5 & 5.10 &1.66 &3.18$\times 10^{-4}$ & 1.40$\times
                                             10^{-3}$&1.04$\times
                                                       10^{-2}$&1.21&
                                                                      0.00&-1.87&- &-&1.35&3.06\\
  \hline
  SSDR7 & 2.88 & 4.96 & 1.51$\times 10^{-4}$& 3.35$\times 10^{-4}$&3.73$\times 10^{-3}$&0.42&6.10$\times 10^{-3}$&-1.20&-&-&0.900&4.05\\
  SSDR9 & 7.99 & 1.20 & 7.25$\times 10^{-5}$& 1.54$\times 10^{-3}$&9.36$\times 10^{-3}$&1.10&1.19$\times 10^{-3}$&-0.1&- &-&1.65&0.774\\
\hline
  SSDR8 & 3.49& 3.93 & 3.09$\times 10^{-5}$& 1.99$\times 10^{-3}$&2.21$\times 10^{-3}$&0.42&0.00&-1.66&- &-&9.02&1.40\\
  SSDR10 & 2.61 & 3.64&2.29$\times 10^{-4}$ &2.88$\times 10^{-3}$ &3.83$\times 10^{-3}$&0.69&1.06$\times 10^{-2}$&-2.13&- &-&7.52&5.98\\
  DEML10 & 4.07 &2.42 &3.09$\times 10^{-4}$ &4.66$\times 10^{-3 }$&7.34$\times 10^{-3}$&1.23&0.00&-2.33&- &-&6.35&4.21\\
  DEML34 & 2.25 & 6.54 & 2.77$\times 10^{-4}$ & 3.19$\times 10^{-3}$&6.71$\times 10^{-3}$&1.02&9.65$\times 10^{-3}$&-2.72&- &-&4.75&4.13\\
  DEML86 & 2.76& 8.25 & 3.95$\times 10^{-4}$ &3.03$\times 10^{-3}$ &6.84$\times 10^{-3}$&1.03&0.00&-2.54&- &-&4.43&5.77\\
  DEML323 & 3.74& 6.10&9.96$\times 10^{-5}$ &1.52$\times 10^{-3}$ &2.37$\times 10^{-3}$&0.40&5.18$\times 10^{-3}$&-2.15&- &-&6.41&4.20\\
  \hline
  \hline
\multicolumn{13}{c}{DBP90 ($\alpha=-2.6, a_{min}, a_{max}$) - Mathis RF}\\
\hline
 Region  & $\chi^2/dof$ &$X_{ISRF}$  &$Y_{PAH^0}$  &$Y_{VSG}$
                                                              &$Y_{BG}$& $Y_{dust,tot}$
                                                                        &
                                                                          $I_{NIR\,cont.}$
                                                                                            &
                                                                                              $\alpha$&
                                                                                                        $a_{min}$ & $a_{max}$ &$\frac{Y_{VSG}}{Y_{BG}}$&$\frac{Y_{PAH}}{Y_{BG}}$\\
    &&&&&&($10^{-2}$)&&&&($\times 10$)&($10^{-1}$) &($10^{-2}$)\\
  \hline
  SSDR1 &1.67 & 1.71&3.01$\times 10^{-4}$ &1.03$\times 10^{-3}$ &3.53$\times 10^{-3}$&0.48&3.36$\times 10^{-3}$&-2.6 & 5.87 &2.46&2.92&8.53\\
  SSDR5 & 4.92& 1.96&2.73$\times 10^{-4}$ & 1.46$\times
                                            10^{-3}$&8.25$\times
                                                      10^{-3}$&1.00&1.04$\times
                                                                     10^{-4}$&-2.6&2.40&2.11&1.77&3.31\\
  \hline
  SSDR7 &2.92 &5.11 &1.49$\times 10^{-4}$ &3.11$\times 10^{-4}$
                                                              &3.66$\times
                                                                10^{-3}$&0.55&6.07$\times
                                                                          10^{-3}$&-2.6&4.34&1.75&
                                                                                                             0.850&4.07\\
  SSDR9 &7.29 &1.90 &6.06$\times 10^{-5}$&1.61$\times 10^{-3}$
                                                              &4.95$\times 10^{-3}$&0.66&8.52$\times 10^{-4}$&-2.6&6.96&2.30& 3.25&1.22\\
\hline
  SSDR8 &3.53 &2.08 &8.44$\times 10^{-5}$ &3.29$\times 10^{-3}$ &3.62$\times 10^{-3}$&0.70&6.50$\times 10^{-4}$&-2.6&3.18&1.60&9.09&2.33\\
  SSDR10 &2.62 &2.95 &3.26$\times 10^{-4}$ &3.33$\times 10^{-3}$ &4.92$\times 10^{-3}$&0.86&1.05$\times 10^{-2}$&-2.6&2.79&1.49&6.77&6.63\\
  DEML10 & 3.87& 4.46&1.86$\times 10^{-4}$& 2.55$\times 10^{-3}$& 5.27$\times 10^{-3}$&0.80&0.00&-2.6&1.82&1.63&4.84&3.78\\
  DEML34 & 1.32& 3.94& 6.30$\times 10^{-4}$& 4.27$\times 10^{-3}$&1.32$\times 10^{-2}$&1.81&1.03$\times 10^{-2}$&-2.6&3.09&0.88&3.23&4.77\\
  DEML86 &2.08 & 4.52& 9.09$\times 10^{-4}$& 4.75$\times 10^{-3}$&1.35$\times 10^{-2}$&1.92&0.00&-2.6&3.43&1.00&3.52&6.73\\
  DEML323 & 3.36&3.66 &2.07$\times 10^{-4}$ &2.29$\times 10^{-3}$ &4.30$\times 10^{-3}$&0.68&5.52$\times 10^{-3}$&-2.6&3.29&1.27&5.33&4.81\\
  \hline
  $<$All$>$ & & & & &&& & &3.72 & 1.65&& \\
  $<$Diff.$>$ & & &&& & & & &5.65& 2.03&&\\
  $<$Mol.$>$ & & &&& & &  & &4.13 & 2.29&&\\
  $<$HII$>$ & & &&& & &  & &2.93 & 1.31&&\\

  \hline
  \hline
  \multicolumn{13}{c}{DBP90 ($\alpha=-2.6, a_{min}, a_{max}$) - 4-Myr RF}\\
\hline
 Region  & $\chi^2/dof$ &$X_{ISRF}$  &$Y_{PAH^0}$  &$Y_{VSG}$
                                                              &$Y_{BG}$& $Y_{dust,tot}$
                                                                        &
                                                                          $I_{NIR\,cont.}$
                                                                                            &
                                                                                              $\alpha$&
                                                                                                        $a_{min}$ & $a_{max}$ &$\frac{Y_{VSG}}{Y_{BG}}$&$\frac{Y_{PAH}}{Y_{BG}}$\\
    &&&&&&($10^{-2}$)&&&&($\times 10$)&($10^{-1}$)&($10^{-2}$)\\
  \hline
  SSDR1 &1.65 & 0.089 &1.13$\times 10^{-4}$ &6.07$\times 10^{-4}$ &3.36$\times 10^{-3}$&0.40&2.44$\times 10^{-3}$&-2.6 & 5.26 &3.51&1.81&3.36\\
  SSDR5 & 4.91& 0.090&1.06$\times 10^{-4 }$& 9.55$\times
                                                           10^{-4}$&9.01$\times
                                                                     10^{-3}$&1.00&0.00&-2.6&2.42&2.66&1.06&1.18\\
  \hline
  SSDR7 &2.93 &0.26&5.53$\times 10^{-5}$ &1.83$\times 10^{-3}$
                                                              &3.77$\times 10^{-3}$&0.57&5.10$\times 10^{-3}$&-2.6&3.32&2.45&
                                                                                                     4.85&1.46\\
 SSDR9 &7.27 &0.15 &1.62$\times 10^{-5}$&7.23$\times 10^{-4}$
                                                              &2.93$\times 10^{-3}$&0.37&3.64$\times 10^{-4}$&-2.6&6.90&3.73& 2.47&0.552\\
 \hline
  SSDR8 &3.33 &0.20 &1.61$\times 10^{-5}$ &1.12$\times 10^{-3}$ &2.04$\times 10^{-3}$&0.32&1.07$\times 10^{-4}$&-2.6&3.26&2.32&5.49&0.789\\
 SSDR10 &2.59 &0.13 &1.30$\times 10^{-4}$ &2.24$\times 10^{-3}$ &5.27$\times 10^{-3}$&0.76&9.07$\times 10^{-3}$&-2.6&2.72&1.98&4.25&2.47\\
  DEML10 & 3.92& 0.19&7.41$\times 10^{-5}$& 1.90$\times 10^{-3}$& 5.89$\times 10^{-3}$&0.79&0.00&-2.6&1.93&2.11&3.23&1.26\\
  DEML34 & 1.32& 0.22& 1.91$\times 10^{-4}$& 2.27$\times 10^{-3}$&1.26$\times 10^{-2}$&1.51&7.50$\times 10^{-3}$&-2.6&2.77&1.25&1.80&1.52\\
  DEML86 &1.96 & 0.22& 3.18$\times 10^{-4}$& 2.91$\times 10^{-3}$&1.41$\times 10^{-2}$&1.73&0.00&-2.6&3.08&1.39&2.06&2.26\\
  DEML323 & 3.34&0.16 &8.60$\times 10^{-5}$ &1.53$\times 10^{-3}$ &4.72$\times 10^{-3}$&0.63&4.16$\times 10^{-3}$&-2.6&3.37&1.70&3.24&1.82\\
  \hline
  $<$All$>$ & & & && && & &3.50& 2.31&&\\
  $<$Diff.$>$ & && && & & & &5.11 & 3.09&&\\
  $<$Mol.$>$ & && && & &  & &3.84& 3.09&&\\
  $<$HII$>$ & & &&& & &  & &2.86& 1.79&&\\
  \hline
  \hline
 \end{tabular}}
\end{center}
\tablefoot{Value
  of the reduced $\chi^2$ is given in column 2, intensity
  of the RF in column 3,
  abundances of different dust components in columns 4 to 6, total dust
  abundance in column 7, intensity
  of the NIR continuum in column 8, small grain size
  parameters in columns 9 to 11, ratio of the VSG
  abundance over the BG grain abundance in column 12, and
    ratio of the PAH abundance over the total Big
    grain abundance in column 13. Average values of
  $a_{min}$ and $a_{max}$ in nm, deduced from all
  regions, from diffuse (SSDR7 and SSDR9), molecular (SSDR1 and SSDR5) and
  ionized region (SSDR8, SSDR10, DEML10, DEML34, DEML86 and DEML323)
  are also given.  A null abundance is not possible for
  computational reason, and is set to \textit{1.00$\times$10$^6$} instead. Standard parameters for our Galaxy are also given for comparison. }
\end{table*}%[!h]}

%\section{Discussion}
%\label{sec_discussion}
\subsection{Dust over gas mass ratio}
For all models, we compute the dust over gas mass ratio, which
corresponds to the total dust abundance ($Y_{dust,tot}$, see
Tables \ref{table_chi2_AJ13} to \ref{table_chi2_DBP90}  and Tables \ref{table_60Myr_AJ13}  to \ref{table_60Myr_DBP90}) to check the reliability of the
models. Below are the reference values for our Galaxy:
0.74$\times$10$^{-2}$ (AJ13), 1.02$\times$10$^{-2}$ (MC11),
1.10$\times$10$^{-2}$ (DL07) and 0.73$\times$10$^{-2}$ (DBP90). In our
study we do not observe any inconsistent values. The dust over gas
mass can be increased by a maximal factor of 3 to 4 in some HII
regions or slightly decreased in some other regions. Statistically,
this study shows that the total dust accounts for $\sim$0.2$\%$ to $\sim$4$\%$ of the
total mass of the interstellar medium in the
LMC, depending on the region and the model. \citet{Roman-Duval22} observed variations by a factor of 4 of the
dust over gas mass, from low to high column densities, derived from metal depletions
in the LMC (see their Table 5). However, their ratio does not exceed
0.34$\%$, with an integrated value over all column densities of
0.23$\%$. 
One more time,
the RF has only a small impact on the derived total dust abundance. 

\section{Discussion}
\label{sec_discussion}
In this section we discuss the VSG population in the LMC, in term of
its relative abundance, its contribution to
the submm emission, as well as its possible link to the 70 $\mic$
excess observed in previous studies. We also evidence a different
behavior of the VSG size distribution depending on the environment,
that we explain by several possible scenarios of dust
 evolution. Different extinction curves calculated for each model and
regions are presented.

\subsection{VSG population in the LMC}
\subsubsection{Increase of the relative abundance}
\label{sec_increase_vsg_mass}
Results of the modeling indicate a clear increase of the VSG
abundance relative to the BG component compared to our Galaxy (see the last column of Tables
\ref{table_chi2_MC11} to \ref{table_chi2_DBP90}). For comparison, we give below the mass ratios of
the VSGs over the Big Grain component in our diffuse Galaxy derived from the
original models: 2.02$\times 10^{-2}$ (MC11), 1.69$\times 10^{-2}$ (DL07), 7.34$\times 10^{-2}$ (DBP90) and 1.30$\times 10^{-1}$ and 1.36$\times 10^{-1}$ in the MW Plane and MW Diffuse
medium (DBP90) from \citet{Bernard08}. 

\begin{figure*}
  \begin{center}
\includegraphics[width=16cm]{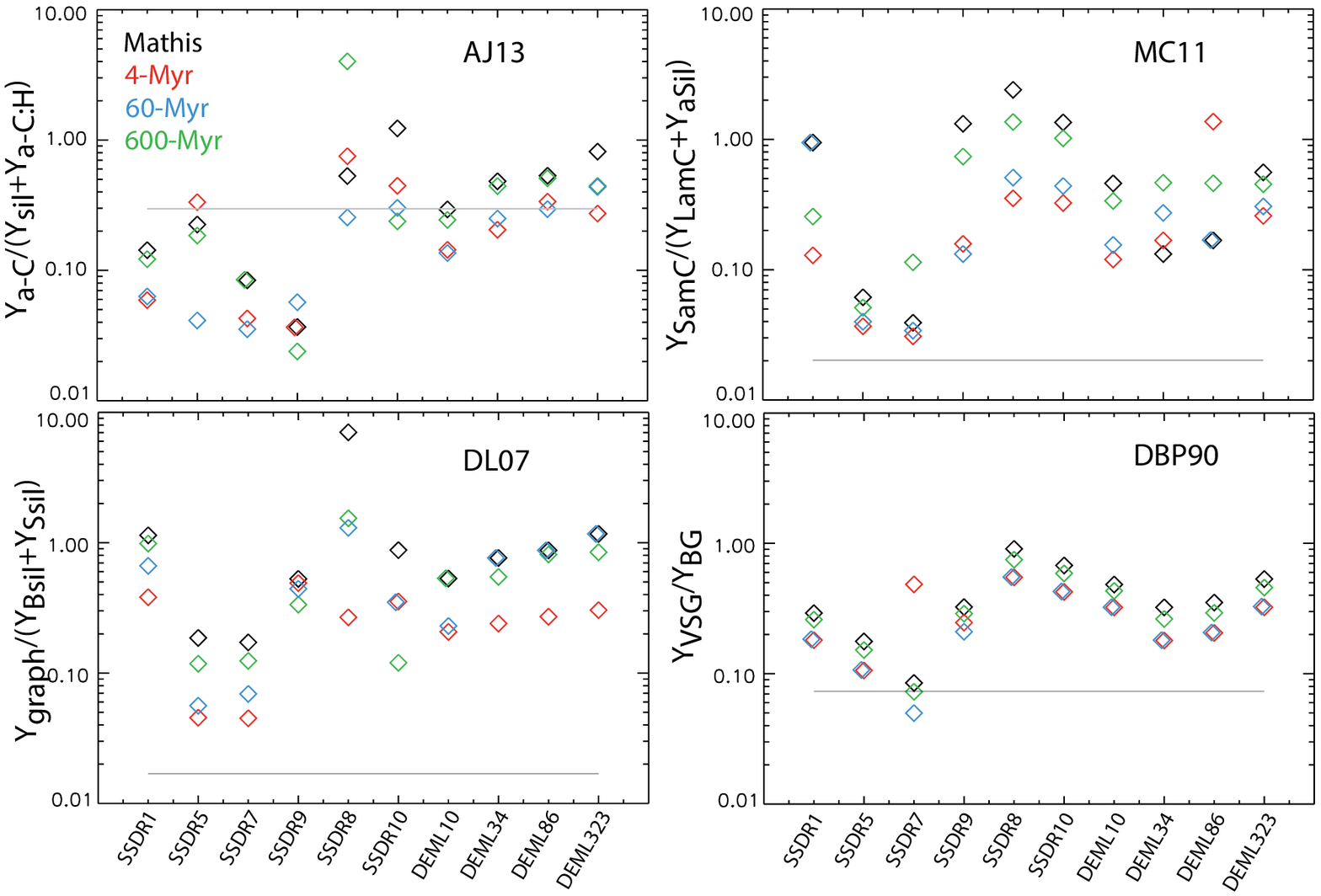}
\caption{Ratio of the small to the big grain component, for the
  different models and for the ten regions, using different RFs (Mathis in black, 4-Myr in
  red, 60-Myr in blue and 600-Myr in green). The Galactic value is
  indicated with the gray line.\label{fig_ratio_vsg}}
\end{center}
\end{figure*}

In our sample of regions, this ratio is significantly increased regardless of the
model (with a factor of 1.9 to 120 with MC11, 10 to 416 with DL07 and 1.2
to 12 with DBP90, see Fig. \ref{fig_ratio_vsg}). When changing the RF, the ratio is directly
affected and mainly decreases. However,
the VSG relative abundance is always higher than the Galactic diffuse
value (except for SSDR7 with DBP90 with the use of the 60 and 600-Myr
RFs). The SSDR5 (molecular) and SSDR7 (diffuse) regions show the lowest values when
compared to the other regions, with all the models and  RFs (except
with DBP90 when using the 4-Myr RF). The HII regions seem to have an
enhanced VSG relative abundance in comparison to diffuse/molecular
environments, even if the trend is not significative.

For AJ13, the MIR domain is
dominated by the a-C component which also describes the PAH
component. In some cases, this component can also dominate the FIR to
submm range. Even if it is not possible to directly compare with the other models, we
have computed the ratio between the abundance of 
the a-C component and the abundance of the other dust components (Olivine, Pyroxene and a-C:H, see Table \ref{table_chi2_AJ13}.  The ratio values are not systematically larger than the Galactic ratio (2.97$\times
10^{-2}$). However, the trend is similar to that of the other models: larger ratio values in the HII regions than in the diffuse and molecular regions.

\citet{Lisenfeld02} studied the dwarf galaxy
NGC1569 using SCUBA and IRAS data and used DBP90 as a model. They observed an increase of the
very small grain abundance relative to the large grains, by a factor
of 2 to 7 (depending on the radiation field and the dust size
distribution in the modeling) compared to the Solar Neighborhood. %Again, the
%LMC is not an unique galaxy with an enhancement of VSGs.
The authors
interpreted  this as the result of large grain shattering due to shocks (turbulence and supernovae remnants) in
the interstellar medium. 
%In the LMC, %\citet{Kim03} evidenced the
%fractal nature of HI that could result from supersonic turbulence
%\citep[see for instance][]{Elmegreen01, Miller21}.  

% If dust mainly originates
%from carbon stars, the large amount of carbon dust relative to the
%silicate dust could explain the emissivity behavior observed in the
%Magellanic Clouds.
%In addition, we know that the UV bump in the NIR-UV dust extinction is
%less important in the LMC and tends to disappear in the SMC \citep{Gordon03}. The small
%grains have not impact at all on the UV bump and we have seen that in
%some cases the abundance of large grains is even lower than that of
%VSGs (or at least significantly more important than in our
%Galaxy). That could explain the decrease and/or disappearance of the UV bump in some
%lines of sights of the Magellanic Clouds.

\subsubsection{Contribution to the submm emission}
The original dust models that describe the dust emission at high Galactic latitudes, show
a low contribution of the VSG component emission at long
wavelengths (see Table \ref{table_vsg}). For instance, depending on the model, the
contribution of the VSG emission to the total emission is in the range
1.4$\%$ -4.3$\%$ at 250 $\mic$, and between 1.2$\%$ and 13.7$\%$ at 1.1
mm in our Galaxy. DBP90 gives the highest VSG contribution in the
submm-mm, with values as high as 13.7 $\%$ at 1.1 mm . \citet{Compiegne11} studied the diffuse dust emission of the
inner Galactic plane using Herschel and Spitzer data. The authors
identified large variations of the VSG abundances. The
contribution of the VSGs in the FIR does not exceed 7$\%$ at 160
$\mic$. They computed the averaged SED into two extreme regions, in
terms of dust property variations. Looking at their Fig. 2, the maximum contribution of the VSGs at 500 $\mic$ is around
3-4$\%$.

In our study, which brings new constrains on the VSG component, we
already identified an increase of the VSG relative abundance (see
Section \ref{sec_increase_vsg_mass}) and changes in the dust size
distribution (see Section \ref{sec_vsgsize}) resulting in a significant increase of
the VSG contribution in the FIR-submm compared to the values observed in our Galaxy and
using Mathis RF (see
Table \ref{table_vsg}). This result is valid for all
models. For instance, depending on the region, at 500 $\mic$, the
contribution of the VSGs can be increased by a factor $\sim$5 to $\sim$28 with
MC11, $\sim$4 to $\sim$46 with DL07, and $\sim$3 to $\sim$10 with
DBP90. In the case of AJ13, except for SSDR1 and SSDR9 which evidence a decrease
of the contribution of the VSGs (represented by $I_{\nu}^{a-C}/I_{\nu}^{tot}$), the
increase goes from a factor of $\sim$1.4 to $\sim$11. As a consequence, one should be
careful when trying to model FIR to mm data by the use of big grains
only, especially when deriving dust mass or column density. However, there is clearly no way to 
predict the contribution of the VSG to the SED fit 
depending on the type of regions. Indeed, diffuse and HII regions
can have a wide range of values of the VSG contribution in each
type of regions.

The values of the VSG contribution are
significantly different (mainly lower) when changing the RFs (see
Tables \ref{table_vsg} and \ref{table_vsg_annexe}). This behavior is expected
because the adopted RFs are bluer than the one of Mathis. As opposed to BGs
which absorb energy over the whole RF spectrum, PAHs and VSGs
absorb most of their energy at short wavelengths. As a consequence, PAHs and VSGs absorb and reemit
more energy with a bluer RF. To produce the same IR
brightness with a bluer RF the model roughly needs
less PAH and VSG abundances. However, a few cases do not follow this
expected behavior because other parameters, such as VSG size
distribution, are at play and are linked to the abundances. 

The VSG contributions are nevertheless most of the time higher than the Galactic
values, DL07 having the lowest contribution, except for SSDR1 and 
SSDR9, compared to the other models. These contributions vary from region to region, and from model to model
with no clear trend with the environment, making
predictions impossible. % Est-ce que je parle des PAHs .????On the opposite, the 

%RAJOUTER VALEURS DANS TABLE A-C/total EMISSION POUR AJ13 ET DISCUTER
%LES VALEURS DANS LE TEXTE OU PAS ... A VOIR

%By analyzing the contribution
%between 500 $\mic$ and 1.1 mm it is possible to see if the VSG
%emission can explain any flattening of the dust emission in the submm
%to mm wavelengths. With DL07 and MC11 values stay reasonably stable in
%this wavelength range which 

\begin{table*}[!h]
\caption{Ratio of the small grain component emission to the total dust
  emission for each best-fit model using the Mathis (top table) and
  4-Myr RF (bottom table)
  , at 250, 500, 850 and 1100
  $\mic$. \label{table_vsg}}
\begin{center}
\resizebox{\textwidth}{!}{%
\begin{tabular}{lcccc|cccc|cccc|cccc}
\hline
 \hline
%\multicolumn{10}{c}{DBP90} \\ 
% \hline
 &&\multicolumn{13}{c}{Mathis RF}&&\\
  &\multicolumn{4}{c}{ AJ13} & \multicolumn{4}{c}{ MC11} & \multicolumn{4}{c}{ DL07}& \multicolumn{4}{c}{ DBP90} \\
 Best modeling & \multicolumn{4}{c}{ ($\alpha$, $a_{min}$, $a_{max}$)} & \multicolumn{4}{c}{ ($a_0$)} & \multicolumn{4}{c}{ ($a_0$)} &
                                                          \multicolumn{4}{c}{ ($\alpha=-2.6$,
                                                           $a_{min}$, $a_{max}$)} \\ 
& \multicolumn{4}{c}{$I^{a-C}_{\nu}/I^{tot}_{\nu}$ ($\%$)} &
                                         \multicolumn{4}{c}{$I^{SamC}_{\nu}/I^{tot}_{\nu}$ ($\%$)} &\multicolumn{4}{c}{$I^{graph.}_{\nu}/I^{tot}_{\nu}$ ($\%$)} & \multicolumn{4}{c}{$I^{VSG}_{\nu}/I^{tot}_{\nu}$ ($\%$)} \\
Wavelengths & 250 & 500 & 850 & 1100 & 250 & 500 & 850 & 1100 & 250 & 500 & 850 & 1100 & 250 &
                                                                500 &
                                                                850 &
                                                                1100 \\
  \hline
  Milky-Way & 7.1 & 7.9 & 8.9 & 9.4 & 1.4 & 1.4 & 1.7 & 1.8 & 1.7 &
                                                                    1.5 & 1.3&  1.2 & 4.3 &
                                                              6.4 &
                                                              10.6 &
                                                              13.7 \\
  \hline
  SSDR1 & 3.4 & 2.7 & 2.5 & 2.4  & 42.0 & 39.8 & 39.6 & 39.5& 75.5& 69.3 & 63.4 & 60.4 & 36.8
                                                                 &
                                                                 53.0
                                                                 &
                                                                 67.5
                                                                 &
                                                                 73.6 \\
 SSDR5 & 31.7 & 36.9 & 42.6 & 45.6 & 6.8 & 6.8 & 7.5 & 7.9 & 6.1 & 5.8 & 5.9 & 6.0& 21.2 & 33.7 &
                                                        48.1 & 55.6 \\
  \hline
  SSDR7 &  9.5 & 11.4 & 13.1 & 13.9 & 10.3 & 10.7 &12.2 & 13.1& 21.1 & 19.1 & 17.0 & 16.0& 13.9
                                                                  &
                                                                  24.4
                                                                  &
                                                                  37.2
                                                                  &
                                                                  44.2
  \\ SSDR9 & 2.0 & 1.6 &1.5 & 1.4 & 50.8 & 48.5 & 48.2 & 48.2 & 57.9 & 51.2 & 44.7 & 41.7& 39.2 & 55.5&
                                                        69.6 & 75.5 \\
\hline
  SSDR8 & 86.1 & 85.6 & 85.7 &85.7 & 53.3 & 52.2 & 53.0 & 53.3 & 72.3 & 69.8 & 70.2 & 70.5& 48.4 & 62.2 &
                                                        74.8 & 80.0
  \\
   SSDR10 & 60.9 & 63.4 & 66.1 & 67.2 & 39.8 & 39.2 & 40.1& 40.5& 34.8 & 32.5 & 31.1 & 30.3& 41.6 &
                                                                56.2&
                                                                70.0 &
                                                                75.8 \\
  DEML10 & 14.1 & 12.9 & 12.4 & 12.2  & 13.5 & 13.7 & 14.4 & 14.6& 15.7 & 15.2 & 15.3 & 15.3& 41.2
                                                                 &
                                                                 57.6
                                                                 &
                                                                 71.3
                                                                 &
                                                                 76.9 \\
  DEML34 &  30.5 & 37.9 & 45.6 & 49.7 & 17.4 &17.8 & 18.6 & 18.9& 13.7 & 13.8 & 14.7 & 15.1& 13.5
                                                                 &
                                                                 20.5
                                                                 &
                                                                 31.3
                                                                 &
                                                                 37.9 \\
  DEML86 & 42.1 & 49.8 &57.8 &61.8 & 20.9 & 22.4 & 25.6 & 27.4& 19.6 & 19.4 & 20.1 & 20.4& 19.4&
                                                                29.5&
                                                                42.8 &
                                                                50.1 \\
  DEML323 &52.2 &57.2& 62.1 & 64.4 & 32.9 & 33.1 & 34.8 & 35.6& 25.2 & 24.5 & 25.3 &25.7& 32.9
                                                                  &
                                                                  46.7
                                                                  &
                                                                  61.3
                                                                  &
                                                                  68.1 \\
\hline
  \hline
 &&\multicolumn{13}{c}{4-Myr RF}&&\\
  SSDR1 & 2.3& 1.6 & 1.4 &1.4  & 11.9 & 8.5 & 8.5 & 8.7 & 32.9
                                                                  &
                                                                  21.9
                                                                  &
                                                                  19.3
                                                                  &
                                                                  18.6
        & 17.5 & 20.0& 29.4 & 35.8
                                                                 \\
 SSDR5 & 46.2 & 50.1 & 57.8 & 62.0 & 2.6 & 1.7 & 1.8 & 2.0 & 0.8
                                                                   &
                                                                   0.6
                                                                   &
                                                                   0.6
                                                                   &
                                                                   0.7
        & 7.6 & 7.7 & 11.5 & 14.6 \\
  \hline
  SSDR7 & 6.0& 7.3& 9.0 & 10.0 & 6.2 &4.5 & 5.1 & 5.7 & 2.5 &
                                                               2.0 &
                                                               1.9 &
                                                               1.9 &
                                                                     4.2 
                                                                  &
                                                                     5.6 
                                                                  &
                                                                   9.1 
                                                                  &
                                                                  11.9
  \\
  SSDR9 & 4.6 & 3.7& 3.4 & 3.3 & 9.6 & 7.4 & 7.4 &7.4 & 66.1
                                                                &
                                                                45.4
                                                                &
                                                                36.2
                                                                &
                                                                32.9 &
                                                                       26.3
                                                                       &
                                                                       33.9
                                                                       &
                                                                       47.5 
                                                                       & 55.0\\
\hline
  SSDR8 & 47.4 & 41.8 & 42.0 & 42.3 & 11.6 & 9.3 & 9.6 & 10.0 &
                                                                      6.2
                                                                      &
                                                                      4.8
                                                                      &
                                                                      5.0
                                                                      &
                                                                      5.2
        & 30.1& 34.5 & 46.5 & 53.6 \\
   SSDR10 & 33.0 & 32.0 & 34.9 & 36.3 & 9.9 & 8.1 & 8.5 & 8.8 &
                                                                      7.5
                                                                      &
                                                                      5.3
                                                                      &
                                                                      5.5
                                                                      &
                                                                      5.8
        & 18.5 & 18.6 & 26.0 & 31.4  \\
  DEML10 & 8.8& 7.9 &8.7 &9.1 & 2.7 & 2.4 & 2.6 & 2.7 & 3.7
                                                                 &
                                                                 2.8
                                                                 &
                                                                 3.0
                                                                 &
                                                                 3.1 & 
                                                                 17.0
                                                                       &18.9
                                                                 &
                                                                       27.3 
                                                                 & 33.2
                                                                 \\
  DEML34 &  11.7 & 13.4 & 18.6 & 22.0 & 7.2& 5.8 &6.6 & 7.2 &
                                                                     2.9
                                                                     &
                                                                     2.3
                                                                     &
                                                                     2.6
                                                                     &
                                                                     2.8
        & 4.7 & 4.5 & 6.6 & 8.4 \\
  DEML86 & 27.5 & 30.9 & 38.9 & 43.5 & 16.1 & 11.9 & 13.9 & 15.6
  & 4.3 & 3.5 & 3.8 & 4.0 & 6.5 & 6.5 & 9.6
                                                                & 12.1
                                                                \\
  DEML323 & 21.0 & 20.8 & 23.9 & 25.5 & 10.9 & 8.8 & 9.5 & 10.0
  & 5.1 & 3.9 & 4.2 & 4.5 & 12.6 & 12.6 & 18.0 & 22.3 \\
\hline
  \hline

\end{tabular}}%[!h]}
\end{center}

   %    \resizebox{\textwidth}{!}{%
\tablefoot{The second line indicates the dust size parameters of
  the small grain component used in the modeling.}
\end{table*}

\subsubsection{70 $\mic$ excess and submm flattening}

\citet{Bernard08} and \citet{Paradis11a} evidenced a 60-70 $\mic$
excess mainly in the neutral medium of the LMC as well as in the
diffuse ionized gas, with the use of standard parameters in the
DBP90 model. Nevertheless, it does not exclude the fact that this excess is
sometimes observed in the highy ionized gas or molecular gas
of some regions. According to the authors the OI (63 $\mic$) or OIII (58 and and 88 $\mic$)
emission lines are not responsible for this excess even though, \citet{Oliveira19}
observed an enhancement of these lines with respect to the dust
continuum in photo-dissociated regions
with Young Stellar Objects (YSO) when compared to Galactic YSOs. 
By changing the slope of the VSG dust size
distribution \citet{Bernard08} were able to reasonably reproduce the data with a
power-law distribution with an arbitrary value of $\alpha$=-1
(instead of 2.6). 
Figure \ref{fig_all_noslope} shows the MIPS 70 $\mic$
photometric data normalized to the integrated flux in the
MIPS-SED band (see Section \ref{sec_sed_construction}) as orange
diamonds. The integrated brigthness in
the MIPS 70 $\mic$ band derived from DBP90 is shown with the blue
asterisks at the same wavelength. We observe a significant excess at 70 $\mic$ in several regions
such as SSDR1, SSDR8, and SSDR9, and a softened excess in SSDR5,
and SSDR10. This excess tend to disappear in all DEML
regions. DBP90 with standard parameters is clearly not able to reproduce the 70 $\mic$ data in
some cases. In addition, the influence of the Oxygen lines in the 70
$\mic$ could be negligible to the total flux since the integrated flux in the MIPS-SED band does not deviate from the
spectroscopic data even if some gas lines are observed. \\
By changing the VSG dust size distribution ($\alpha$, $a_{min}$ and $a_{max}$), we can see that the
excess does not appear anymore (see Fig. \ref{fig_all_best}). This study reinforces the idea that in
the framework of DBP90, the VSG size distribution
in some regions of the LMC is different than that in our
Galaxy. The analysis of the dust size distribution for all models is
further discussed in the following Section \ref{dustsize_vsg}.
 
It is known that the dust emission spectrum of the LMC 
shows a flattening in the submm compared to that of our Galaxy, which
 is even more pronounced in the SMC. At the same time,
the SMC also shows a much larger excess of emission at 70 $\mic$ than
 the LMC \citep{Bernard08,Bot10}. We can therefore examine wether there is a
possible link between the 70 $\mic$ excess and the submm
flattening. For instance, in the LMC, this 70 $\mic$ excess and 
the submm flattening tend to disappear in most of the molecular gas phase,
 \citep{Paradis19}.  Note that this is not necessarily the case in molecular regions 
 that may mix HI and CO gas phases. However, changing the VSG size distribution does not
seem to have an impact on  the submm flattening as the diffuse and HII regions 
have a distinct size distribution while exhibiting submm flattening \citep{Paradis19} (see Section \ref{dustsize_vsg}). 
In the same way, the
increase in the VSG relative abundance in the ionized gas of the LMC
highlighted by \citet{Paradis19} could explain the submm flattening
observed in the ionized regions, but not in the atomic ones.
In conclusion, a change in the size distribution or the relative 
abundance of VSGs or a combination of both could explain the difference in the submm 
emission observed in the LMC compared to our Galaxy.
 %In conclusion, either the size distribution in some cases, \textbf{or} the VSG relative
%abundance in some others, or the combination of both could explain
%the submm behavior observed in the LMC, when compared to our Galaxy. 
Therefore one could expect that the adequate change in the VSG size
distribution and abundance in the SMC could help reproducing the 70 $\mic$
excess identified in \citet{Bot04}, and could result in a large contribution of the VSG
emission in the submm-mm. %However, the use of the 70 $\mic$ alone only give an information
%on the fact that a change in the VSG size distribution is required but
%does not allow to give strong constraints on the size distribution.

%In all models, VSG grains are mainly composed of carbons. Carbon
%grains have a flatter spectrum than silicate grains and can explain
%spectral indices of 1 in the submm-mm range.
If dust mainly originates
from carbon stars \citep{Boyer12}, the large amount of carbon dust relative to the
silicate dust could explain the emissivity behavior observed in the
LMC. In other words, small carbon grains (or a combination
of small and big carbon grains) could be responsible for the
general behavior of the submm-mm flattening in the emission
spectrum. For the SMC, the flattening seems to be too pronounced to be
explained by carbon grains only. Indeed, in a previous study
(unpublished) we found a submm emissivity spectral index of 0.9 in the SMC
using IRIS \citep[new processing of IRAS data][]{Miville05} and Planck data.
In addition, variations of the submm
flattening have been observed in the diffuse medium of our Galaxy
whereas the amount of
VSG does not have any impact on the Galactic submm emission due to its
low contribution at long wavelengths. The negligible submm emission
 from VSGs in our Galaxy, which shows a submm excess, indicates
  that the VSG component alone cannot be responsible of the submm
  excess observed in our Galaxy. Therefore, other processes might be at play, such
as TLS (Two-Level-System) processes proposed by \citet{Meny07},
describing the amorphous state of large dust grains to explain the submm 
behavior observed in our Galaxy. The TLS model is able to reproduce 
the different dust emission behavior observed in our
Galaxy \citep{Paradis11b, Paradis12, Planck14XIV} and in the Magellanic Clouds \citep{Planck14XVII}. 
More recently, the TLS model 
  was also fully able to reproduce observations in molecular complexes of
  our Galaxy such as the Perseus molecular cloud and W43 \citep{Nashimoto20} 
%Moreover, \citet{Nashimoto20} studied intensity and
%polarization SEDs of the Perseus molecular cloud and W43 and concluded
%that the TLS model was fully able to explain anomalous microwave
%emission without introducing new species.

To summarize, since in our Galaxy the VSG component emission is negligible
in the submm range, the VSG contribution alone cannot be the origin of the
submm excess. It is therefore most likely that the very pronounced and important 
submm flattening evidenced in the Magellanic Clouds originates from a combination of at least two
emission processes: the emission from the VSG component plus the 
TLS processes in large grains;  whereas only the TLS processes could be responsible for the  
local variations observed in the diffuse regions of the MW. 

\subsection{VSG size distribution: Diffuse versus HII regions}
\label{dustsize_vsg}

First, we compare the results obtained with the different 
models using the Mathis RF. For the same reason as in Section \ref{sec_increase_vsg_mass}, AJ13 is not
discussed here since the a-C component includes both PAHs and small
grains. We observe that the fits are
significantly improved when changing some
parameters of the VSG dust size distribution. The values of $a_0$
presented in Tables \ref{table_chi2_MC11} and
\ref{table_chi2_DL07} show some variations from one model to the other. 
We observe values of $a_0$ going from 2 nm to 4.6 nm
for MC11 and from 1 nm to 3.5 nm for DL07. Values
for MC11 are close to the Galactic value of 2 nm in HII
regions, whereas the values systematically increase in the diffuse
regions. For DL07 all the values decrease compared to the Galactic
ones. We observe the same trend for MC11: the values in HII
regions are significantly lower than those in diffuse
regions. Indeed, the mean value of $a_0$ in each type of environment
is equal to 2.47 nm/1.34 nm in all HII regions with MC11/DL07, whereas the
value is 4.17 nm/3.11 nm in diffuse regions with
MC11/DL07. Molecular regions evidence intermediate values between
these two extreme environments (diffuse and HII regions). We caution
however that the mean value for each type of environment has been
obtained with only two values for the diffuse and the molecular medium,
and with six values for the ionized medium. This shift in the central value of the log-normal VSG size
distribution in the different type of environments shows that HII regions
contain mostly smaller VSGs (and less large grains in this component),
and also an increase in the amount of small
VSGs in comparison with the diffuse regions. 
For DBP90, the fits show lower values for both 
$a_{min}$ and $a_{max}$ (see Table \ref{table_chi2_DBP90}) in HII regions (mean values
$a_{min}$=2.93 nm and $a_{max}$=13.1 nm) when  compared to diffuse
regions (mean values $a_{min}$=5.65 nm and $a_{max}$=20.3 nm), showing
again smaller VSGs in HII regions than in the
diffuse ones. 
The modeling with different RFs (Tables \ref{table_60Myr_AJ13} to
\ref{table_60Myr_DBP90}) does not change these conclusions and confirms the trend observed with the use of the Mathis RF.

%Figure \ref{fig_alpha} shows the values of $a_0$ or $a_{min}$/$a_{max}$
%obtained with the four distinct interstellar RFs, for each
%model. With the exception of SSDR5, we observe higher values of the
%parameters for
%diffuse/molecular regions as compared to the ionized regions. The
%modeling with different RFs (see also Tables \ref{table_60Myr_AJ13} to \ref{table_60Myr_DBP90}) does not change the conclusions but either
%confirm the trend observed with the use of the Mathis RF. 

%the slope of the
%power-law distribution is different when fitting with or with the
%$a_{min}$ and $a_{max}$ parameters. However, absolute values of
%$\alpha$ go to higher values in UCHII regions than in diffuse
%regions. Again, this result indicate that the VSG size distribution needs
%mainly more smaller grains and less bigger grains in UCHII regions. In
%addition, values of $a_{min}$ and $a_{max}$ are in the same order of
%magnitude as in our Galaxy. 

To summarize, our results indicate the same trend for all models,
regardless the RF : the size distribution of VSGs is different in
HII and diffuse regions with an increase in the quantity of small VSGs (and fewer large VSGs) in HII regions when compared to diffuse LMC regions. %Again, since our diffuse
%regions are mainly molecular clouds (except SSDR7), atomic and
%molecular gas phases are mixed.
In this analysis, the SEDs represent the mixture of all gas phases
(except for SSDR7 and SSDR9 that are almost pure atomic regions). For
instance, SSDR1 and SSDR5 have a high level of HI emission, and most
of the ionized regions of our sample have large column densities
 in the ionized gas as well as in other gas phases. We therefore
expect a larger dispersion in the VSG size parameters, i. e. even more
pronounced results, when analyzing
dust emission associated with each gas phase independently. For
instance, in SSDR8, DEML10 and DEML323 regions, the amount of Hydrogen
column density in the HI gas phase is higher than in the H$\alpha$ gas phase. We therefore expect lower values of amin/amax or a0 (depending on the model) in these three regions, when looking at the H$\alpha$ gas phase only. 
%\citet{Bernard08} and \citet{Paradis11} identified a change in the dust size
%distribution mainly in the atomic phase, not in the molecular one. We
%therefore expect to observe more larger VSGs in the atomic phase. 

%We can extract the mean parameters of the size distribution for each
%environment (see Tables \citet{table_chi2_MC11},
%\citet{table_chi2_DL07}, and \citet{table_chi2_DBP90}). For DL07, we
%only consider the $a_0$ parameter, because the modeling with $a_{min}$ and
%$a_{max}$ is slightly improved for only three regions and does not
%have any impact between 20 and 90 $\mic$. We therefore favor a model
%with the minimal number of free parameters.
%DONNER LES VALEURS DE LA TABLE QQ PART DANS LE TEXTE (peut etre
%enlever table DL07 (a0,amin,amax)

%In addition, values of $a_{min}$ are
%significantly higher (by a factor 2 to 4) than the Galactic value
%in UCHII regions, whereas values of $a_{max}$ are more concentrate
%around the Galactic value (1.5$\times 10^6$), with values fluctuating
%between 1.1 and 2.5$\times 10^6$) for all envinronments. As a consequence, the range in the
%size of the VSG grains could be reduced in UCHII regions, as compared
%to our Galaxy, by
%rejecting very small VSG grains. 

%To summarize, our results indicate that the VSG size distribution could
%favor the presence of more small grains (and less big grains) in UCHII
%regions as compared to diffuse regions. In addition, the minimal size
%of VSG in UCHII regions could be slightly higher than the Galactic
%value. 

\subsection{VSG lifecycle}

\begin{figure}
  \begin{center}
\includegraphics[width=8.5cm]{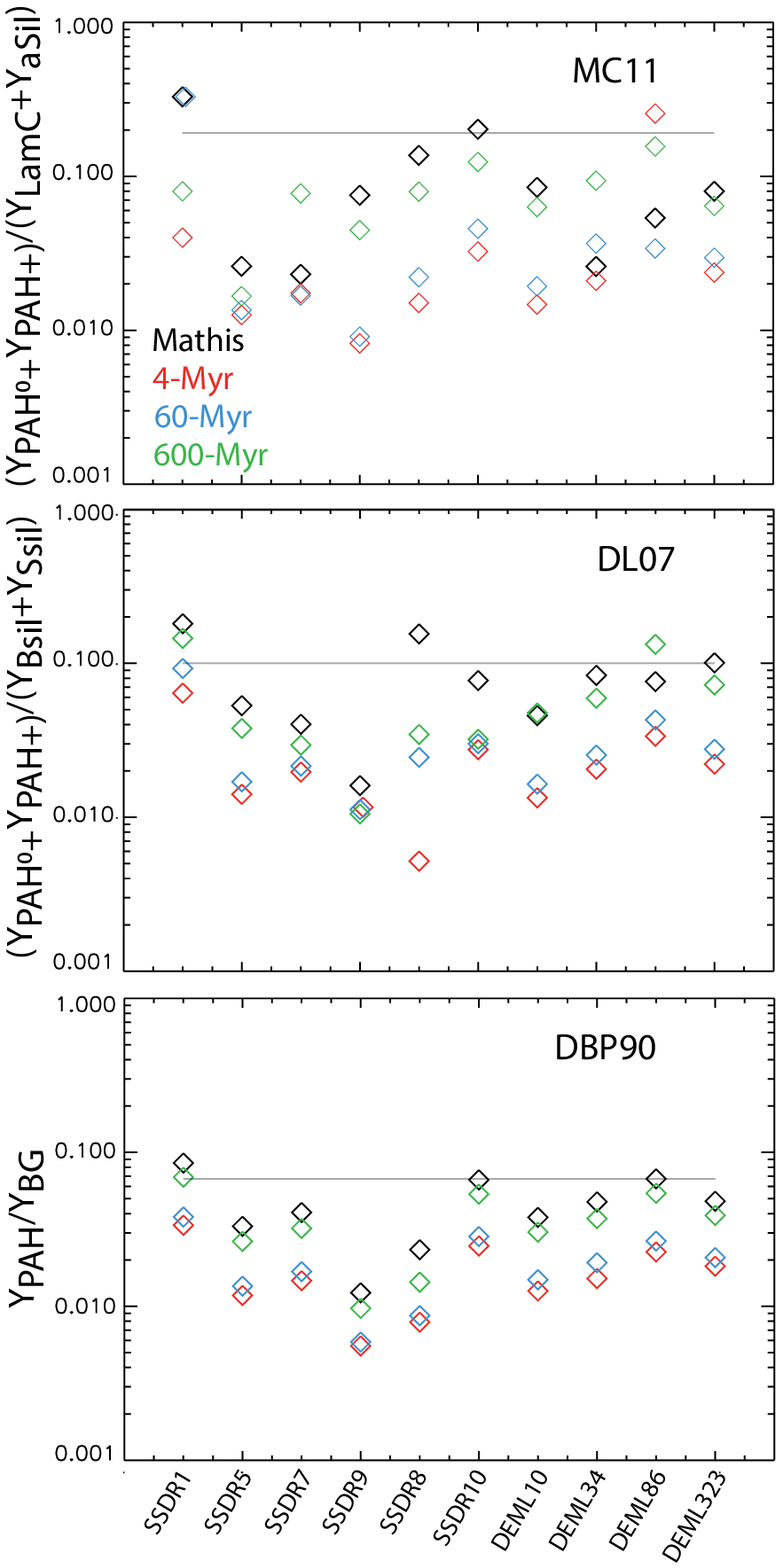}
\caption{Ratio of the PAHs to the big grain component abundance, for the
  different models and for the ten regions, using different RFs (Mathis in black, 4-Myr in
  red, 60-Myr in blue and 600-Myr in green). The Galactic value is
  indicated with the gray line.\label{fig_ratio_pah}}
\end{center}
\end{figure}

The results of this analysis show a significant increase 
in the VSG relative abundance compared to the MW. They also show an increase in smaller VSGs, compared to larger ones, in HII regions compared to the atomic medium.
Some scenarios emphasize that, while Galactic dust could
mainly be produced by O-rich AGB stars \citep[67$\%$ and 20$\%$ from O-rich
and C-rich AGBs,][]{Gehrz89}, most of the dust in the Magellanic Clouds
could originate from carbon stars (extreme AGB, i. e. mostly embedded
carbon stars) with a dust production reaching 61$\%$ and 66$\%$ in the
LMC and SMC \citep{Boyer12}. The dust produced in such environments could consist in small carbon grains. Additional sources of dust production are
supernovae, although their dust-destruction rates remain poorly
constrained, and dust growth in the ISM.
Another explanation to the enhancement of small carbon grains could be the destruction of large grains into smaller
grains. 

Regarding the enhancement of small VSGs in HII regions, two
options could explain this behavior.  First, (small) VSGs could be formed 
in HII regions rather than in molecular clouds, as proposed by
\citet{Paradis19}. Indeed, in their study they have shown that the
relative abundance of VSGs is enhanced in the ionized phase of the
gas, whatever the nature of the clouds, ie quiescent or forming stars,
and independently of the intensity of the radiation field. Such VSGs could
be large PAH clusters or cationic PAH clusters
\citep{Rapacioli05, Rapacioli11, Roser20}.
Then, whatever the nature of these grain species,
grain growth could occur in the atomic and molecular regions, via accretion or
coagulation, and could explain the presence of larger VSGs in the
molecular and diffuse environments.  
To examine the hypothesis that PAH clusters could be responsible
  of the VSG increase in the LMC, we present in Fig. \ref{fig_ratio_pah} the PAH relative abundance
  for each model (except AJ13) and region. First,
for all models, the PAH relative abundance is lower than the  Galactic value in agreement with the hypothesis of the presence of PAH clusters. 
However, Fig. \ref{fig_ratio_pah} does not show any trend with the
nature of the region, unlike in the case of the VSGs. This 
absence of real trend suggests that other processes might occur. 
For instance, the BG component could be also affected by
strong shocks and turbulence, and contribute to increase the relative
abundance of VSGs, as observed in the LMC.  In this second option, 
the largest VSGs could be destroyed in HII regions in shocks,
 resulting in an increase of the amount of smallest VSGs. This effect could be the
result of supernova explosions or turbulence. This hypothesis could explain the
changes in size of the VSG population.  
As these two processes most likely both occur, the
VSG component could include two distinct populations: small VSGs
originating from large PAH clusters or cationic PAHs clusters, mainly
formed in HII regions, and large VSGs resulting from BG destruction.   

%The second option could be the largest VSGs that could be destroyed in UCHII regions in shocks,
%producing an increase of the smallest VSGs. This effect could be the
%result of supernova explosions or turbulence. This hypothesis only concerns the
%changes in size of the VSG
%population. In addition, the BG component could be also affected by
%strongs shocks and turbulence, and contribute to increase the abundance of VSGs,
%observed in the LMC. In that case, the large VSGs would definitively not result from
%PAH clusters.  \\
%If this two options play a role, the VSG component could account for two distinct populations: small VSGs
%coming from large PAH clusters or cationic PAHs clusters, mainly
%formed in UCHII regions, and large VSGs resulting from BG destruction.     
\citet{Heiles00} analyzed the Barnard's Loop HII region and evidenced
an increase of the 60 $\mic$ emission. They showed that this increase
is not the consequence of the presence of warm big grains and proposed that it could be due to an
enhancement of the VSG population relative to BGs in the ionized
region compared to the global neutral medium. As a consequence it
appears that the increase in the VSG relative abundance in the ionized
medium could be a more general result than just an isolated result
concerning the LMC. 
\citet{Jones96} have developed an analytical
model to derive the fragment size distribution as well as the final
crater mass in grain-grain collisions depending on different
parameters (grain properties, sizes and collision
velocity). They found that grain shattering leads to the redistribution of the dust
mass from large grains into smaller grains.   
More recently, \citet{Hirashita10} has theoretically studied
interstellar shattering of large grains (a$\sim$0.1 $\mic$) to explain
the production of small grains. They were able to reproduce the small
grain abundance derived by \citet{Mathis83} in the warm neutral
medium. They also showed that additional shattering in the warm
ionized medium could destroy carbonaceous grains with a size of
$\sim$0.01 $\mic$ and generate smaller grains. On the opposite
silicate grains are harder to shatter than graphite. However, in this
study (see Table \ref{table_chi2_DL07}), we observe in four regions high abundances
of small silicates compared to large silicates with DL07. 
According to the theory of 
\citet{Hirashita10}, we cannot explain this
enhancement of small silicates as resulting from large silicate
destruction nor from another source of production.

A recent analysis of the turbulence in the LMC \citep{Szotkowski19}
evidenced spatial variations of HI turbulent properties. The turbulence
is often characterized by estimating spatial power spectrum (SPS) of
intensity fluctuations. The authors pointed out several localized
steepening of the small-scale SPS slope around HII regions, and around 30 Doradus
in particular, in agreement with numerical simulations \citep{Grisdale17, Grisdale19} 
suggesting steepening of the SPS slope due to stellar feedback eroding 
and destroying small clouds. This study is in agreement with the possible additional
grain shattering in the ionized medium, i. e. in HII regions, where we
observe the smallest VSG populations when compared to the atomic and
molecular environments.

\subsection{Extinction}
For the four tested models, it is always possible to find a set of parameters
that correctly fits the emission part of the SED. Fitting the dust emission from NIR to FIR wavelength 
therefore does not allow us to unambiguously determine which dust models best reproduce the 
observations. In the context it is interesting to check wether these best-fit models also 
show similar extinction curves or whether extinction data could help to differentiate between 
the different grain models. 

Comparing the modeled extinction curves with 
observations is not easy because we do not have extinction
curves associated to the studied regions. In the Milky Way, a wide variety of possible shapes of these curves has
been observed \citep{Papaj91, Megier97, Barbaro01, Wegner02, Fitzpatrick07, Gordon21} and 
we can similarily expect the extinction curves in the LMC have different profiles. 
This was observed by \citet{Gordon03} who found that the LMC averaged extinction curve is characterized 
by a stronger far-UV rise than the one of the MW, and by a weaker 2175 Å
bump (see for instance LMC2-supershell, near the 30 Doradus 
starbursting region).
The authors argue that the difference between the extinction curves of the
Magellanic Clouds and the MW could be due to the fact that the observed environments are different. 
For the Magellanic Clouds, the extinction
curves are observed in active star formation regions where large grains could
be destroyed by strong shocks and/or UV photons. Nevertheless, the sample of LMC extinction
curves is quite limited compared to that of our Galaxy.
\begin{figure*}
 \begin{center}
\includegraphics[width=14cm]{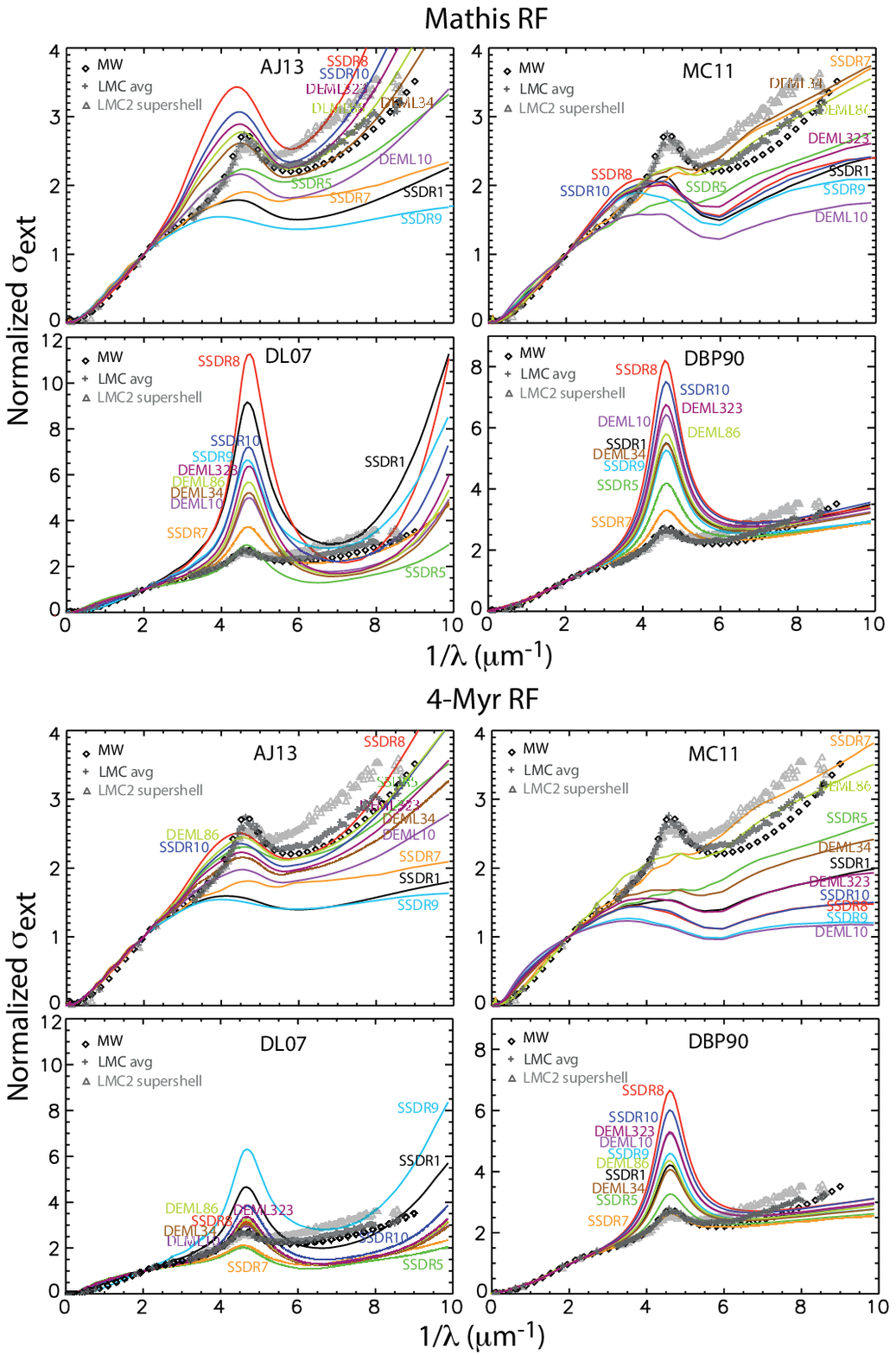}
\caption{Extinction curves for each region, derived from the different
  dust models, with Mathis RF (top) and 4-Myr RF (bottom). The
  averaged Galactic and LMC extinction curves are given in black
  diamonds and dark gray crosses for comparison. The LMC2 supershell
  extinction curve is presented with light gray triangles. Caution: the scales have been chosen to show the difference between
the different models in a clear way, and one should take this into account in the comparisons.\label{fig_ext}}
\end{center}
\end{figure*}

The extinction curves calculated by DustEM for 
the different dust best-fit models are shown in Figs. \ref{fig_ext} and \ref{fig_ext_annexe} 
as well as, for comparison, the average LMC and LMC2-supershell extinction curves \citep{Gordon03} and the average Galactic 
extinction curve  representative of the interstellar medium \citep{Cardelli89}. 
We notice that the predicted extinction curves are highly model dependent, 
are also different from one region to another and are sensitive to the RF.
According to the model predictions computed using the Mathis RF (Fig. \ref{fig_ext})
 the extinction curves of DL07 and DBP09 models show
a prominent 2175 $\AA$ bump followed by a fast far-UV rise in the case of DL07 for all regions. 
Such strong bumps are not observed in
our Galaxy or in the Magellanic Clouds. On the other hand, the extinction curves of 
all regions modeled with MC11 have large and flattened bumps (except for SSDR7 
that does not show any bump at all). When the DBP90 model is used we notice a 
significant increase in the bump from
diffuse/molecular to HII regions. The extinction curves calculated with
the AJ13 model show different behaviors depending on the region, with a weaker 2175
$\AA$ bump in diffuse/molecular regions than in
ionized regions. The UV bumps are significantly stronger in the extinction curves calculated with  DL07 and DBP90
than with AJ13 which are the closest to the observed extinction curves.

Changing the RF has a strong effect on the extinction curves, and
more specifically on the 2175 $\AA$ bump whose strength decreases
when the strength of the RF increases (see Figs. \ref{fig_ext} and \ref{fig_ext_annexe}). 
However, for some regions and models the bump remains unchanged, e.g. for SSDR9 and DL07. 
The extinction curves calculated with the 4-Myr RF are in better agreement with the Galactic and 
LMC behaviors, in contrast with those calculated using Mathis RF and 60-600 Myr RFs. 
In models using the 4-Myr RF, the extinction curves calculated for DL07 and DBP90 models
evidence the strongest 2175 $\AA$ bumps,
whereas MC10 shows very flat bumps and UV-rise for almost all the
regions. AJ13 model seems to minimize the discrepancies between
predictions and observations, which makes this model more compatible with
LMC curves than the other models. 

Even if it is hard to interpret such results since we do not have
the extinction curve associated to the different studied regions, no model seems to
give a satisfactory prediction. However, we now know that the dust emission
models give very distinct behaviors in terms of extinction
predictions. This point is important and should be used to refine
constraints on the dust models themselves. For instance, maybe if models consider a two-population of 
VSG component, as discussed in Section 7.3, with only one carbonaceous 
component such as large PAH clusters or cationic PAHs impacting the 
2175 $\AA$ bump, maybe they would better reconcile with extinction curves.
Moreover, it appears that the extinction
curves can also play an important role to better constrain the RF of each
environments. These preliminary results, in terms of extinction
description, seem to indicate that the Mathis RF is not the
best to be used in the Magellanic Clouds, and suggest that maybe
stronger RFs are necessary.  

Recently, new cosmic dust models have been or will soon be made available \citep{Hensley22, Siebenmorgen22, Ysard20}. It would be interesting to continue this study by using these models to fit the LMC observations, both in emission and extinction. Finally, this work highlights the fact that future studies should, if possible, simultaneously fit dust emission and extinction.

\section{Conclusion}
\label{sec_conclusions}
Using Spitzer IRS, MIPS SED and photometric data combined with Herschel
data, we performed modeling of the spectra in diffuse, molecular and
HII regions of the LMC. We compared four distinct dust models available
in the DustEM package: \citet{Jones13} (AJ13), \citet{Compiegne11}
(MC11), \citet{Draine07} (DL07) and an updated version of
\citet{Desert90} (DBP90). To check the robustness of the results, we
adopted four different radiation fields
(interstellar RF or Mathis, stellar clusters with various ages: 4-Myr,
60-Myr and 600-Myr). None of the models is able to reproduce the
MIR-to-FIR emission using the Galactic standard parameters
 even when the abundances of the dust components and 
 the radiation field strength are allowed to vary. Changes in the size distribution and abundances of the
dust component that dominates the MIR-to-FIR emission (commonly referred to as very
small grains or VSGs) are needed to reasonably fit the dust
emission spectra.

One of the first results of this analysis is the significant increase of the VSG
abundance relative to the big grain (BG) component in the LMC 
compared to the Milky-Way. Changes in the size distribution and
 abundance of dust have a clear impact on the contribution of the VSG
emission in the submm. Depending on the model, the VSG component can strongly dominate
the submm emission, especially when using the standard Mathis
RF. Although no correlation could be shown between 
this strong VSG emission in the submm and the type of environment, 
care should be taken when analysing the FIR to submm emission
in the LMC using only the big grain component.
Small carbon grains could be partly responsible for the global submm-mm
flattening observed in the LMC, even if other processes might be at play
to explain local variations observed in the LMC and the Milky-Way.
The 70 $\mic$ emission excess evidenced in previous studies of the
Magellanic Clouds could result from distinct VSG properties (size
distribution and abundances) compared to our Galaxy.

Another important result is an increase in the amount of small VSGs (and a decrease of 
big VSGs) in HII regions when compared to diffuse regions of the LMC. 
In contrast to our Galaxy, where dust could mainly be produced by O-rich
AGB stars, some dust in the LMC could come from C-rich AGB stars 
(extreme AGB, i. e. mosty embedded carbon stars). The
presence of small VSGs in HII regions could be explained by:\\
- the formation of small VSGs in HII regions (rather than in
molecular clouds); grain growth could occur in the diffuse and
molecular medium via  accretion or coagulation processes. \\
- the destruction of the largest VSGs and BG component in HII regions by shocks
resulting from supernova explosions or turbulence. \\
If these two scenarios take place, the VSG component could include two populations : small
VSGs resulting from large PAH clusters or cationic PAH clusters, and
large VSGs resulting from BG destruction.

The extinction curves calculated by DustEM show a great diversity of
behaviors according to dust models and radiation fields. The
AJ13 model shows reasonable predictions when compared to the
"usual" behaviors. 
In the LMC stronger RFs seem to reproduce the shape of the extinction curve better, 
in particular by reducing the strong  2175 \AA bump predicted by the models.
 Observations in the LMC are in that
sense important to better constrain the dust models (and more specifically
the VSG component) but also to better
constrain the RFs. Further studies simultaneously fitting 
dust emission and extinction and/or using the latest grain models should provide better constraints 
on the properties of grains in the LMC and other galaxies.

\begin{acknowledgements}
  We acknowledge the use of the DustEM package.
\end{acknowledgements}

\begin{appendix}
  \section{Additional material}
  \begin{figure*}%[!h]
  \begin{center}
\includegraphics[width=18cm]{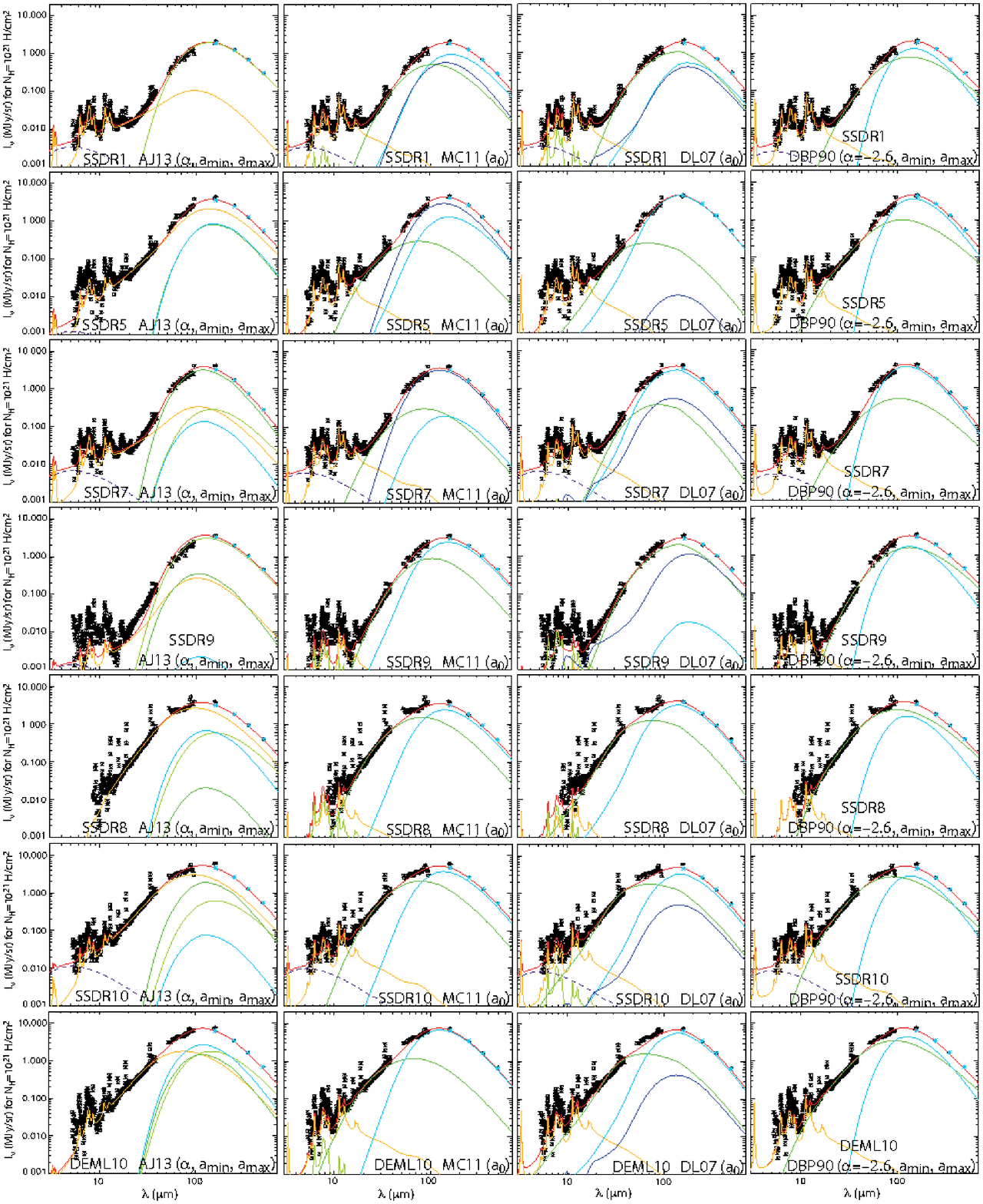}
 \caption{Modeling of the SEDs of the ten regions with different dust models and free
  parameters ($\rm X_{ISRF}$, dust abundances and
  small grains dust size distribution), using the
  4-Myr RF. The
  observations (Spitzer IRS SS and LL, MIPS SED, MIPS 160
  $\mic$, Herschel Photometric PACS 160 $\mic$ and SPIRE 250 $\mic$, 350 $\mic$
  and 500 $\mic$ data) are shown in black. The total modeled SED is shown
  as a red line. The other colored lines correspond to
  the different dust components of the models (see
  Fig. \ref{fig_all_noslope} or \ref{fig_all_best}). The dashed line
  represents the additional NIR continuum. Blue asterisks show
  the color-corrected brightness derived from the models. Each column shows the fit using different dust models
(from left to right: AJ13, MC11, DL07 and DBP90). Each row presents a
different region. The figure continues on the next page. \label{fig_all_best_4Myr}}

\end{center}
\end{figure*}

\addtocounter{figure}{-1}
\begin{figure*}
  \addtocounter{subfigure}{1}
  \begin{center}
\subfigure{\includegraphics[width=18cm]{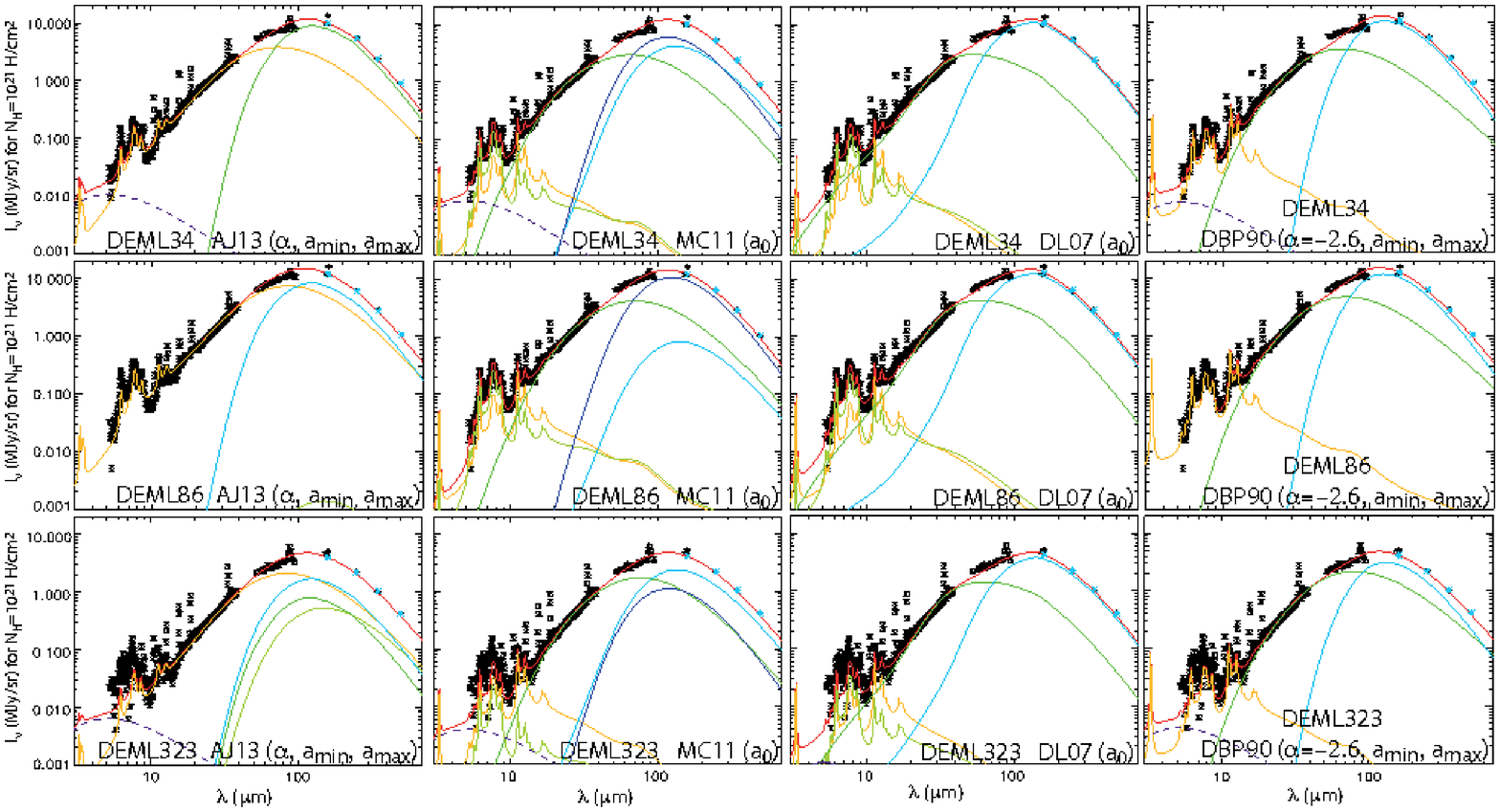}}
\end{center}
\caption{Continued}
\end{figure*}

 \begin{figure*}%[!h]
  \begin{center}
\includegraphics[width=18cm]{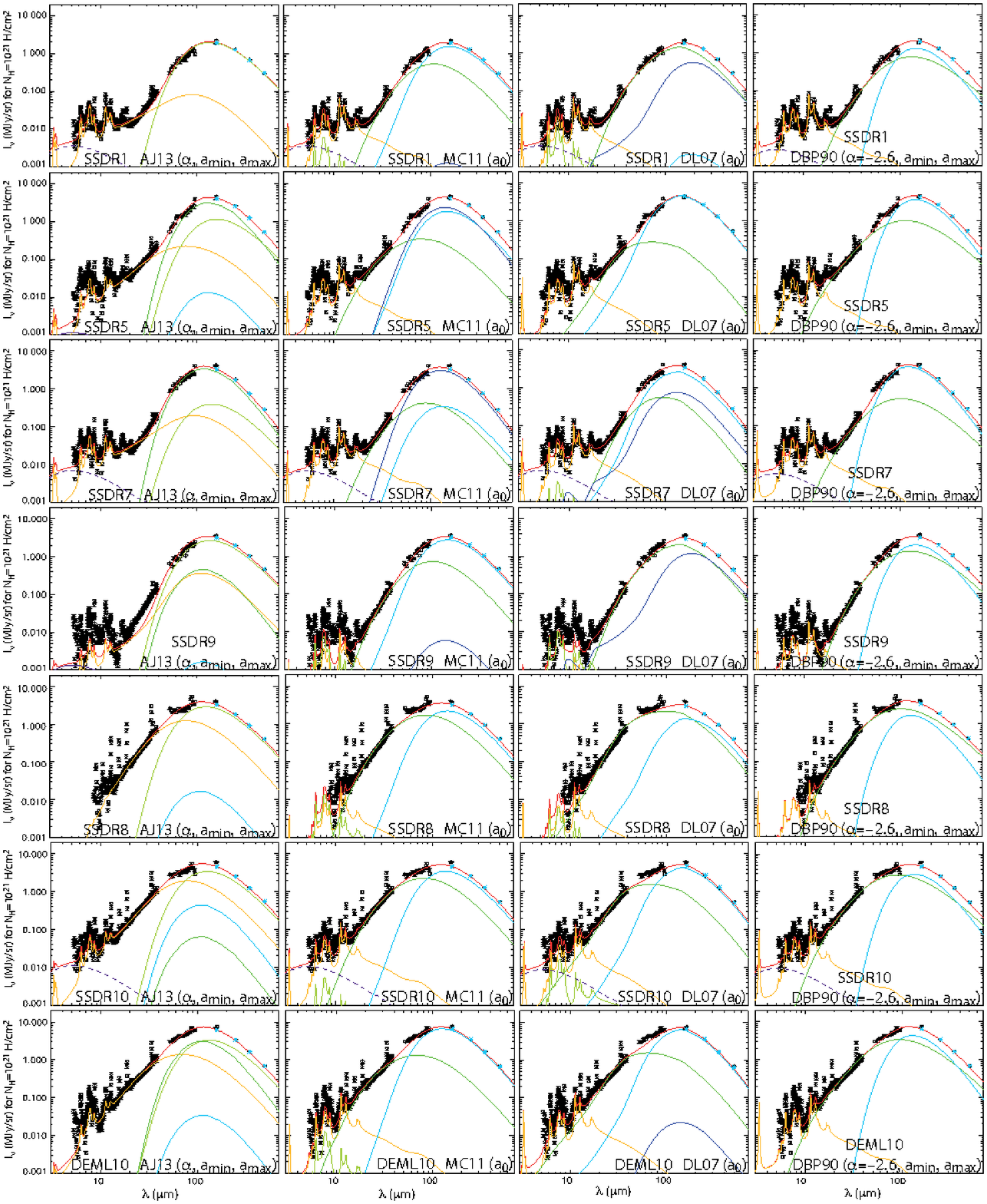}
 \caption{Modeling of the SEDs of the ten regions with different dust models and free
  parameters ($\rm X_{ISRF}$, dust abundances and
  small grains dust size distribution), using the
  60-Myr RF. The
  observations (Spitzer IRS SS and LL, MIPS SED, MIPS 160
  $\mic$, Herschel Photometric PACS 160 $\mic$ and SPIRE 250 $\mic$, 350 $\mic$
  and 500 $\mic$ data) are shown in black. The total modeled SED is shown
  as a red line. The other colored lines correspond to
  the different dust components of the models (see
  Fig. \ref{fig_all_noslope} or \ref{fig_all_best}). The dashed line
  represents the additional NIR continuum. Blue asterisks show
  the color-corrected brightness derived from the models. Each column shows the fit using different dust models
(from left to right: AJ13, MC11, DL07 and DBP90). Each row presents a
different region. The figure continues on the next page. \label{fig_all_best_6E7yr}}

\end{center}
\end{figure*}

\addtocounter{figure}{-1}
\begin{figure*}
  \addtocounter{subfigure}{1}
  \begin{center}
\subfigure{\includegraphics[width=18cm]{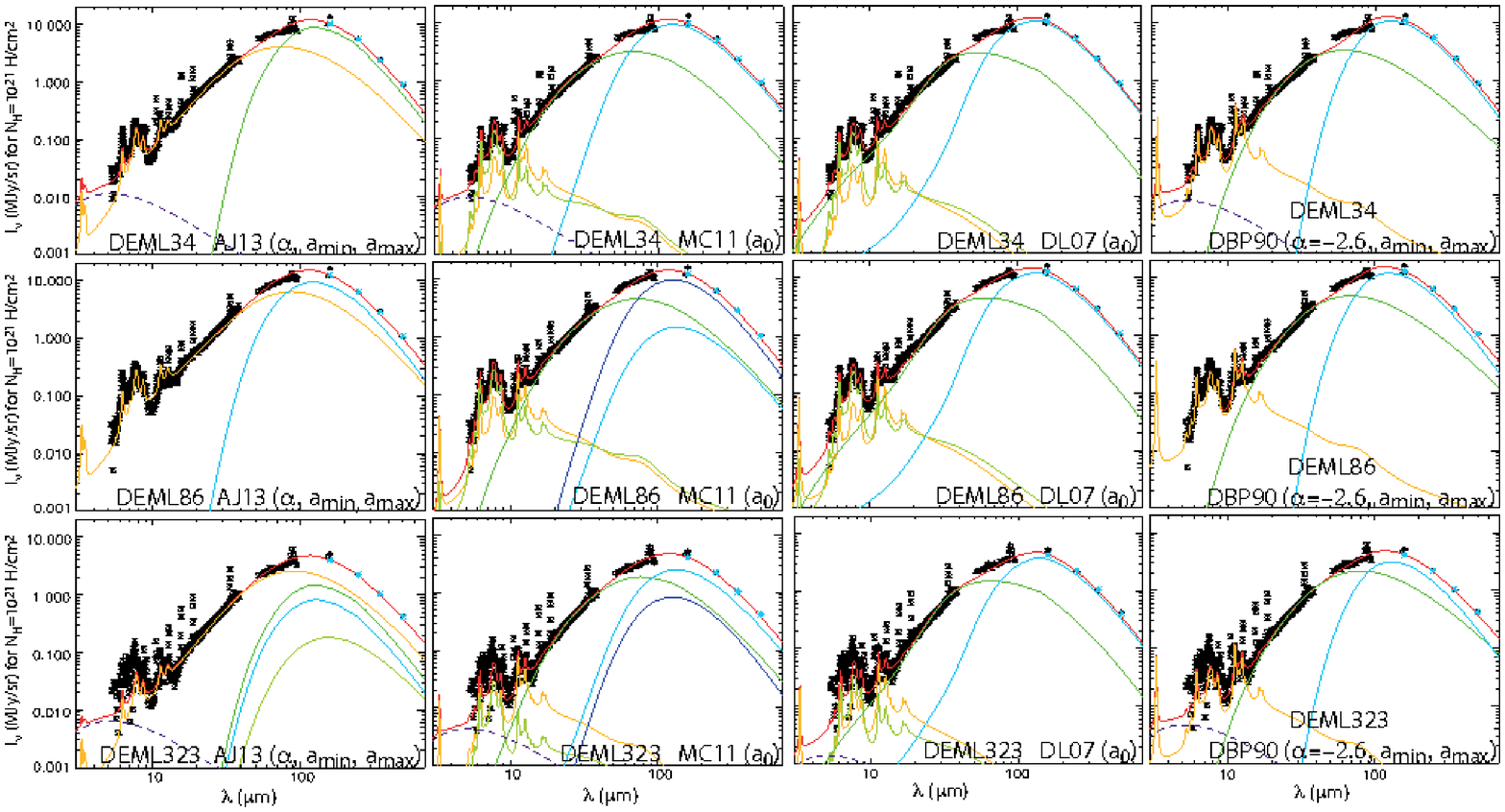}}
\end{center}
\caption{Continued}
\end{figure*}

\begin{figure*}%[!h]
  \begin{center}
\includegraphics[width=18cm]{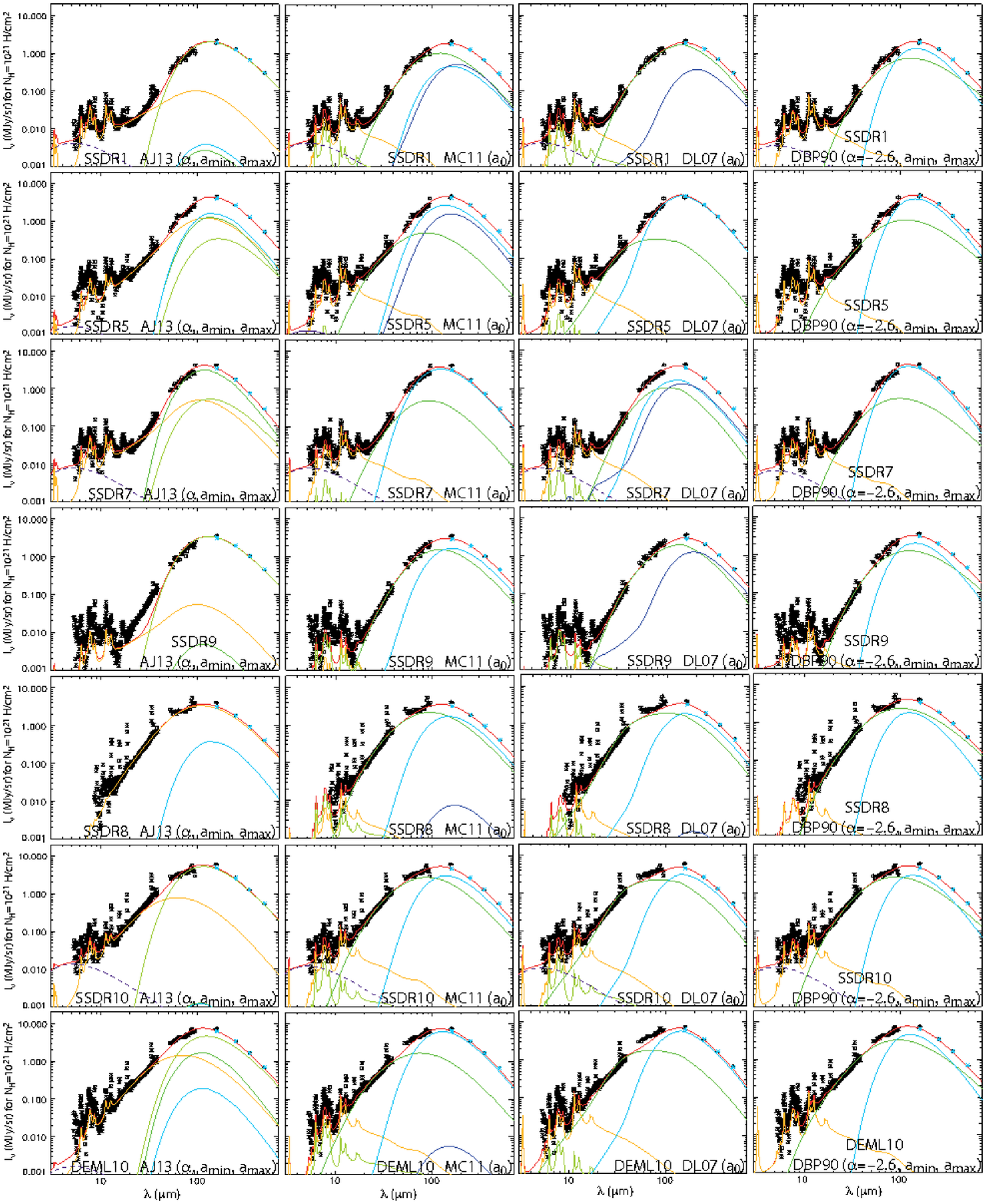}
 \caption{Modeling of the SEDs of the ten regions with different dust models and free
  parameters ($\rm X_{ISRF}$, dust abundances and
  small grains dust size distribution), using the
  600-Myr RF. The
  observations (Spitzer IRS SS and LL, MIPS SED, MIPS 160
  $\mic$, Herschel Photometric PACS 160 $\mic$ and SPIRE 250 $\mic$, 350 $\mic$
  and 500 $\mic$ data) are shown in black. The total modeled SED is shown
  as a red line. The other colored lines correspond to
  the different dust components of the models (see
  Fig. \ref{fig_all_noslope} or \ref{fig_all_best}). The dashed line
  represents the additional NIR continuum. Blue asterisks show
  the color-corrected brightness derived from the models. The orange
  diamonds that are visible in the DBP90 panels show the MIPS 70 $\mic$
photometric data normalized to the integrated flux in the
MIPS-SED band. Each column shows the fit using different dust models
(from left to right: AJ13, MC11, DL07 and DBP90). Each row presents a
different region. The figure continues on the next page. \label{fig_all_best_6E8yr}}

\end{center}
\end{figure*}

\addtocounter{figure}{-1}
\begin{figure*}
  \addtocounter{subfigure}{1}
  \begin{center}
\subfigure{\includegraphics[width=18cm]{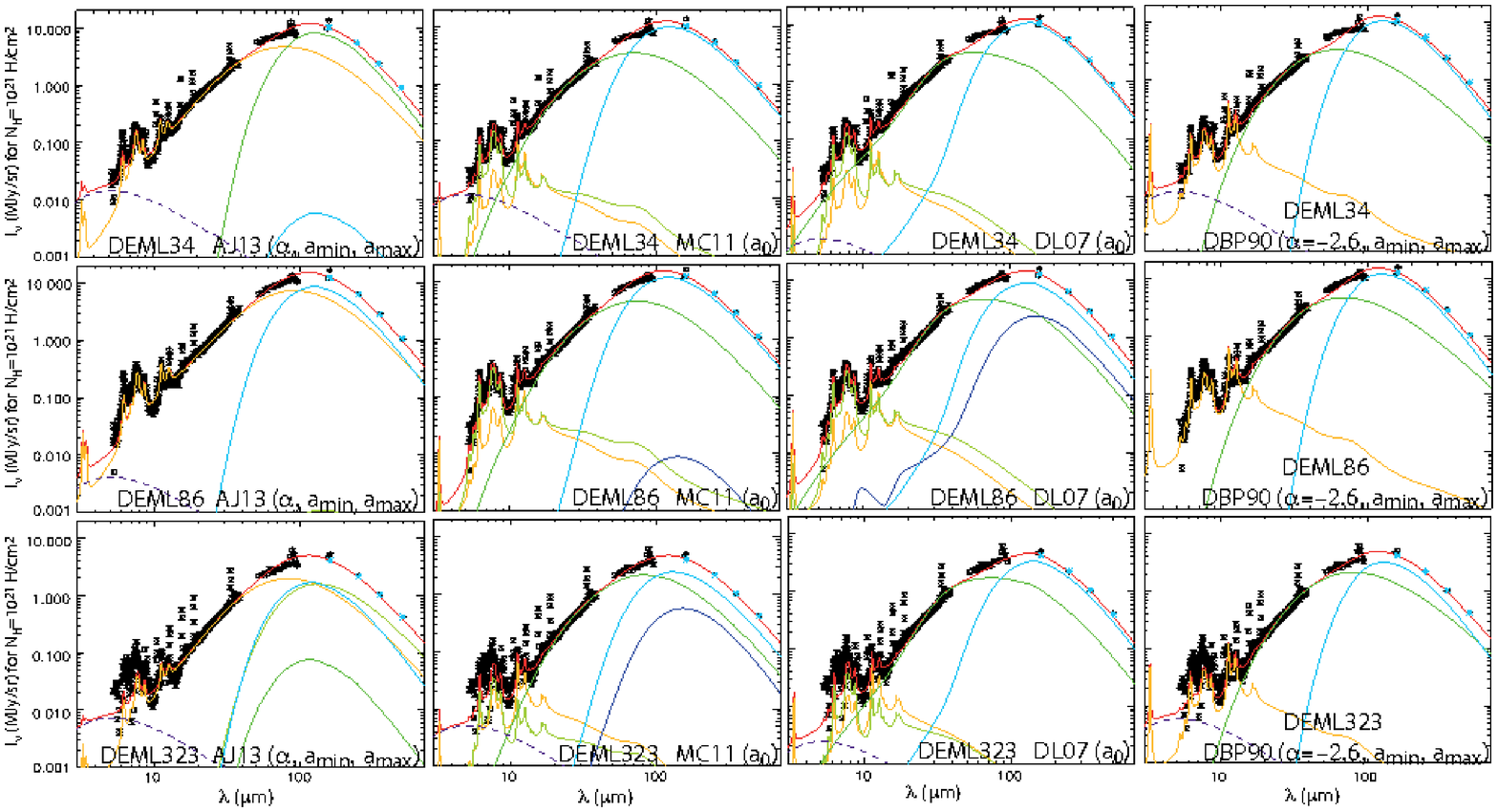}}
\end{center}
\caption{Continued}
\end{figure*}

   \begin{table*}%[!h]}
\caption{Same table as Table \ref{table_chi2_AJ13} but using a 60-Myr
  (top) and 600-Myr RF (bottom).\label{table_60Myr_AJ13}}
\begin{center}
%\resizebox{\textwidth}{!}{%
\begin{tabular}{m{1.3cm}m{0.8cm}m{0.6cm}m{1.4cm}m{1.4cm}m{1.4cm}m{1.4cm}m{0.8cm}m{0.8cm}m{0.8cm}m{0.4cm}m{0.4cm}c}
\hline
\hline
  \multicolumn{13}{c}{AJ13 ($\alpha, a_{min}, a_{max}$) - 60-Myr RF}\\
\hline
  Region              & $\chi^2/dof$ &$X_{ISRF}$  &$Y_{a-C}$  & $Y_{a-C:H}$
                 &$Y_{Pyr.}$ &$Y_{Oliv.}$ & $Y_{dust,tot}$& $I_{NIR\,cont.}$ &
                                                          $\alpha$ &
                                                                       $a_{min}$&$a_{max}$&$\frac{Y_{a-C}}{Y_{Sil.}+Y_{a-C:H}}$
  \\
      &&&&&&&($10^{-2}$)&($10^{-3}$)&&($10^{-1}$)&&($10^{-1}$)\\

  \hline
   SSDR1 & 2.42&0.59 &1.08$\times 10^{-4}$ & 1.71$\times 10^{-3}$
                 &\textit{1.00$\times 10^{-6}$} &\textit{1.00$\times 10^{-6}$} &0.18 &3.59 &-4.34&4.00&13.3 &0.631\\
  SSDR5 & 6.01& 0.25&4.83$\times 10^{-4}$  &1.96$\times 10^{-3}$ 
                 &9.70$\times 10^{-3}$ &4.15$\times 10^{-5}$ &1.22
                                                         &1.16
                                                                            &-3.07&4.00 &5.57 &0.413 \\
  \hline
  SSDR7 & 3.53& 0.47&
                      2.36$\times 10^{-4}$ &4.07$\times 10^{-4}$
                 &6.26$\times 10^{-3}$ &\textit{1.00$\times 10^{-6}$} &0.69 &7.00 &-4.03&4.00&11.7&0.354 \\
     SSDR9 & 10.6 &0.58&1.71$\times 10^{-4}$  &2.30$\times 10^{-3}$  &6.97$\times 10^{-4}$ 
                             &2.53$\times 10^{-6}$ &0.32 &1.33 &-3.31&4.18&3830& 0.570\\
          
  \hline
    SSDR8 & 3.71 &0.97 & 4.28$\times 10^{-4}$  &1.66$\times 10^{-3}$
                 &\textit{1.00$\times 10^{-6}$}  &1.70$\times 10^{-5}$
                                         &0.21&0.00&-2.00&14.7 &6.25 & 2.55\\
  SSDR10 & 2.71&0.86&8.31$\times 10^{-4}$  &2.18$\times 10^{-3}$  &7.19$\times 10^{-5}$ &4.96$\times 10^{-4}$ &0.36&11.2 &-2.44&4.00 &6.64&3.02\\
  DEML10 & 4.34 & 0.67&9.30$\times 10^{-4}$ & 2.54$\times 10^{-3}$ &4.23$\times 10^{-3}$ &4.68$\times 10^{-5}$ &0.77&0.558 &-2.12&5.93 &5.00 &1.36 \\
  DEML34 & 1.56&0.30 &5.80$\times 10^{-3}$  & \textit{1.00$\times 10^{-6}$} &2.32$\times 10^{-2}$ &1.00$\times 10^{-6}$ &2.90&11.6&-2.09&4.00 &5.09 &2.50 \\
 DEML86 & 2.18& 0.40&5.87$\times 10^{-3}$  &\textit{1.00$\times 10^{-6}$}  &\textit{1.00$\times 10^{-6}$} &1.99$\times 10^{-2}$&2.58&0.00&-2.70&4.00 &6.90 &2.95 \\
   DEML323 & 3.80& 0.29& 2.89$\times 10^{-3}$ & 2.79$\times 10^{-4}$ &4.05$\times 10^{-3}$ &2.28$\times 10^{-3}$ &0.95&6.45 &-2.00&4.00 &6.47 &4.37 \\
  \hline
\hline
\multicolumn{13}{c}{AJ13 ($\alpha, a_{min}, a_{max}$) - 600-Myr RF}\\
\hline
  Region              & $\chi^2/dof$ &$X_{ISRF}$  &$Y_{a-C}$  & $Y_{a-C:H}$
                 &$Y_{Oliv}$ &$Y_{Pyr.}$ & $Y_{dust,tot}$ &$I_{NIR\,cont.}$ &
                                                          $\alpha$ &
                                                                       $a_{min}$&$a_{max}$&$\frac{Y_{a-C}}{Y_{Sil.}+Y_{a-C:H}}$
  \\
      &&&&&&&($10^{-2}$)&($10^{-3}$)&&($10^{-1}$)&&($10^{-1}$)\\

  \hline
   SSDR1 & 2.44&3.17 &1.96$\times 10^{-4}$  &
                                1.59$\times 10^{-3}$ &4.25$\times 10^{-6}$ &6.42$\times 10^{-6}$& 0.18&4.18&-4.45&4.00 &13.3  &1.22 \\
  SSDR5 & 5.72& 1.13&2.17$\times 10^{-3}$  &6.32$\times 10^{-4}$
                 &4.75$\times 10^{-3}$ &6.35$\times 10^{-3}$
                                         &1.39&1.63&-3.12&4.00 &10.4
                                                                                          &1.85 \\
  \hline
  SSDR7 & 3.57& 3.31& 4.35$\times 10^{-4}$ &4.00$\times 10^{-4}$ &4.73$\times 10^{-3}$ &\textit{1.00$\times 10^{-6}$} &0.56&7.88&-4.10&4.00&180 &0.848\\
     SSDR9 & 11.2 &4.03&5.42$\times 10^{-5}$ &2.26$\times 10^{-3}$  &6.27$\times 10^{-6}$  &\textit{1.00$\times 10^{-6}$} &0.23&0.439 &-4.37&4.68 &4680 &0.239\\
  \hline
   SSDR8 & 3.65 &1.26 & 5.43$\times 10^{-3}$  &\textit{1.00$\times 10^{-6}$}  &\textit{1.00$\times 10^{-6}$} &1.35$\times 10^{-3}$&0.68&0.00&-2.00&13.7 &6.70 &40.2\\
  SSDR10 & 3.01&7.28&5.10$\times 10^{-4}$  &2.14$\times 10^{-3}$  &\textit{1.00$\times 10^{-6}$} &\textit{1.00$\times 10^{-6}$} &0.27&13.7 &-2.29&4.00 &3.84 &2.38\\
  DEML10 & 4.25 & 5.04&1.11$\times 10^{-3}$ & 2.52$\times 10^{-3}$ &1.81$\times 10^{-3}$ &2.16$\times 10^{-4}$ &0.57&1.63 &-2.15&6.16 &4.26 &2.44 \\
  DEML34 & 1.57&1.89&8.88$\times 10^{-3}$  & \textit{1.00$\times 10^{-6}$} &2.00$\times 10^{-2}$ &1.50$\times 10^{-5}$ &2.89&14.5 &-2.04&4.00 &4.43 &4.44 \\
 DEML86 & 2.26& 2.70&8.29$\times 10^{-3}$  &\textit{1.00$\times 10^{-6}$}  &\textit{1.00$\times 10^{-6}$} &1.63$\times 10^{-2}$ &2.46&4.12 &-2.69&4.00 &5.95 &5.09 \\
   DEML323 & 3.79& 3.74& 1.61$\times 10^{-3}$ & 1.10$\times 10^{-3}$ &1.04$\times 10^{-4}$ &2.41$\times 10^{-3}$ &0.52&7.35 &-2.00&4.00 &5.08 &4.45 \\
  \hline
\hline

\end{tabular}
\end{center}
\end{table*}%[!h]}

  \begin{table*}%[!h]}
\caption{Same table as Table \ref{table_chi2_MC11} but using a 60-Myr
  (top) and 600-Myr RF (bottom).\label{table_60Myr_MC11}}
\begin{center}
%\resizebox{\textwidth}{!}{%
%\begin{tabular}{m{1.4cm}m{0.8cm}m{0.6cm}m{1.4cm}m{1.4cm}m{1.4cm}m{1.4cm}m{1.4cm}m{0.8cm}m{0.8cm}m{0.4cm}m{0.8cm}c}
\resizebox{\textwidth}{!}{
%\begin{tabular}{m{1.3cm}m{0.8cm}m{0.6cm}m{1.4cm}m{1.4cm}m{1.4cm}m{1.4cm}m{1.4cm}m{1.cm}m{0.7cm}m{0.4cm}m{0.8cm}c}
 \begin{tabular}{*{13}{V{12cm}}}
\hline
  \hline
\multicolumn{13}{c }{MC11 ($a_0$) - 60-Myr RF } \\
 \hline
   Region  & $\chi^2/dof$ &$X_{ISRF}$  &$Y_{PAH^0}$  & $Y_{PAH^+}$
                 &$Y_{SamC}$ &$Y_{LamC}$& $Y_{aSil}$  & 
                                                          $Y_{dust,tot}$ &
                                                 $I_{NIR\,cont.}$ &
                                                                    $a_0$&
                                                                           $\frac{Y_{SamC}}{(Y_{LamC}+Y_{aSil})}$&$\frac{(Y_{PAH^0}+Y_{PAH^+})}{(Y_{LamC}+Y_{aSil})}$\\
     &&&&&&&&($10^{-2}$)&($10^{-3}$)&&($10^{-1}$)& ($10^{-2}$)\\
\hline
  
  SSDR1 & 1.60 & 0.14& 8.74$\times 10^{-4}$ &
                                2.83$\times 10^{-4}$ &3.33$\times 10^{-3}$ &3.50$\times 10^{-3}$ &1.92$\times 10^{-5}$ &0.80&4.06 &4.51 &9.46&32.9\\
  SSDR5 & 4.96& 0.18& 2.42$\times 10^{-4}$ &
                               \textit{1.00$\times 10^{-6}$} &7.17$\times
                                                      10^{-4}$
                             &4.96$\times 10^{-3}$ &1.30$\times
                                                     10^{-2}$ &1.89
                                                                      &0.558 & 2.83&0.399& 1.35\\
  \hline
  SSDR7 & 2.88&0.34 &1.78$\times 10^{-4}$  &
                               \textit{1.00$\times 10^{-6}$} &3.63$\times
                                                      10^{-4}$
                             &5.60$\times 10^{-4}$ &1.01$\times
                                                     10^{-2}$ & 1.12 &6.73&3.80 &0.341&1.68\\
   SSDR9  &7.41 &0.19 &
                               2.68$\times 10^{-5}$ &3.83$\times 10^{-5}$ &9.46$\times 10^{-4}$ &7.12$\times 10^{-3}$ &3.14$\times 10^{-5}$ 
   & 0.82 &0.289 &4.58 &1.32&0.910\\
\hline
  SSDR8 & 3.48&0.23 &5.13$\times 10^{-5}$  &
                               5.82$\times 10^{-5}$ &2.52$\times
                                                      10^{-3}$
                             &4.95$\times 10^{-3}$ &1.25$\times
                                                     10^{-6}$ &0.76 &0.00&3.21 &5.09&2.21 \\
  SSDR10 & 2.58& 0.28&2.81$\times 10^{-4}$  &1.78$\times 10^{-5}$  &2.87$\times 10^{-3}$ &6.54$\times 10^{-3}$ &\textit{1.00$\times 10^{-6}$} &0.97&0.105 &3.00 &4.39&4.57\\
  DEML10 & 3.89& 0.48& 1.30$\times 10^{-4}$ & 2.54$\times 10^{-5}$ &1.25$\times 10^{-3}$ &8.04$\times 10^{-3}$ &\textit{1.00$\times 10^{-6}$} &0.94&0.00&2.38 &1.55&1.93\\
  DEML34 & 1.07 &0.53 &1.35$\times 10^{-4}$  &2.53$\times 10^{-4}$ 
                  &2.89$\times 10^{-3}$ &1.06$\times 10^{-2}$ &1.83$\times 10^{-6}$ &1.39&9.60 &2.29 & 2.73&3.66\\
  DEML86 & 1.08 &0.33&4.05$\times 10^{-4}$  &7.79$\times 10^{-4}$  &5.88$\times 10^{-3}$ &2.42$\times 10^{-3}$ &3.24$\times 10^{-2}$ &4.19&0.00&2.52 &1.69&3.40\\
  DEML323 & 3.22& 0.34&1.24$\times 10^{-4}$  &7.43$\times 10^{-5}$  &2.05$\times 10^{-3}$ &3.99$\times 10^{-3}$ &2.70$\times 10^{-3}$ &0.89&4.32 &2.84&3.06&2.96\\
  \hline
  $<$All$>$ & & & &&&&&&&3.20&&  \\
  $<$Diff.$>$ & & & &&&&&&&4.19&&\\
  $<$Mol.$>$ & & & &&&&&&&3.67&& \\
  $<$HII$>$ & & & &&&&&&&2.71&& \\
  \hline
  \hline
\multicolumn{13}{c }{MC11 ($a_0$) - 600-Myr RF } \\
 \hline
    Region  & $\chi^2/dof$ &$X_{ISRF}$  &$Y_{PAH^0}$  & $Y_{PAH^+}$
                 &$Y_{SamC}$ &$Y_{LamC}$& $Y_{aSil}$  & 
                                                          $Y_{dust,tot}$ &
                                                 $I_{NIR\,cont.}$ &
                                                                    $a_0$&
                                                                           $\frac{Y_{SamC}}{(Y_{LamC}+Y_{aSil})}$&$\frac{(Y_{PAH^0}+Y_{PAH^+})}{(Y_{LamC}+Y_{aSil})}$\\
      &&&&&&&&($10^{-2}$)&($10^{-3}$)&&($10^{-1}$) &($10^{-2}$)\\

  \hline
  SSDR1 & 1.65 & 0.47& 6.41$\times 10^{-4}$ &
                                2.45$\times 10^{-4}$ &2.84$\times 10^{-3}$ &1.65$\times 10^{-3}$ &9.44$\times 10^{-3}$ &1.48&4.27 &4.58  &2.56&7.99\\
  SSDR5 & 5.00& 1.14& 3.61$\times 10^{-4}$ &
                               1.43$\times 10^{-5}$ &8.77$\times
                                                      10^{-4}$
                             &4.32$\times 10^{-3}$ &1.28$\times
                                                     10^{-2}$
                                                      &1.84&1.21& 2.64
                                                                                 &0.512&1.67\\
  \hline
  SSDR7 & 3.15&2.31 &2.29$\times 10^{-4}$  &
                               1.73$\times 10^{-5}$ &3.64$\times 10^{-4}$ &3.18$\times 10^{-3}$ &\textit{1.00$\times 10^{-6}$} &0.38&7.96&3.41 &1.14&7.74 \\
  SSDR9  &7.36 &0.48 &
                               7.17$\times 10^{-5}$ &1.89$\times 10^{-4}$ &4.31$\times 10^{-3}$ &5.83$\times 10^{-3}$ &\textit{1.00$\times 10^{-6}$} 
   & 1.04&0.458 &4.45 &7.39&4.47\\
 \hline
  SSDR8 & 3.43&0.77 &1.33$\times 10^{-4}$  &
                               1.90$\times 10^{-4}$ &5.53$\times 10^{-3}$ &3.97$\times 10^{-3}$ &8.93$\times 10^{-5}$ &0.99&0.00&2.86 &13.6&7.96\\
 SSDR10 & 2.58& 1.02&6.73$\times 10^{-4}$  &1.56$\times 10^{-5}$  &5.66$\times 10^{-3}$ &5.55$\times 10^{-3}$ &\textit{1.00$\times 10^{-6}$} &1.19&1.19&2.65 &10.2&12.4\\
  DEML10 & 3.87& 1.69& 3.43$\times 10^{-4}$ & 1.35$\times 10^{-4}$ &2.55$\times 10^{-3}$ &7.53$\times 10^{-3}$ &3.46$\times 10^{-5}$ &1.06&0.00&2.16&3.37& 6.32\\
  DEML34 & 1.09 &2.29 &2.23$\times 10^{-4}$  &6.60$\times 10^{-4}$ 
                  &4.38$\times 10^{-3}$ &9.42$\times 10^{-3}$ &\textit{1.00$\times 10^{-6}$} &1.47&11.8 &1.95 & 4.65&9.37 \\
  DEML86 & 1.22 &2.50&3.98$\times 10^{-4}$  &1.25$\times 10^{-3}$  &4.86$\times 10^{-3}$ &1.05$\times 10^{-2}$ &3.90$\times 10^{-5}$ &1.70&0.487&2.13 &4.61 &15.6\\
  DEML323 & 3.20& 1.56&2.22$\times 10^{-4}$  &2.20$\times 10^{-4}$  &3.12$\times 10^{-3}$ &3.19$\times 10^{-3}$ &3.69$\times 10^{-3}$ &1.04&5.11 &2.13 &4.53&6.42\\
  \hline
  $<$All$>$ & & & &&&&&&&2.90&& \\
  $<$Diff.$>$ & & & &&&&&&&3.93 && \\
  $<$Mol.$>$ & & & &&&&&&&3.61&&  \\
  $<$HII$>$ & & & &&&&&&&2.31&& \\
  \hline
  \hline
 \end{tabular}
 }
\end{center}
\end{table*}%[!h]}

\begin{table*}%[!h]}
\caption{Same table as Table \ref{table_chi2_DL07} but using a 60-Myr
  (top) and 600-Myr RF (bottom).\label{table_60Myr_DL07}}
\begin{center}
%\resizebox{\textwidth}{!}{%
%\begin{tabular}{m{1.4cm}m{0.8cm}m{0.7cm}m{1.4cm}m{1.4cm}m{1.4cm}m{1.4cm}m{1.4cm}m{0.7cm}m{0.9cm}m{0.4cm}m{0.9cm}c}
\resizebox{\textwidth}{!}{
%\begin{tabular}{m{1.3cm}m{0.8cm}m{0.6cm}m{1.4cm}m{1.4cm}m{1.4cm}m{1.4cm}m{1.4cm}m{1.cm}m{0.7cm}m{0.4cm}m{0.8cm}c}
 \begin{tabular}{*{13}{V{12cm}}}
  \hline
 \hline
\multicolumn{13}{c }{DL07 ($a_0$) - 60-Myr RF} \\
 \hline
 Region  & $\chi^2/dof$ &$X_{ISRF}$  &$Y_{PAH^0}$  & $Y_{PAH^+}$
                 &$Y_{graph.}$ &$Y_{Bsil}$ & $Y_{Ssil}$& $Y_{dust,tot}$&
                                                         $I_{NIR\,cont.}$
                                                                           &
                                                                             $a_0$&
                                                                                    $\frac{Y_{graph.}}{(Y_{Bsil}+Y_{Ssil})}$&$\frac{(Y_{PAH^0}+Y_{PAH^+})}{(Y_{Bsil}+Y_{Ssil})}$\\
       &&&&&&&&($10^{-2}$)&($10^{-3}$)&&($10^{-1}$)&($10^{-2}$) \\
  \hline
  SSDR1 & 1.60& 0.031 & 1.12$\times 10^{-3}$ &3.56$\times 10^{-4}$ 
                 &1.06$\times 10^{-2}$ &3.29$\times 10^{-5}$
                                           &1.59$\times 10^{-2}$ &2.80
                                                                       &3.54&3.75 &6.65&9.26 \\
  SSDR5 & 5.18&0.16 &2.84$\times 10^{-4}$  & 1.53$\times 10^{-6}$
                 &9.44$\times 10^{-4}$ &1.68$\times 10^{-2}$
                                           &1.08$\times 10^{-6}$
                                                       &1.80&0.136
                                                                           &1.45 &0.562 &1.70\\
  \hline
  SSDR7 & 2.84& 0.34& 1.73$\times 10^{-4}$ &
                               \textit{1.00$\times 10^{-6}$} &5.60$\times 10^{-4}$ &5.29$\times 10^{-3}$ &2.80$\times 10^{-3}$ &0.88&6.61 &2.64 &0.692&2.15 \\
   SSDR9 & 7.18& 0.043 & 2.89$\times 10^{-5}$ & 2.55$\times 10^{-4}$ &1.12$\times 10^{-2}$ &\textit{1.00$\times 10^{-6}$} &2.53$\times 10^{-2}$ &3.68&0.208 &3.72 &4.43&1.12 \\
\hline
  SSDR8 & 4.22& 0.066& 1.76$\times 10^{-4}$ & 9.85$\times 10^{-5}$ &1.46$\times 10^{-2}$ &1.12$\times 10^{-2}$ &\textit{1.00$\times 10^{-6}$} &2.61&0.00&1.91 &13.0&2.45\\
  SSDR10 & 2.77& 0.19& 3.32$\times 10^{-4}$ & 9.17$\times 10^{-5}$ &4.93$\times 10^{-3}$ &1.41$\times 10^{-2}$ &\textit{1.00$\times 10^{-6}$} &1.95&8.77 &1.36 &3.50&3.00\\
  DEML10 & 4.70&0.24 &2.80$\times 10^{-4}$  &\textit{1.00$\times 10^{-6}$}  &3.93$\times 10^{-3}$ &1.70$\times 10^{-2}$ &1.11$\times 10^{-4}$ &2.13&0.00&1.30 &2.30&1.64 \\
  DEML34 & 1.24 &0.36 &1.64$\times 10^{-4}$  &3.71$\times 10^{-4}$  &5.86$\times 10^{-3}$ &2.11$\times 10^{-2}$ &\textit{1.00$\times 10^{-6}$} &2.75&0.00&1.06&7.65&2.54\\
  DEML86 & 1.50& 0.34&3.28$\times 10^{-4}$  &7.23$\times 10^{-4}$  &7.95$\times 10^{-3}$ &2.45$\times 10^{-2}$ &1.74$\times 10^{-6}$ &3.35&0.00&1.26 &8.75&4.29 \\
  DEML323 & 3.42& 0.24&1.31$\times 10^{-4}$  & 1.54$\times 10^{-4}$
                 &3.78$\times 10^{-3}$ &1.03$\times 10^{-2}$
                                           &\textit{1.00$\times
                                             10^{-6}$} &1.44&1.31&1.34
                                                                                  &11.7 &2.77\\
  \hline
  $<$All$>$ &&&&& &&&&&1.98 &&\\
   $<$Diff.$>$ &&&&& &&&&&3.18 &&\\
  $<$Mol.$>$ &&&&& &&&&&2.60 &&\\
    $<$HII$>$ &&&&& &&&&&1.37 &&\\
  \hline
  \hline
  \multicolumn{13}{c }{DL07 ($a_0$) - 600-Myr RF} \\
 \hline
 Region  & $\chi^2/dof$ &$X_{ISRF}$  &$Y_{PAH^0}$  & $Y_{PAH^+}$
                 &$Y_{graph.}$ &$Y_{Bsil}$ & $Y_{Ssil}$&
                                                         $Y_{dust,tot}$ &
                                                         $I_{NIR\,cont.}$
                                                                           &
                                                                             $a_0$&
                                                                                    $\frac{Y_{graph.}}{(Y_{Bsil}+Y_{Ssil})}$&$\frac{(Y_{PAH^0}+Y_{PAH^+})}{(Y_{Bsil}+Y_{Ssil})}$\\
        &&&&&&&&($10^{-2}$)&($10^{-3}$)&&($10^{-1}$) &($10^{-2}$)\\
  \hline
  SSDR1 & 1.66& 0.24& 1.28$\times 10^{-3}$ &6.36$\times 10^{-4}$ 
                 &1.30$\times 10^{-2}$ &\textit{1.00$\times 10^{-6}$}
                                           &1.32$\times 10^{-2}$
                                                       &2.81&4.14
                                                                           &3.40 &9.85 &14.5\\
  SSDR5 & 5.21&0.85 &5.07$\times 10^{-4}$  & 4.49$\times 10^{-5}$
                 &1.72$\times 10^{-3}$ &1.46$\times 10^{-2}$
                                           &\textit{1.00$\times
                                             10^{-6}$} &1.69&0.688
                                                                           &1.55 &1.18 &3.78\\
  \hline
  SSDR7 & 2.85& 2.15& 2.50$\times 10^{-4}$ &
                               3.99$\times 10^{-5}$ &1.22$\times
                                                      10^{-3}$
                               &2.42$\times 10^{-3}$ &7.42$\times
                                                       10^{-3}$
                                                       &1.13&7.50
                                                                           &2.71 &1.24 &2.95\\
    SSDR9 & 7.38& 0.43& 2.92$\times 10^{-5}$ & 2.61$\times 10^{-4}$ &9.26$\times 10^{-3}$ &1.29$\times 10^{-6}$ &2.76$\times 10^{-2}$ &3.72&0.391 &3.16 &3.35 &1.05\\
\hline
  SSDR8 & 4.32& 0.55& 1.76$\times 10^{-4}$ & 1.22$\times 10^{-4}$
                 &1.33$\times 10^{-2}$ &8.61$\times 10^{-3}$
                                           &2.48$\times 10^{-5}$
                                                       &2.22&0.00&1.60
                                                                                  &15.4 &3.45\\

  SSDR10 & 2.79& 0.64& 1.04$\times 10^{-4}$ & 3.04$\times 10^{-4}$
                 &1.53$\times 10^{-3}$ &1.27$\times 10^{-2}$
                                           &1.03$\times 10^{-6}$
                                                       &1.46&10.4&1.40
                                                                                  &1.20 &3.21\\
  DEML10 & 4.67&1.03 &6.71$\times 10^{-4}$  &7.60$\times 10^{-5}$
                 &8.39$\times 10^{-3}$ &1.57$\times 10^{-2}$
                                           &\textit{1.00$\times
                                             10^{-6}$} &2.48&0.00&1.23
                                                                                  &5.34 & 4.76\\
  DEML34 & 1.26 &1.73 &3.07$\times 10^{-4}$  &8.21$\times 10^{-4 }$
                 &1.04$\times 10^{-2}$ &1.90$\times 10^{-2}$
                                           &1.07$\times 10^{-6}$
                                                       &3.05&1.84
                                                                           &0.967 &5.47 & 5.94\\
  DEML86 & 1.42& 1.90&5.03$\times 10^{-4}$  &1.49$\times 10^{-3}$
                 &1.22$\times 10^{-2}$ &1.50$\times 10^{-2}$
                                           &1.50$\times 10^{-6}$
                                                       &2.92&0.00&1.15
                                                                                  &8.13 & 13.3\\
  DEML323 & 3.44& 1.05&2.73$\times 10^{-4}$  & 3.83$\times 10^{-4}$ &7.67$\times 10^{-3}$ &9.06$\times 10^{-3}$ &1.82$\times 10^{-6}$ &1.74&2.43 &1.26 &8.46&7.24 \\
  \hline
  $<$All$>$ &&&& &&&&&&1.84 &&\\
   $<$Diff.$>$ &&&& &&&&&&2.94 &&\\
  $<$Mol.$>$ &&&& &&&&&&2.48 &&\\
    $<$HII$>$ &&&& &&&&&&1.27&&\\
  \hline
  \hline

 \end{tabular}
 }
\end{center}
\end{table*}%[!h]}

\begin{table*}
\caption{Same table as Table \ref{table_chi2_DBP90} but using a 60-Myr
  (top) and 600-Myr RF (bottom).\label{table_60Myr_DBP90}}
\begin{center}
%\begin{tabular}{m{1.4cm}m{0.8cm}m{0.9cm}m{1.4cm}m{1.4cm}m{1.4cm}m{1.cm}m{1.4cm}m{0.6cm}m{0.7cm}m{0.7cm}cc}
\resizebox{\textwidth}{!}{
%\begin{tabular}{m{1.3cm}m{0.8cm}m{0.6cm}m{1.4cm}m{1.4cm}m{1.4cm}m{1.4cm}m{1.4cm}m{1.cm}m{0.7cm}m{0.4cm}m{0.8cm}c}
 \begin{tabular}{*{13}{V{12cm}}}
  \hline
  \hline
\multicolumn{13}{c}{DBP90 ($\alpha=-2.6, a_{min}, a_{max}$) - 60-Myr RF}\\
\hline
 Region  & $\chi^2/dof$ &$X_{ISRF}$  &$Y_{PAH^0}$  &$Y_{VSG}$
                                                              &$Y_{BG}$
  & $Y_{dust,tot}$
                                                                        &
                                                                          $I_{NIR\,cont.}$
                                                                                            &
                                                                                              $\alpha$&
                                                                                                        $a_{min}$
  & $a_{max}$ &$\frac{Y_{VSG}}{Y_{BG}}$&$\frac{Y_{PAH}}{Y_{BG}}$ \\
        &&&&&&($10^{-2}$)&&&&($\times 10$)&($10^{-1}$)&($10^{-2}$) \\

  \hline
  SSDR1 &1.65 & 0.23&1.32$\times 10^{-4}$  &6.40$\times 10^{-4}$
                                                              &3.47$\times
                                                                10^{-3}$
  &0.42&2.76$\times 10^{-3}$ &-2.6 & 5.50  &3.26 &1.84 &3.80 \\
  SSDR5 & 4.90& 0.25&1.19$\times 10^{-4}$  & 9.45$\times 10^{-4}$
                                                              &8.80$\times
                                                                10^{-3}$
  &0.99&0.00&-2.6&2.43 &2.57 &1.07&1.35 \\
  \hline
  SSDR7 &2.92 &0.68 &6.33$\times 10^{-5}$  &1.88$\times 10^{-4}$ 
                                                              &3.77$\times 10^{-3}$ &0.40&5.45$\times 10^{-3}$ &-2.6&3.68 &2.31 &    0.499&1.68 \\
   SSDR9 &7.19 &0.25 &2.85$\times 10^{-5}$ &1.02$\times 10^{-3}$ 
                                                              &4.86$\times 10^{-3}$ &0.59&5.41$\times 10^{-4}$ &-2.6&7.00 &3.05 & 2.10&0.586\\
\hline
  SSDR8 &3.33 &0.55 &1.75$\times 10^{-5}$  &1.11$\times 10^{-3}$  &2.01$\times 10^{-3}$ &0.31&1.79$\times 10^{-4}$ &-2.6&3.15 &2.22&5.52&0.871 \\
   SSDR10 &2.59 &0.36 &1.48$\times 10^{-4}$  &2.23$\times 10^{-3}$  &5.21$\times 10^{-3}$ &0.76&9.62$\times 10^{-3}$ &-2.6&2.74 &1.89 &4.28&2.84\\
  DEML10 & 3.89& 0.51&8.59$\times 10^{-5}$ & 1.87$\times 10^{-3}$ & 5.77$\times 10^{-3}$ &0.77&0.00&-2.6&1.90&2.02 &3.24&1.49 \\
  DEML34 & 1.22& 0.52& 2.67$\times 10^{-4}$ & 2.51$\times 10^{-3}$ &1.39$\times 10^{-2}$ &1.67&8.17$\times 10^{-3}$ &-2.6&3.04 &1.10 &1.81&1.92\\
  DEML86 &1.98 & 0.59& 3.74$\times 10^{-4}$ & 2.92$\times 10^{-3}$ &1.41$\times 10^{-2}$ &1.74&0.00&-2.6&3.16 &1.31&2.07 &2.65\\
  DEML323 & 3.34&0.44 &9.73$\times 10^{-5}$ &1.54$\times 10^{-3}$  &4.69$\times 10^{-3}$ &0.63&4.64$\times 10^{-3}$ &-2.6&3.33 &1.61 &3.28&2.07\\
  \hline
  $<$All$>$ & & & && && & &3.59 & 2.13& & \\
  $<$Diff.$>$ & && && & & & &5.34  & 2.68 && \\
  $<$Mol.$>$ & && && & &  & &3.97 & 2.92 && \\
  $<$HII$>$ & & &&& & &  & &2.89  & 1.69 &&\\
  \hline
  \hline
\multicolumn{13}{c}{DBP90 ($\alpha=-2.6, a_{min}, a_{max}$) - 600-Myr RF}\\
\hline
 Region  & $\chi^2/dof$ &$X_{ISRF}$  &$Y_{PAH^0}$  &$Y_{VSG}$
                                                              &$Y_{BG}$
                                                                        & $Y_{dust,tot}$ &
                                                                          $I_{NIR\,cont.}$
                                                                                            &
                                                                                              $\alpha$&
                                                                                                        $a_{min}$ & $a_{max}$ &$\frac{Y_{VSG}}{Y_{BG}}$&$\frac{Y_{PAH}}{Y_{BG}}$\\
         &&&&&&($10^{-2}$)&&&&($\times 10$)&($10^{-1}$) &($10^{-2}$) \\

  \hline
  SSDR1 &1.68 & 1.09&2.51$\times 10^{-4}$  &9.47$\times 10^{-4}$  &3.64$\times 10^{-3}$ &0.48&3.62$\times 10^{-3}$ &-2.6 & 5.98 &2.62 &2.60 &6.90\\
  SSDR5 & 4.92& 1.30&2.22$\times 10^{-4}$  & 1.28$\times 10^{-3}$
                                                              &8.41$\times
                                                                10^{-3}$
  &0.99&3.13$\times 10^{-4}$ &-2.6&2.44 &2.26&1.52 &2.64\\
  \hline
  SSDR7 &2.91 &3.42 &1.20$\times 10^{-4}$  &2.72$\times 10^{-4}$ 
                                                              &3.74$\times 10^{-3}$ &0.41&6.40$\times 10^{-3}$ &-2.6&4.45 &1.87 &
                                                                                                     0.727&3.21\\
   SSDR9 &7.30 &1.25 &4.80$\times 10^{-5}$ &1.43$\times 10^{-3}$ 
                                                              &4.95$\times 10^{-3}$ &0.64&9.71$\times 10^{-4}$ &-2.6&6.77&2.52 & 2.89&0.970\\
\hline
  SSDR8 &3.36 &3.17 &2.77$\times 10^{-5}$  &1.36$\times 10^{-3}$  &1.93$\times 10^{-3}$ &0.33&4.12$\times 10^{-4}$ &-2.6&2.89 &1.91 &7.5 &1.44\\
   SSDR10 &2.61 &1.92&2.68$\times 10^{-4}$  &2.95$\times 10^{-3}$  &5.01$\times 10^{-3}$ &0.82&1.11$\times 10^{-2}$ &-2.6&2.76 &1.61 &5.89&5.35\\
  DEML10 & 3.86& 2.83&1.65$\times 10^{-4}$ & 2.35$\times 10^{-3}$ & 5.44$\times 10^{-3}$ &0.80&0.00&-2.6&1.87 &1.76&4.32 &3.03\\
  DEML34 & 1.37& 2.64& 5.05$\times 10^{-4}$ & 3.58$\times 10^{-3}$ &1.36$\times 10^{-2}$ &1.77&1.12$\times 10^{-2}$ &-2.6&2.99 &0.947 &2.63&3.71\\
  DEML86 &2.13 & 2.97& 7.59$\times 10^{-4}$ & 4.10$\times 10^{-3}$ &1.40$\times 10^{-2}$ &1.89&0.00&-2.6&3.38 &1.09 &2.93 &5.42\\
  DEML323 & 3.37&2.32 &1.76$\times 10^{-4}$  &2.07$\times 10^{-3}$  &4.52$\times 10^{-3}$ &0.68&5.95$\times 10^{-3}$ &-2.6&3.25&1.38 &4.58 &3.89\\
  \hline
  $<$All$>$ & & & && && & &3.68  & 1.80 && \\
  $<$Diff.$>$ & && && & & & &5.61 & 2.20&& \\
  $<$Mol.$>$ & & &&& & &  & &4.21 & 2.44&&\\
  $<$HII$>$ & & &&& & &  & &2.86 & 1.45&&  \\
  \hline
  \hline

\end{tabular}}
\end{center}
\end{table*}%[!h]}

\begin{table*}%[!h]
\caption{Ratio of the small grain component emission to the total dust
  emission for each best-fit model using the 60-Myr (top table) and 600-Myr
  RF (bottom table), at 250, 500, 850 and 1100
  $\mic$. \label{table_vsg_annexe}}
\begin{center}
\resizebox{\textwidth}{!}{%
\begin{tabular}{lcccc|cccc|cccc|cccc}
\hline
 \hline
%\multicolumn{10}{c}{DBP90} \\ 
% \hline
 &&\multicolumn{13}{c}{60-Myr RF}&&\\
   &\multicolumn{4}{c}{AJ13} & \multicolumn{4}{c}{MC11} & \multicolumn{4}{c}{DL07}& \multicolumn{4}{c}{DBP90} \\
 Best modeling & \multicolumn{4}{c}{($\alpha$, $a_{min}$, $a_{max}$)} & \multicolumn{4}{c}{($a_0$)} & \multicolumn{4}{c}{($a_0$)} &
                                                           \multicolumn{4}{c}{($\alpha=-2.6$,
                                                           $a_{min}$, $a_{max}$)} \\ 
& \multicolumn{4}{c}{$I^{a-C}_{\nu}/I^{tot}_{\nu}$ ($\%$)} &
                                         \multicolumn{4}{c}{$I^{SamC}_{\nu}/I^{tot}_{\nu}$ ($\%$)} &\multicolumn{4}{c}{$I^{graph.}_{\nu}/I^{tot}_{\nu}$ ($\%$)} & \multicolumn{4}{c}{$I^{VSG}_{\nu}/I^{tot}_{\nu}$ ($\%$)} \\
Wavelengths & 250 & 500 & 850 & 1100 & 250 & 500 & 850 & 1100 & 250 & 500 & 850 & 1100 & 250 &
                                                                500 &
                                                                850 &
                                                                1100 \\
  \hline
%  Milky-Way & 7.1 & 7.9 & 8.9 & 9.4 & 1.4 & 1.4 & 1.7 & 1.8 & 1.7 & 1.5 & 1.3 & 1.2 & 4.3 &
%                                                              6.4 &
%                                                              10.6 &
%                                                              13.7 \\
%  \hline
  SSDR1 & 1.6 & 1.2 & 1.0 &1.0  &  10.3 & 8.7 & 8.7 & 8.8 &
                                                                   69.2
                                                                   &
                                                                   53.3
                                                                   &
                                                                   44.7
                                                                   &
                                                                   41.3& 
                                                                20.0
                                                                         &
                                                                           27.1 &                                                               
                                                               39.7 & 47.1
                                                                 \\
 SSDR5 & 1.9 & 1.9 &2.1 & 2.2 & 2.9 & 2.4 & 2.7 & 2.9 & 1.2 &
                                                               1.0 &
                                                               1.1 &
                                                               1.1 &
                                                                     9.2
                                                                     &
                                                                     12.1
                                                                     &19.0
                                                                     & 23.9 \\
  \hline
  SSDR7 &  2.8 & 3.4 & 4.1 & 4.5 & 7.5 & 6.6 & 7.5 & 8.2 & 5.2
                                                                 &
                                                                 4.4
                                                                 &
                                                                 4.1
                                                                 &
                                                                 4.0 & 
                                                                 5.5
                                                                       &
                                                                       8.7 
                                                                  & 14.6
                                                                  
                                                                  & 18.8                                                               
  \\
  SSDR9 & 7.1 & 6.0 & 5.5 & 5.4 & 8.1 & 7.0 & 7.0 & 7.0 & 59.1
                                                                &
                                                                44.0
                                                                &
                                                                36.0
                                                                &
                                                                33.0 &
                                                                       21.8
                                                                       &
                                                                       29.1
                                                                       & 42.0
                                                        & 49.4 \\
\hline
  SSDR8 & 12.0 & 9.4 & 8.6 & 8.4 & 17.1 & 15.1 &15.7 & 16.0 &
                                                                     28.6
                                                                     &
                                                                     22.6
                                                                     &
                                                                     22.9
                                                                     &
                                                                     23.5
        & 35.8 & 46.1 & 60.2 &67.1 
  \\
   SSDR10 & 14.6 & 11.9 & 11.0 &10.7 & 14.2 & 12.8 & 13.3 &
                                           13.7 & 6.4 & 5.5 & 5.9 &
                                                  6.1 & 22.0 & 26.8 
                                                                & 38.0
                                                                & 45.0
                                                               \\
  DEML10 & 6.4 & 5.7 & 5.7 & 5.7 & 4.2 & 3.9 & 4.2 & 4.4 & 
                                                                 4.2
                                                                 &
                                                                 3.6
                                                                 &
                                                                 3.9
                                                                 &4.1
                                                                 & 
                                                                 21.0
                                                                   &
                                                                   27.8 
                                                                 & 40.0
                                                                 & 47.3
                                                                 \\
  DEML34 &  15.3 & 19.2 & 25.5 & 29.3 & 6.7 & 6.4 & 6.9 & 7.2 &
                                                                      3.7
                                                                      &
                                                                      3.4
                                                                      &
                                                                      3.8
                                                                      &
                                                                      4.0
        & 5.1 & 5.9 & 9.1 & 11.6 \\
  DEML86 & 26.1 & 31.1 & 38.9 & 43.2 & 16.0 & 14.5 & 17.0 & 18.6
  & 5.9 & 5.3 & 5.7 & 6.0 & 8.1 & 10.1 & 15.6 & 19.7
                                                                \\
  DEML323 & 34.0 & 38.3 & 44.7 & 47.9 & 13.9 & 12.7 & 13. 6 &
                                           14.0 & 6.8 & 5.9 & 6.4 &
                                                  6.7 & 15.4 & 18.9
                                                        & 27.9 & 33.9 \\
\hline
  \hline
 &&\multicolumn{13}{c}{600-Myr RF}&&\\
  SSDR1 & 2.7 &2.1 & 1.9 &1.9  & 37.4 & 34.9 & 36.6 & 37.8 & 
                                                                 72.5
                                                                     &
                                                                     64.8                                                                 
                                                                  & 58.3
                                                                 & 55.2
                                                                  
        & 31.9 & 46.6 & 61.6 & 68.4 \\
 SSDR5 & 23.9 & 28.0 & 32.4 & 34.7 & 4.9& 3.9 & 5.4 & 5.7 & 
                                                                  3.0
                                                                   &
                                                                   2.9 
                                                                  & 3.0
                                                                   & 3.1
                                                                   
        & 17.8 & 28.2 & 41.8 &49.2 \\
  \hline
  SSDR7 & 10.8 & 13.1 &15.3 & 16.3 & 7.3 & 7.0 & 7.0 & 7.0  &
                                                                     15.7
                                                                     &
                                                                     14.0 
                                                               & 12.3 
                                                               &11.6
                                                              & 11.3
                                                                &
                                                                19.9 
                                                                &
                                                                31.3 
                                                                  & 37.9
  \\
  SSDR9 & 0.9 & 0.7 & 0.7 &0.7 & 34.9 & 32.6 & 32.5 & 32.5 & 
                                                               45.5 &
                                                                    38.3 
                                                               & 32.4
                                                                & 29.7
                                                                &34.8
                                                                  & 50.1
                                                                       & 64.9
                                                                       &
                                                                  71.4
  \\
\hline
  SSDR8 & 87.4 & 90.3 & 92.8 & 93.4 & 37.5 & 36.3 &37.2 & 37.7
  &
    33.2 & 31.1 &31.7 & 32.2                                                              
        & 53.6 & 69.1 & 80.4 & 84.7 \\
   SSDR10 & 4.6 & 3.5 & 3.1 & 2.9 & 29.5 & 28.8 & 29.6 & 30.0 &
                                                                      24.4
                                                                      &
                                                                      23.0 
                                                                     & 23.9
                                                                       & 24.4
       & 36.8 & 50.2 & 64.5&71.0 \\
  DEML10 & 8.5 & 7.3 &6.8 & 6.6 & 10.1 &10.2 &10.7 & 11.0 & 
                                                                11.3
                                                                     & 10.9
                                                                  & 11.5
                                                                 & 11.9
                                                                 & 36.3
                                                                   & 51.7
                                                                 & 66.1
                                                               & 72.4
                                                                 \\
  DEML34 &  25.4 & 31.8 & 39.1 & 43.1 & 12.3& 12.6 & 13.4 &
                                           13. 7 & 8.8 & 8.9&9.7
                                                                    & 10.0
         & 10.6 & 15.8 & 24.6 & 30.\\
  DEML86 & 38.3 & 45.6 & 53.6 & 57.7 &  14.3 & 14.5 & 15.1 &
                                          15.3  
  & 13.2 & 12.8 & 12.8 &12.7 & 15.7 & 23.7& 35.5 
                                                                & 42.5
                                                                \\
  DEML323 & 24.6 & 23.8 &23.9 & 23.9 & 24.3 & 24.3 & 25.7 &
                                           26.3  
  & 17.5 & 16.9 & 17.7& 18.1 & 27.9 & 39.7 & 54.2 & 61.4 \\
\hline
  \hline

\end{tabular}}%[!h]}
\end{center}
\tablefoot{The second line indicates the dust size parameters of
  the small grain component used in the modeling. }
%\resizebox{\textwidth}{!}{%
\end{table*}

\begin{figure*}%[!h]
 \begin{center}
\includegraphics[width=18cm]{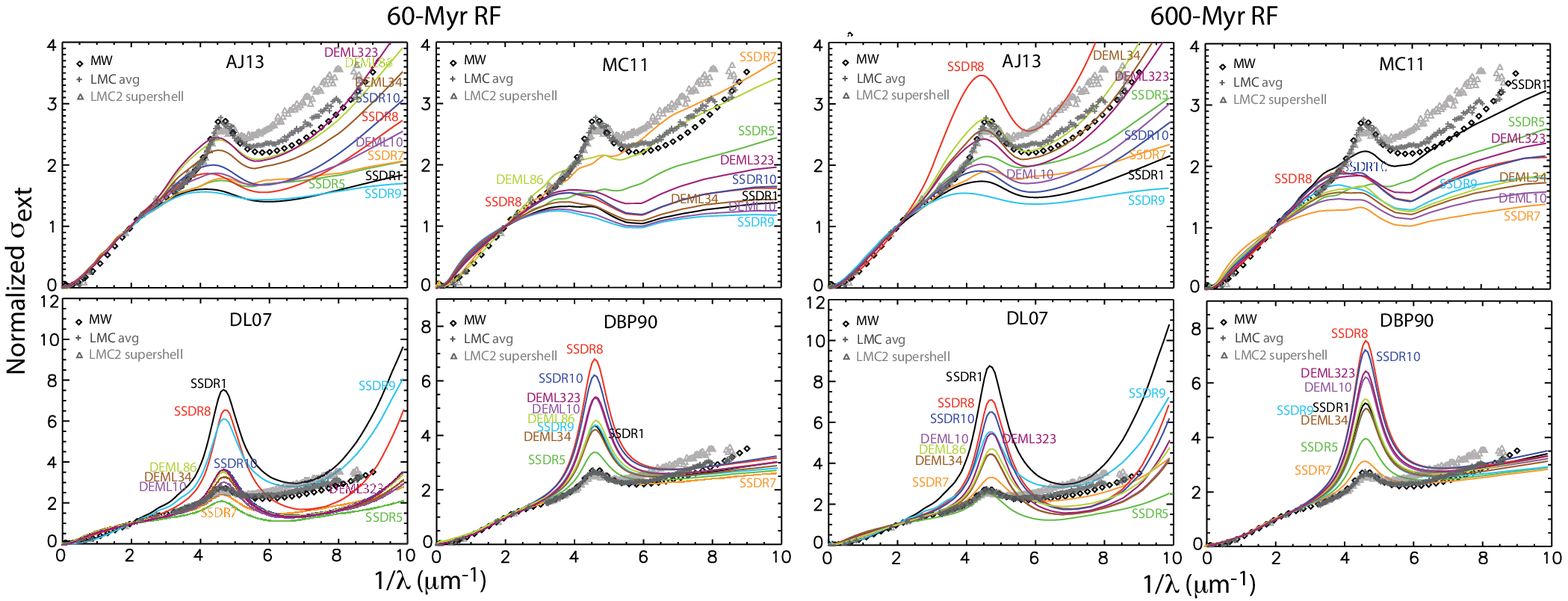}
\caption{Extinction curves for each region, derived from the different
  dust models, with 60-Myr RF (left) and 600-Myr (right). The
  averaged Galactic and LMC extinction curves are given in black
  diamonds and dark gray crosses for comparison. The LMC2 supershell
  extinction curve is presented with light gray triangles. Caution: the scales have been chosen to show the difference between
the different models in a clear way, and one should take this into account in the comparisons.\label{fig_ext_annexe}}
\end{center}
\end{figure*}

   \end{appendix}

 \end{document}